\documentclass[prd,reprint,superscriptaddress,altaffillsymbol,amsmath,nofootinbib,longbibliography]{revtex4-1}
\pdfoutput=1
\usepackage{lipsum}
\usepackage{ifthen}
\usepackage{epsfig}
\usepackage{booktabs} 
\usepackage{natbib}
\usepackage{wasysym}

\usepackage{slashed} 
\usepackage{bbm}
\usepackage{mathrsfs}
\usepackage{amssymb}
\usepackage{dsfont}
\usepackage[export]{adjustbox}

\usepackage{comment}
\usepackage{glossaries}

\usepackage{xcolor}

\definecolor{OurBlue}{RGB}{13,115,178}
\definecolor{OurRed}{RGB}{213,91,1}     
\definecolor{OurOrange}{RGB}{222,143,6}    
\definecolor{OurGreen}{RGB}{19,158,114}   
\definecolor{OurPurple}{RGB}{204,120,188}  
\definecolor{OurBrown}{RGB}{202,145,96}   
\definecolor{OurLightBlue}{RGB}{86,180,233}   
\definecolor{OurPink}{RGB}{251,175,228}  
\definecolor{OurGray}{RGB}{148,148,148}  
\definecolor{OurYellow}{RGB}{236,225,50} 

\usepackage[
colorlinks=true, 
linkcolor=OurBlue,
citecolor=OurRed,
urlcolor=OurBlue,
]{hyperref}

\allowdisplaybreaks

\newcommand{\erf}{\mathop{\mathrm{erf}}}
\newcommand{\beqra}{\begin{eqnarray}}
\newcommand{\eeqra}{\end{eqnarray}}
\newcommand{\beq}{\begin{equation}}
\newcommand{\eeq}{\end{equation}} 
\newcommand{\dd}{\mathrm{d}}

\newcommand{\boldell}{\boldsymbol{\ell}}
\newcommand{\ket}[1]{\left| #1 \right\rangle}

\newcommand{\braket}[2]{\left\langle #1 \middle| #2 \right\rangle}

\newcommand{\Real}[1]{\textrm{Re}\left[#1\right]}

\newcommand{\angstrom}{\text{\normalfont\AA}}

\renewcommand{\epsilon}{\varepsilon}

\renewcommand{\vec}[1]{\mathbf{#1}}
\renewcommand{\bar}{\overline}

\newcommand{\braOket}[3]{\left\langle #1 \middle| #2 \middle| #3 \right\rangle}

\newcommand{\vPerpEl}{\mathbf{v}_{\rm el}^\perp}

\newcount\colveccount

\newcommand*\columnvector[1]{
        \global\colveccount#1
        \begin{pmatrix}
        \colvecnext
}
\def\colvecnext#1{
        #1
        \global\advance\colveccount-1
        \ifnum\colveccount>0
                \\[2mm]
                \expandafter\colvecnext
        \else
                \end{pmatrix}
        \fi
}

\begin{document}

\title{Direct searches for general dark matter-electron interactions with graphene detectors: \\Part I.   Electronic structure calculations}

\author{Riccardo Catena}
\email{catena@chalmers.se}
\affiliation{Chalmers University of Technology, Department of Physics, SE-412 96 G\"oteborg, Sweden}

\author{Timon Emken}
\email{timon.emken@fysik.su.se}
\affiliation{The Oskar Klein Centre, Department of Physics, Stockholm University, AlbaNova, SE-10691 Stockholm, Sweden}

\author{Marek Matas}
\email{marek.matas@mat.ethz.ch}
\affiliation{Department of Materials, ETH Z\"urich, CH-8093 Z\"urich, Switzerland}

\author{Nicola A. Spaldin}
\email{nicola.spaldin@mat.ethz.ch}
\affiliation{Department of Materials, ETH Z\"urich, CH-8093 Z\"urich, Switzerland}

\author{Einar Urdshals}
\email{urdshals@chalmers.se}
\affiliation{Chalmers University of Technology, Department of Physics, SE-412 96 G\"oteborg, Sweden}

\begin{abstract}
We develop a formalism to describe electron ejections from graphene-like targets by dark matter (DM) scattering for general forms of scalar and spin 1/2 DM-electron interactions and compare their applicability and accuracy within the density functional theory (DFT) and tight binding (TB) approaches. This formalism allows for accurate prediction of the daily modulation signal expected from DM in upcoming direct detection experiments employing graphene sheets as the target material.  A key result is that the physics of the graphene sheet and that of the DM and the ejected electron factorise, allowing for the rate of ejections from all forms of DM to be obtained with a single graphene response function. We perform a comparison between the TB and DFT approaches to modeling the initial state electronic wavefunction within this framework, with DFT emerging as the more self-consistent and reliable choice due to the challenges in the embedding of an appropriate atomic contribution into the TB approach.
\end{abstract}
\maketitle

\section{Introduction}
\label{sec:introduction}

Dark Matter (DM) plays a key role in simultaneously explaining otherwise anomalous physical phenomena that occur on extremely different astronomical length scales~\cite{Bertone:2016nfn}.~For example, 
it provides the initial density fluctuations that trigger the formation of cosmic structures and generate the anisotropy pattern observed in the cosmic microwave background temperature and polarisation maps~\cite{Planck:2018vyg}.
It bends the light emitted by distant astrophysical sources, giving rise to spectacular gravitational lensing events~\cite{Clowe:2006eq} and, furthermore, it provides the mass required to support the flat rotation curves of spiral galaxies~\cite{Persic:1995ru}.

Despite these remarkable observations, we still do not know what DM is made of.~The leading hypothesis in astroparticle physics is that DM is made of unidentified, yet-to-be-discovered particles~\cite{Bertone:2016nfn}.~While this simple assumption can collectively explain all phenomena listed above, the hypothetical particles forming our universe's DM component have so far escaped detection.~This dichotomy between solid gravitational evidence and lack of microscopic description makes the search for the ``DM particle'' a top priority.  

A prominent class of experiments searching for the DM particle relies on the direct detection technique~\cite{Drukier:1983gj,Goodman:1984dc}.~This technique seeks for rare interactions between DM particles from the Milky Way and detector materials located deep underground in low background environments. As far as the DM-material interaction is concerned, DM direct detection experiments have until recently focused on the search for nuclear recoil events induced by the scattering of Weakly Interacting Massive Particles (WIMPs) in crystals, or liquid noble gases~\cite{Schumann:2019eaa}.~Consequently, direct detection experiments have so far only probed DM particles of mass above about 1~GeV, as lighter particles would not be able to cause an observable nuclear recoil.~However, the lack of detection of WIMPs has recently motivated the exploration of alternative experimental approaches that are better suited to probe DM particles of sub-GeV mass~\cite{Battaglieri:2017aum}.~It is in this context that DM direct detection experiments sensitive to DM-induced electronic transitions or electron ejections in materials play a central role.

Materials that have been proposed to search for sub-GeV DM particles via DM-electron interactions include liquid argon~\cite{Agnes:2018oej} and xenon~\cite{Essig:2012yx,Essig:2017kqs,Aprile:2019xxb}, semiconductor crystals~\cite{Essig:2011nj,Graham:2012su,Lee:2015qva,Essig:2015cda,Crisler:2018gci,Agnese:2018col,Abramoff:2019dfb,Aguilar-Arevalo:2019wdi,Amaral:2020ryn,Andersson:2020uwc}, 3D Dirac materials~\cite{Hochberg:2017wce,Geilhufe:2019ndy}, graphene~\cite{Hochberg:2016ntt,Geilhufe:2018gry} and carbon nanotubes~\cite{Capparelli:2014lua, Cavoto:2016lqo, Cavoto:2017otc, Cavoto:2019flp}, to name a few.~In this context, anisotropic media, particularly materials with anisotropic Fermi velocities such as graphene and carbon nanotubes, are interesting, as the associated rate of DM-induced electron ejections exhibits an enhanced daily modulation.
This enhancement is caused by the structural anisotropy of the target material in combination with its relative orientation to the DM~wind.~Given that such a daily modulation is not present in typical experimental backgrounds, it would thus be a smoking gun for DM signal.~A proposed experiment to search for DM-induced electron ejections from graphene sheets or arrays of carbon nanotubes, which is currently in the conceptual design stage, is the Princeton Tritium Observatory for Light, Early-Universe, Massive-Neutrino Yield, or PTOLEMY~\cite{Betts:2013uya,PTOLEMY:2018jst,PTOLEMY:2022ldz}.

PTOLEMY's experimental design employs a large area surface-deposition tritium target coupled to a graphene substrate to detect the cosmic neutrino background via the observation of single electrons produced in the neutrino absorption by tritium atoms~\cite{Betts:2013uya}.~The coupling between tritium and graphene reduces the energy dispersion of the final state electrons by about an order of magnitude compared to the case of molecular tritium~\cite{PTOLEMY:2018jst}.~A small electron energy dispersion allows for better discrimination between electrons produced by neutrino absorption and electrons populating the tail of the tritium $\beta$-decay spectrum~\cite{PTOLEMY:2018jst}.~For this experimental setup to work, it is crucial to experimentally validate the use of graphene as a substrate by accurately measuring the electron-graphene interaction properties.~In this intermediate stage of the PTOLEMY experimental program, when tritium target and graphene substrate are still decoupled, PTOLEMY can also operate as a directional MeV-scale DM detector.~Specifically, two experimental configurations have been proposed.~In the first one, a sample of stacked graphene sheets is considered (PTOLEMY-G$^3$)~\cite{Hochberg:2016ntt}.~Once an electron is ejected from one of the graphene sheets, it drifts in an external electric field until it reaches a calorimeter at the edge of the detector volume.~This configuration allows for a full reconstruction of the final state electron kinematics.~In a second experimental configuration (PTOLEMY-CNT)~\cite{Capparelli:2014lua, Cavoto:2016lqo, Cavoto:2017otc, Cavoto:2019flp,PTOLEMY:2022ldz}, an array of single- or multi-wall metallic carbon nanotubes is positioned in vacuum.~When an electron is ejected from one of the nanotubes, it is driven by an electric field to the detection region, and recorded by a single electron sensor.~The idea of adopting graphene sheets and carbon nanotubes as targets for directional, light DM detection has recently been further developed by the ``Graphene-FET'' and ``dark-PMT'' projects, respectively~\cite{Apponi:2021lyd}.

The possibility of using graphene or carbon nanotubes as directional detectors sensitive to DM-induced electron ejections motivates an accurate and comprehensive modelling of DM scattering by electrons bound in this class of anisotropic materials.~In contrast, the rate of DM-induced electron ejections from graphene sheets~\cite{Hochberg:2016ntt} and from carbon nanotubes~\cite{Capparelli:2014lua, Cavoto:2016lqo, Cavoto:2017otc, Cavoto:2019flp} has so far been computed assuming that the amplitude for DM-electron interactions depends on the momentum transfer only, and that the DM-electron interaction is spin-independent.~This is a rather restrictive assumption, which can easily be violated, e.g.~in models where DM has a non negligible magnetic or anapole moment~\cite{Catena:2019gfa}.~Furthermore, current estimates rely on the tight-binding approximation and have not been validated against first-principles calculations.

The purpose of this work is to extend and improve the formalism currently used to model the scattering of DM particles by electrons bound to graphene sheets.~First, we extend the current formalism to virtually arbitrary DM-electron interactions by using the non-relativistic effective theory framework we developed in~\cite{Catena:2019gfa, Catena:2022fnk}, and recently applied by the XENON group in an analysis of the electron recoil data reported in~\cite{XENON:2021qze}.~Second, we improve the existing formalism by performing state-of-the art density functional theory (DFT) calculations in order to accurately model the electronic properties of graphene.

We expect that the formalism and findings we present here will be useful in the design of the PTOLEMY detector, as well as for the development of the Graphene-FET and dark-PMT projects.~However, the relevance of our formalism goes beyond its application to these experimental concepts, as it can also be straightforwardly used to study the ejection of electrons in other experimental settings, where the final state is a free electron that can be described by a plane wave.~We leave this exploration for future work.

This paper is the first of a two part series studying DM-electron scatterings in graphene targets. In this paper (Paper I), we lay the theoretical foundations. In the companion Paper II, we will focus on more explicit experimental setups and sensitivity studies~\cite{PaperII}.
In addition, the software tools \texttt{Darphene} and \texttt{QEdark-EFT} developed for TB and DFT calculations respectively are publicly available~\cite{Darphene,QEdark-EFT}.

This article is organized as follows.~In Sec.~\ref{sec:formalism} we introduce our general formalism for modeling the ejection of electrons by the scattering of DM particles in two- and three-dimensional periodic systems.~In Sec.~\ref{sec:electronic}, we describe the detailed electronic structure calculations we performed for graphene, both within the tight-binding-approximation and within DFT.~We apply these results to study the daily modulation of the DM-induced electron ejection rate for a hypothetical graphene detector in Sec.~\ref{sec:results} and conclude in Sec.~\ref{sec:conclusions}.~We complement this work with appendices where we provide analytic formulae that are useful for evaluating our general electron ejection rates.

\section{Rate of electron ejection caused by general dark matter-electron interactions}
\label{sec:formalism}
In this section, we derive an expression for the rate of electron ejection caused by general DM-electron interactions in periodic systems.~In Sec.~\ref{sec:electronic}, we will perform the detailed electronic structure calculations that will enable us in Sec.~\ref{sec:results} to apply this general expression to the specific, experimentally relevant case of graphene.

\subsection{General formalism}

We are interested in processes in which a DM~particle~$\chi$ of mass~$m_\chi$, initial velocity in the detector rest frame $\mathbf{v}$, and momentum~$\mathbf{p} = m_\chi \mathbf{v}$ is scattered by an electron in initial state~$\ket{\mathbf{e}_1}$. During the interaction, the DM~particle transfers momentum~$\mathbf{q} = \mathbf{p}-\mathbf{p}^\prime$ to the electron, where~$\mathbf{p}^\prime$ is the final DM momentum, and causes an electronic transition from $\ket{\mathbf{e}_1}$ to the final state~$\ket{\mathbf{e}_2}$.~In the notation of~\cite{Catena:2019gfa,Catena:2021qsr}, the rate $R_{1\rightarrow 2}$ for this transition is 
\begin{align}
R_{1\rightarrow 2}&=\frac{n_{\chi}}{16 m^2_{\chi} m^2_e} \,
\int \frac{{\rm d}^3 \mathbf{q}}{(2 \pi)^3} \int {\rm d}^3 \mathbf{v}\, f_{\chi}(\mathbf{v})  \nonumber \\
&\times  (2\pi) \delta(E_f-E_i) \overline{\left| \mathcal{M}_{1\rightarrow 2}\right|^2}\, , 
\label{eq:transition rate}
\end{align}
where $m_e$ is the electron mass, $n_\chi = \rho_\chi / m_\chi$ is the local DM~number density, $\rho_\chi = 0.4\,\text{GeV cm}^{-3}$ is the local DM mass density, and~$f_\chi(\mathbf{v})$ is the local DM velocity distribution boosted to the detector rest frame.~For $f_\chi(\mathbf{v})$, we assume a truncated Maxwell-Boltzmann distribution, as in the so-called Standard Halo Model (SHM)~\cite{Baxter:2021pqo}. Specifically, 
\begin{align}
    f_\chi(\mathbf{v})&= \frac{1}{N_{\rm esc}\pi^{3/2}v_0^3}\exp\left[-\frac{(\mathbf{v}+\mathbf{v}_e)^2}{v_0^2} \right]
    \nonumber\\
    &\times 
    \Theta\left(v_{\rm esc}-|\mathbf{v}+\mathbf{v}_e|\right)\,,
    \label{eq:df}
\end{align}
and we take $v_0 = |\vec{v}_0| =~238$~km~s$^{-1}$~\cite{Baxter:2021pqo} for the local standard of rest speed, and $v_\mathrm{esc}=544$~km~s$^{-1}$~\cite{Baxter:2021pqo} for the galactic escape speed.~Following~\cite{Geilhufe:2019ndy}, we express the Earth's velocity with respect to the galactic centre, $\mathbf{v}_e$, in a coordinate system with $z$-axis in the $\mathbf{v}_0+\mathbf{v}_{\odot}$ direction, $\mathbf{v}_{\odot}$ the Sun's peculiar velocity and $v_e=|\mathbf{v}_0+\mathbf{v}_{\odot}|\simeq 250.5$~km~s$^{-1}$ ~\cite{Baxter:2021pqo},
\begin{align}
\mathbf{v}_e = v_e \left(
\begin{array}{c}
\sin\alpha_e \sin\beta \nonumber\\
\sin\alpha_e \cos\alpha_e (\cos\beta -1) \nonumber\\
\cos^2\alpha_e + \sin^2\alpha_e\cos\beta
\end{array}
\right) \,,
\end{align}
where $\alpha_e=42^\circ$, $\beta = 2\pi\, t/{\rm day}$, and $t$ is the time variable. Finally, we also introduced the normalization constant,
\begin{align}
N_{\rm esc}&\equiv \erf(v_{\rm esc}/v_0)-{2\over \sqrt{\pi}} {v_{\rm esc}\over v_0}\exp\left(-\frac{v_{\rm esc}^2}{v_0^2}\right)\, .
\end{align}
The total initial (final) energy~$E_i$ ($E_f$) in Eq.~(\ref{eq:transition rate}) is the sum of the DM and electronic energies,
\begin{align}
    E_i &= \frac{|\mathbf{p}|^2}{2m_\chi}+E_1\, ,\quad   E_f = \frac{|\mathbf{p}-\mathbf{q}|^2}{2m_\chi}+E_2\, ,
\end{align}
where~$E_1$ ($E_2$) is the energy eigenvalue of the electronic state~$\ket{\mathbf{e}_1} (\ket{\mathbf{e}_2})$. We denote the corresponding wave functions by~$\psi_1$ and $\psi_2$, and their associated Fourier transforms by $\widetilde{\psi}_1$ and $\widetilde{\psi}_2$, respectively.
These electron wave functions enter the electron transition amplitude~$\mathcal{M}_{1\rightarrow 2}$, defined as in Eq.~(14) of~\cite{Catena:2019gfa} by the  integral,
\begin{align}
     \mathcal{M}_{1\rightarrow 2} &=
     \int  
     \frac{{\rm d}^3 \boldell}{(2 \pi)^3} \, \widetilde{\psi}_2^*(\boldsymbol{\ell}+\mathbf{q})  
\mathcal{M}(\boldsymbol{\ell},\mathbf{p},\mathbf{q})
\widetilde{\psi}_1(\boldsymbol{\ell}) \, ,
\label{eq:transition amplitude}
\end{align}
where $\mathcal{M}(\boldsymbol{\ell},\mathbf{p},\mathbf{q})$ is the free electron scattering amplitude, and $\boldell$ the initial state electron momentum.~Here, we use momentum conservation to eliminate explicit dependence on the final state electron momentum from $\mathcal{M}$.~Furthermore, since the scattering of Milky Way DM particles by free electrons is expected to be non-relativistic, we use the Galilean invariance of $\mathcal{M}$ to write $\mathcal{M}(\boldsymbol{\ell},\mathbf{p},\mathbf{q}) = \mathcal{M}(\mathbf{q},\mathbf{v}^\perp_{\rm el})$, where $\mathbf{v}^\perp_{\rm el}=\mathbf{v}- \mathbf{q}/(2\mu_{\chi e}) -\boldsymbol{\ell}/m_e$ and $\mu_{\chi e}$ is the DM-electron reduced mass.~Finally, we expand $\mathcal{M}$ at linear order in $\boldsymbol{\ell}/m_e$, and write it as follows~\cite{Catena:2019gfa}
\begin{align}
    \mathcal{M}(\mathbf{q},\mathbf{v}^\perp_{\rm el}) \approx \left. \mathcal{M}(\mathbf{q},\mathbf{v}^\perp_{\rm el}) \right|_{\boldell=\mathbf{0}} + \boldell \cdot \left.\nabla_{\boldell} \mathcal{M}(\mathbf{q},\mathbf{v}^\perp_{\rm el})  \right|_{\boldell=\mathbf{0}}\, .
\end{align}
This expansion allows us to express the transition amplitude as
\begin{align}
    \mathcal{M}_{1\rightarrow 2} &= \left. \mathcal{M}(\mathbf{q},\mathbf{v}^\perp_{\rm el}) \right|_{\boldell=\mathbf{0}} f_{1\rightarrow 2}(\mathbf{q}) \nonumber\\
    &+ m_e \left.\nabla_{\boldell} \mathcal{M}(\mathbf{q},\mathbf{v}^\perp_{\rm el})  \right|_{\boldell=\mathbf{0}} \cdot \mathbf{f}_{1\rightarrow 2}(\mathbf{q}) \,,
    \label{eq:M_expansion}
\end{align}
where we introduce the scalar and vectorial overlap integrals,
\begin{align}
       f_{1\rightarrow 2}(\mathbf{q}) &\equiv \int\dd^3 \textbf{x}\, \psi_2^*(\mathbf{x})\,e^{i\mathbf{q}\cdot\mathbf{x}}\,\psi_1\left( \mathbf{x} \right)\, , \label{eq: scalar atomic form factor}\\
    \mathbf{f}_{1\rightarrow 2}(\mathbf{q}) &\equiv \int\dd^3 \textbf{x}\, \psi_2^*(\mathbf{x})\,e^{i\mathbf{q}\cdot\mathbf{x}}\,\frac{i\nabla}{m_e}\psi_1\left( \mathbf{x} \right) \, . \label{subeq: f vectorial}
\end{align}
In order to evaluate the expressions above, we need to specify the initial and final electron wave functions.

We begin by specifying the final-state electron wave function for the case in which the electron is ejected by the DM particle. In this case, the state $\ket{\mathbf{e}_2}$ asymptotically approaches a free particle of momentum~$\mathbf{k}^\prime$.~Consequently, the wave function $\psi_2(\mathbf{x})$ can be approximated by the plane wave
\begin{align}
    \psi_2(\mathbf{x}) \rightarrow \psi_{\mathbf{k}^\prime}(\mathbf{x}) = \frac{1}{\sqrt{V}} e^{i\mathbf{k}^\prime\cdot \mathbf{x}}\, ,
    \label{eq:pw}
\end{align}
which is normalised to one over a finite volume $V$.~For electrons initially bound in graphene, this plane wave approximation has been validated by comparing results from angular-resolved photoemission spectroscopy (ARPES) measurements with simulated photoemission intensity maps, for which excellent agreement was found~\cite{PUSCHNIG2015193}.~Within this plane-wave assumption, we can express the scalar and vectorial overlap integrals in Eq.~(\ref{eq: scalar atomic form factor}) and Eq.~(\ref{subeq: f vectorial}) in terms of the Fourier transform of the initial state electron wave function
\begin{align}
    f_{1\rightarrow 2} &= \frac{1}{\sqrt{V}} \widetilde{\psi}_{1} (\mathbf{k}^\prime-\mathbf{q})\, ,\\
    \mathbf{f}_{1\rightarrow 2} &\equiv \frac{1}{\sqrt{V}} \frac{\mathbf{q}-\mathbf{k}^\prime}{m_e}\, \widetilde{\psi}_{1} (\mathbf{k}^\prime-\mathbf{q}) \, .
\end{align}
Also, using a plane wave as a final state, we find that the square of the transition amplitude in Eq.~\eqref{eq:transition rate} can be written as
\begin{widetext}
\begin{align}
    \overline{\left| \mathcal{M}_{1\rightarrow 2}\right|^2} &= \bigg\{ \overline{\left|\mathcal{M}\right|^2} + 2\; \overline{\Real{\mathcal{M}\nabla_{\boldell}\mathcal{M}\cdot (\mathbf{q}-\mathbf{k}^\prime)}} + \overline{\left| \nabla_{\boldell}\mathcal{M}\cdot(\mathbf{q}-\mathbf{k}^\prime)\right|^2}  \bigg\}\times \frac{1}{V} \left| \widetilde{\psi}_1(\mathbf{k}^\prime-\mathbf{q}) \right|^2\nonumber\\
    &\equiv \underbrace{R_\mathrm{free}(\mathbf{k}^\prime,\mathbf{q},\mathbf{v})}_\text{free electrons} \times \frac{1}{V} \underbrace{\left| \widetilde{\psi}_1(\mathbf{k}^\prime-\mathbf{q}) \right|^2}_\text{material properties}\,,\label{eq: DM material factorisation}
\end{align}
\end{widetext}
where we introduced the free particle response function $R_\mathrm{free}(\mathbf{k}^\prime,\mathbf{q},\mathbf{v})$, for which we give a general expression in Appendix~\ref{app: matrix element}.~In order to understand the physical meaning of $R_\mathrm{free}(\mathbf{k}^\prime,\mathbf{q},\mathbf{v})$, it is instructive to take the limit of a free initial state electron in Eq.~(\ref{eq: DM material factorisation}), and hence replace~$\psi_1(\mathbf{x})$ with a plane wave of linear momentum $\boldsymbol{\ell}$.~In this limit, one finds
\begin{align}
    \overline{\left| \mathcal{M}_{1\rightarrow 2}\right|^2} = R_\mathrm{free}(\mathbf{k}^\prime,\mathbf{q},\mathbf{v})
    \times (2\pi)^3 \delta^{3}(\mathbf{k}^\prime-\boldsymbol{\ell}-\mathbf{q}) \,,\label{eq: DM material factorisation 2}
\end{align}
where all information, besides momentum conservation, is contained in $R_\mathrm{free}$.~This shows that the second factor in Eq.~(\ref{eq: DM material factorisation}) contributes non trivially to the squared transition amplitude only when the initial state electron is bound within a material.~In this latter case, it encodes all relevant material properties via the Fourier transform of the initial state electron wave function.~As we will see in the next section,  this factorization allows us to express the rate of DM-induced electron ejection from materials in terms of a single material response function.~This is in contrast with our previous findings for the cases of atomic ionizations~\cite{Catena:2019gfa} and excitations in crystals~\cite{Catena:2021qsr} where up to five material response functions were required to evaluate the rate of DM-induced electronic transitions between filled valence and empty conduction bands.~One should also note that the results reported in~\cite{Catena:2019gfa,Catena:2021qsr} neglect the directional information of the event rate and assume a simplified treatment of the velocity integral in the transition-rate formula.~By performing this integral exactly, as we do here via Monte Carlo integration (see Sec.~\ref{sec:results}), up to five scalar and two vectorial material response functions are in general expected to contribute to the DM-induced electronic transition rate~\cite{Catena:2021qsr}.

\subsection{Effective theory expansion of the scattering amplitude}
\label{sec:eft}

In order to evaluate our general electron ejection formulae, Eqs.~(\ref{eq:transition rate}) and (\ref{eq: DM material factorisation}), we need to specify the coefficients, $\mathcal{M}(\mathbf{q},\mathbf{v}^\perp_{\rm el})_{\boldell=\mathbf{0}}$ and $\nabla_{\boldell} \mathcal{M}(\mathbf{q},\mathbf{v}^\perp_{\rm el})_{\boldell=\mathbf{0}}$ in the non-relativistic expansion of the scattering amplitude $\mathcal{M}$ in Eq.~(\ref{eq:M_expansion}).~From these coefficients, one can in turn obtain an explicit expression for the free-particle response function $R_{\rm free}$, as shown in App.~\ref{app: matrix element}.~In this work, we calculate these coefficients using effective theory methods.~Specifically, we extract them from the non-relativistic effective theory of spin 0 and spin 1/2 DM-electron interactions~\cite{Catena:2019gfa}, within which 
the scattering amplitude can be written as

\begin{equation}
 \label{eq:Mnr}
\mathcal{M}(\mathbf{q},\mathbf{v}_{\rm el}^\perp) = \sum_i c_i\; F_{\mathrm{DM},i}(q) \,\langle \mathcal{O}_i \rangle  \,.
\end{equation}
Here $c_i$ is the dimensionless effective coupling corresponding to the $i$-th  operator,~$\mathcal{O}_i$, in Tab.~\ref{tab:operators}, angle brackets denote an expectation value between DM-electron spin states, and~$F_{\mathrm{DM},i}(q)$ is the DM~form factor that encapsulates the~$q$-dependence of the amplitude not captured by the operator~$\mathcal{O}_i$ itself~\cite{Essig:2011nj}
~\footnote{If the non-relativistic amplitude $\mathcal{M}(q)$ contains a given operator $\mathcal{O}_i$ within two terms with distinct~$q$-dependencies, i.e. two different DM form factors, one can still use our formalism by replacing $c_i F_{\mathrm{DM},i}(q)$ with the sum $c^{(1)}_i  F^{(1)}_{\mathrm{DM},i}(q)+c^{(2)}_i F^{(2)}_{\mathrm{DM},i}(q)$ for that particular operator.}.
Possible forms of the DM~form factor  include
\begin{align}
    F_{\mathrm{DM},i}(q) &= \begin{cases}
    1  &\text{ for short-range interactions,}\\
    \left(\frac{q_\mathrm{ref}}{q}\right)^2 &\text{ for long-range interactions,}\\
    \left(\frac{q_\mathrm{ref}^2+m_\phi^2}{q^2+m_\phi^2}\right)  &\text{ for a massive mediator~$\phi$\,,}
    \end{cases}
\end{align}
where we introduced an arbitrary reference momentum transfer~$q_\mathrm{ref}$. In the context of sub-GeV~DM searches, $q_\mathrm{ref}$ is usually set to $\alpha m_e$, where $\alpha$ is the fine-structure constant, since this is the typical momentum of an electron in the outer atomic orbitals.

Eq.~(\ref{eq:Mnr}) gives the most general expression for the non-relativistic amplitude for DM-electron scattering that is compatible with momentum conservation and Galilean invariance.~The formalism we develop in this work, and in particular the free-particle response function used in the numerical calculations, relies on the expansion in Eq.~(\ref{eq:Mnr}).

\begin{table}[t]
    \centering
    \begin{tabular*}{\columnwidth}{@{\extracolsep{\fill}}ll@{}}
    \toprule
      $\mathcal{O}_1 = \mathds{1}_{\chi e}$ & $\mathcal{O}_9 = i\mathbf{S}_\chi\cdot\left(\mathbf{S}_e\times\frac{ \mathbf{q}}{m_e}\right)$  \\
        $\mathcal{O}_3 = i\mathbf{S}_e\cdot\left(\frac{ \mathbf{q}}{m_e}\times \mathbf{v}^{\perp}_{\rm el}\right)$ &   $\mathcal{O}_{10} = i\mathbf{S}_e\cdot\frac{ \mathbf{q}}{m_e}$   \\
        $\mathcal{O}_4 = \mathbf{S}_{\chi}\cdot \mathbf{S}_e$ &   $\mathcal{O}_{11} = i\mathbf{S}_\chi\cdot\frac{ \mathbf{q}}{m_e}$   \\                                                                             
        $\mathcal{O}_5 = i\mathbf{S}_\chi\cdot\left(\frac{ \mathbf{q}}{m_e}\times \mathbf{v}^{\perp}_{\rm el}\right)$ &  $\mathcal{O}_{12} = \mathbf{S}_{\chi}\cdot \left(\mathbf{S}_e \times \mathbf{v}^{\perp}_{\rm el} \right)$ \\                                                                                                                 
        $\mathcal{O}_6 = \left(\mathbf{S}_\chi\cdot\frac{ \mathbf{q}}{m_e}\right) \left(\mathbf{S}_e\cdot\frac{{\bf{q}}}{m_e}\right)$ &  $\mathcal{O}_{13} =i \left(\mathbf{S}_{\chi}\cdot  \mathbf{v}^{\perp}_{\rm el}\right)\left(\mathbf{S}_e\cdot \frac{ \mathbf{q}}{m_e}\right)$ \\   
        $\mathcal{O}_7 = \mathbf{S}_e\cdot  \mathbf{v}^{\perp}_{\rm el}$ &  $\mathcal{O}_{14} = i\left(\mathbf{S}_{\chi}\cdot \frac{ \mathbf{q}}{m_e}\right)\left(\mathbf{S}_e\cdot  \mathbf{v}^{\perp}_{\rm el}\right)$  \\
        $\mathcal{O}_8 = \mathbf{S}_{\chi}\cdot  \mathbf{v}^{\perp}_{\rm el}$  & $\mathcal{O}_{15} = i\mathcal{O}_{11}\left[ \left(\mathbf{S}_e\times  \mathbf{v}^{\perp}_{\rm el} \right) \cdot \frac{ \mathbf{q}}{m_e}\right] $ \\       
    \bottomrule
    \end{tabular*}
    \caption{Interaction operators defining the non-relativistic effective theory of spin 0 and 1/2 DM-electron interactions~\cite{Catena:2019gfa}.~$\mathbf{S}_e$ ($\mathbf{S}_\chi$) is the electron (DM) spin, $\mathbf{v}_{\rm el}^\perp=\mathbf{v}-\boldsymbol{\ell}/m_e-\mathbf{q}/(2 \mu_{\chi e})$, where $\mu_{\chi e}$ is the DM-electron reduced mass, $\mathbf{v}_{\rm el}^\perp$ is the transverse relative velocity and $\mathds{1}_{\chi e}$ is the identity in the DM-electron spin space.~In the case of elastic scattering, $\mathbf{v}_{\rm el}^\perp \cdot \mathbf{q}=0$, which explains the notation.}
\label{tab:operators}
\end{table}

\subsection{Benchmark particle physics models}

With Eq.~\eqref{eq:Mnr}, we have a general parametrization of the non-relativistic scattering amplitude that virtually any fundamental DM~particle model can be mapped onto. In the numerical applications presented in Sec.~\ref{sec:results}, we focus on four benchmark models, each of them corresponding to a different linear combination of operators in Tab.~\ref{tab:operators}. 
These models -- briefly reviewed below -- provide interesting examples of DM-electron interactions, and demonstrate both the generality of the effective theory expansion in Eq.~(\ref{eq:Mnr}), as well as how the mapping from fundamental to effective coupling constants works in practice.

\subsubsection{Dark photon model}

Our first benchmark model has guided the direct search for sub-GeV~DM particles over the past few years, and is referred to as the dark photon model.~In this framework, the Standard Model (SM) Lagrangian is extended by a new $U(1)$ gauge group with gauge coupling~$g_D$, and by a massive dark photon~$A^\prime$~\cite{Fayet:1980ad,Fayet:1980rr,Holdom:1985ag,Boehm:2003hm,Fayet:2004bw,Essig:2011nj}. The DM-ordinary matter interaction portal opens via a kinetic mixing between ordinary and dark photons in the interaction Lagrangian, i.e.~$\epsilon F_{\mu\nu}F^{\prime\mu\nu}$, where $F_{\mu\nu}(F^{\prime\mu\nu})$ is the field strength tensor of the ordinary photon (massive dark photon).
The Lagrangian of the dark sector in this model is given by
\begin{align}
    \mathscr{L}_D &= \bar{\chi} (i\gamma^{\mu}D_\mu-m_{\chi}) \chi   -\frac{1}{4}F'_{\mu \nu}F'^{\mu \nu}\nonumber\\ 
    &+ \frac{1}{2} m_{A^\prime}^{2} A'_{\mu}A'^{\mu} - \frac{\varepsilon}{2} F_{\mu \nu}F'^{\mu \nu}\, ,\label{eq: lagrangian dark photon}
    \intertext{with the covariant derivative defined as}
    D_\mu\chi &= \partial_\mu \chi - i g_D A^{\prime}_\mu \chi\, ,
\end{align}
where $g_D$ is the gauge coupling corresponding to the dark $U(1)$ gauge group.

In our general framework, the DM-electron scattering amplitude in the dark photon model can be mapped on to the operator~$\mathcal{O}_1$ if one relates the coupling constants $g_D$ and $\varepsilon$ to the effective coupling $c_1$ as follows,
\begin{subequations}
    \begin{align}
    c_1 &= \frac{4m_\chi m_e g_D\epsilon e}{q_\mathrm{ref}^2+m_{A^\prime}^2}\, , 
    \intertext{with}
    F_{\mathrm{DM},1}(q)&=\frac{q_\mathrm{ref}^2+m_{A^\prime}^2}{q^2+m_{A^\prime}^2} \, .
\end{align}
\end{subequations}

\subsubsection{Electric dipole interactions}
As another less trivial example that illustrates the wide applicability of our framework, we next consider the case of electric dipole DM-electron interactions induced by the interaction Lagrangian 
\begin{align}
\mathscr{L}_{\rm int}&= \frac{g}{\Lambda} \, i\bar{\chi} \sigma^{\mu\nu} \gamma^5 \chi \, F_{\mu\nu} \, .
\label{eq: lagrangian electric dipole}
\end{align}
In~\cite{Catena:2019gfa}, we showed that the scattering amplitude of this model can be mapped to the operator $\mathcal{O}_{11}$ via
\begin{align}
    c_{11} &= \frac{16 e m_\chi m_e^2}{q_{\rm ref}^2}\frac{g}{\Lambda} &\text{ with }  &F_{\mathrm{DM},11}(q)=\left( q_\mathrm{ref}\over q\right)^2\, .
\end{align}

\subsubsection{Magnetic dipole interactions}
Similarly, one can assume an interaction portal via magnetic dipole interactions between DM and electrons by the following interaction term in the Lagrangian,
\begin{align}
\mathscr{L}_{\rm int}&= \frac{g}{\Lambda} \, \bar{\chi}\sigma^{\mu\nu}\chi \, F_{\mu\nu}\,.
\label{eq: lagrangian magnetic dipole}
\end{align}
As shown in~\cite{Catena:2019gfa}, the corresponding scattering amplitude can be identified with a linear combination of four of the operators in Tab.~\ref{tab:operators}, with non-zero effective couplings given by
\begin{subequations}
\begin{align}
c_1 &= 4 e m_e\frac{g}{\Lambda}\, ,                                     & \text{ with } &F_{\mathrm{DM},1}(q)=1\, ,\\    
c_4 &= 16 e m_\chi\frac{g}{\Lambda}\, ,                                 &\text{ with }  &F_{\mathrm{DM},4}(q)=1\, ,\\    
c_5 &= \frac{16 e m_e^2m_\chi}{q_\mathrm{ref}^2}\frac{g}{\Lambda}\, ,   &\text{ with }  &F_{\mathrm{DM},5}(q)=\left( q_\mathrm{ref}\over q\right)^2\, ,\\
c_6 &= -\frac{16 e m_e^2m_\chi}{q_\mathrm{ref}^2}\frac{g}{\Lambda}\, ,  &\text{ with }  &F_{\mathrm{DM},6}(q)=\left(q_\mathrm{ref}\over q\right)^2\, .
\end{align}
\end{subequations}
As one can see, in this case the amplitude $\mathcal{M}$ is a linear combination of ``short-range'' and ``long-range'' contributions.

\subsubsection{Anapole interactions}
Finally, we also study anapole interactions, defined by the interaction Lagrangian
\begin{align}
\mathscr{L}_{\rm int}&= \frac{g}{2\Lambda^2} \, \bar{\chi}\gamma^\mu\gamma^5\chi \, \partial^\nu F_{\mu\nu} \,.
\label{eq: lagrangian anapole}
\end{align}
Just as before, we compare the scattering amplitude with the effective operators and find a correspondence to $\mathcal{O}_8$ and $\mathcal{O}_9$ with the effective couplings
\begin{subequations}
\label{eq: anapole effective couplings}
\begin{align}
    c_8 &= 8 e m_e m_\chi\frac{g}{\Lambda^2}\, , & \text{ with } &F_{\mathrm{DM},8}(q)=1\, ,\\ 
    c_9 &= -8 e m_e m_\chi\frac{g}{\Lambda^2}\, , & \text{ with } &F_{\mathrm{DM},9}(q)=1\, .
\end{align}
\end{subequations}

\subsection{Application to periodic systems}
\label{sec:periodic}

We continue the development of our formalism by specifying the wave function for the initial state electrons, which we assume to be bound within a crystal consisting of periodically repeating atoms.~In this case, the electron eigenfunction is characterized by an energy band, $i$, and a lattice momentum, $\mathbf{k}$,
\begin{align}
    \psi_1(\mathbf{x}) \rightarrow \psi_{i\mathbf{k}}(\mathbf{x})\,,
\end{align}
and has the form of a Bloch state, in which 
\begin{align}
\psi_{i\mathbf{k}}(\mathbf{x}+\mathbf{a}) = e^{i \mathbf{a} \cdot \mathbf{k}  }\,\psi_{i\mathbf{k}}(\mathbf{x}) \,,
\end{align}
with $\mathbf{a}$ being an arbitrary lattice vector.~We refer to App.~\ref{app: tight binding review} for further details on Bloch's theorem and Bloch states.~In this work, we consider two different bases to form the Bloch states of the initial state electrons.~For our tight-binding approximation calculations (Sec.~\ref{ss: general tight binding}) we build the wavefunctions from linear combinations of atomic orbitals, and for our DFT calculations (Sec.~\ref{sec:DFT}) we use linear combinations of plane waves.

Assuming a Bloch state of crystal momentum $\mathbf{k}$ for the initial electron wave function and a plane wave of linear momentum $\mathbf{k}'$ for the final state electron, we can combine Eq.~(\ref{eq: DM material factorisation}) with Eq.~(\ref{eq:transition rate}) to calculate the transition rate~$R_{i\mathbf{k}\rightarrow \mathbf{k}^\prime}$, that is the rate of electron ejections by DM scattering when the initial electron is in  energy band~$i$ with crystal  momentum~$\mathbf{k}$ and the outgoing electron has linear momentum~$\mathbf{k}^\prime$.~The energy difference in Eq.~\eqref{eq:transition rate} now reads
\begin{align}
    E_f-E_i &=\frac{k^{\prime 2}}{2m_e}+\frac{q^2}{2m_\chi}- \mathbf{v}\cdot\mathbf{q} + \Phi - E_i(\mathbf{k})\, , \label{eq: energy difference}
\end{align}
where $E_i(\mathbf{k})$ is the energy of an electron in energy band $i$ with wavevector $\mathbf{k}$, which is negative for bound electrons. $\Phi$ is the (positive) work function, which corresponds to the energy difference between the highest occupied electronic state and the zero-energy unbound free-electron plane-wave state; in this work we take the measured value for graphene of $\Phi = $ 4.3\,eV~\cite{PUSCHNIG2015193, Hochberg:2016ntt}.

Adding the contributions from all occupied initial states yields
\begin{equation}
    R_{\textrm{any}\rightarrow\mathbf{k}^\prime}\equiv 2\sum_i \int_\text{BZ} \frac{N_\text{cell}V_{\text{cell}}\mathrm{d}^3 \mathbf{k}}{(2\pi)^3}R_{i\mathbf{k}\rightarrow \mathbf{k}^\prime}\,,
    \label{eq:Rany}
\end{equation}
where the factor of 2 accounts for the double occupation of each electronic state due to spin degeneracy. We then sum the contributions from all final plane-wave states to obtain the total rate of electron ejections by DM scattering,
\begin{equation}
    R\equiv \int\frac{V\dd^3 \mathbf{k}^\prime}{(2\pi)^3}R_{\textrm{any}\rightarrow\mathbf{k}^\prime}\,.
\end{equation}
Finally, we express the total ejection rate, $R$, as follows
\begin{widetext}
\begin{align}
 R&=\frac{n_\chi N_\text{cell}}{32\pi^2m_\chi^2m_e^2}\int\dd^3 \mathbf{k}^\prime\int \mathrm{d}\, E_e\int\dd^3 \mathbf{q}\int\dd^3 \mathbf{v}\,f_\chi(\mathbf{v})\delta\left(\Delta E_e +\frac{q^2}{2m_\chi}- \mathbf{v}\cdot\mathbf{q}\right)
 R_\mathrm{free}(\mathbf{k}^\prime,\mathbf{q},\mathbf{v})
 \;W(\mathbf{k}^\prime-\mathbf{q},E_e)\, ,
 \label{eq:rate_general}
\end{align}
\end{widetext}
where we define the electron's energy change as~$\Delta E_e\equiv\frac{k^{\prime 2}}{2 m_e}+\Phi- E_e$, and introduce the material-specific response function
\begin{align}
    W\left(\boldell,E_e\right) &=
    \frac{V_\mathrm{cell}}{(2\pi)^3}\sum_i \int_\text{BZ} \frac{\mathrm{d}^3 \mathbf{k}}{(2\pi)^3} \delta\left(E_e -E_i(\mathbf{k})\right) \nonumber\\
    &\times \left| \widetilde{\psi}_{i\mathbf{k}}(\boldell) \right|^2\, . \label{eq: response function general}
\end{align}
As implied by the above $\delta$-function, $E_e$ takes the value of the initial state electron energy where $E_e=0$\,eV for the highest occupied state and $E_e<0$\,eV for other bound states. Eq.~(\ref{eq:rate_general}) shows that both free-electron physics and material properties contribute to the total rate of electron ejections by DM scattering in a factorisable way, and that this factorisation involves a single material response function.
The response function is normalized to
\begin{align}
    \int \dd E_e \dd^3 \boldell\; W(\boldell,E_e)   = N_\mathrm{bands}\, ,
\end{align}
where $N_\mathrm{bands}$ is the number of occupied initial-state electronic bands and we use the normalization of $\widetilde{\psi}_{i\mathbf{k}}(\boldell)$ (see Eq.~\eqref{eq: wavefunction normalization}) and  $V_\mathrm{cell} V_\mathrm{1BZ} = (2\pi)^3$.

In the next two subsections we express the response function $W$ by writing $\psi_{i\mathbf{k}}$ as a linear combination of atomic orbitals and plane waves.~Atomic orbitals are often employed within the tight-binding approximation (see Sec.~\ref{sec:tb_calc}), whereas plane waves are a standard basis in DFT electronic structure calculations (see Sec.~\ref{sec:DFT_calc}).

\subsubsection{Atomic orbital basis}
\label{ss: general tight binding}

Let us now derive a compact expression for the response function $W$ by expanding the initial state electron wave function, $\psi_{i\mathbf{k}}(\mathbf{x})$, in a basis of atomic orbitals,
\begin{subequations}
\label{eq: wave function tb}
\begin{align}
    \psi_{i\mathbf{k}}(\mathbf{x}) &= \mathcal{N}_{\mathbf{k}}\sum_{j=1}^n C_{ij}(\mathbf{k}) \Phi_{j\mathbf{k}}(\mathbf{x})\, ,
    \intertext{where~$j$ runs over the $n$ atomic orbitals present in each unit cell, and  $\Phi_{j\mathbf{k}}(\mathbf{x})$ are the Bloch states corresponding to each atomic orbital,} 
    \Phi_{j\mathbf{k}}(\mathbf{x}) &= \frac{1}{\sqrt{N_\mathrm{cell}}}\sum_{r=1}^{N_\mathrm{cell}} e^{i\mathbf{k}\cdot \mathbf{R}_{jr}}\varphi_j(\mathbf{x}-\mathbf{R}_{jr})\, .
\end{align}
\end{subequations}
Here $\varphi_j$ is an atomic wave function on an atom at position $\mathbf{R}_{jr}$, $N_\mathrm{cell}$ is the number of unit cells and $\mathcal{N}_{\mathbf{k}}$ is a normalisation constant defined in App.~\ref{app: tight binding review}.~Within the tight-binding approximation (introduced below in Sec.~\ref{sec:tb_calc}), the coefficients $C_{ij}$ in Eq.~(\ref{eq: wave function tb}) are computed by solving the secular equation (Eq.~\ref{eq: secular equation}) with band energies extracted from measurements, or calculated using a more sophisticated technique such as DFT.

To evaluate the material response function defined in Eq.~\eqref{eq: response function general}, $W(\boldell,E_e)$, we need to calculate the square of the Fourier transform of~$\psi_{i\mathbf{k}}(\mathbf{x})$. Denoting by $\widetilde{\varphi}_j$ the Fourier transform of $\varphi_j$, for the Fourier transform of $\psi_{i\mathbf{k}}$ we find
\begin{align}
\widetilde{\psi}_{i\mathbf{k}} (\boldell) &={\mathcal{N}_{\mathbf{k}} \over \sqrt{N_\mathrm{cell}}} \sum_{j=1}^n C_{ij}(\mathbf{k}) \widetilde{\varphi}_j(\boldell) \sum_{r=1}^{N_\mathrm{cell}}e^{i (\mathbf{k}-\boldell)\cdot \mathbf{R}_{jr}}\, . \label{eq: wave function tb fourier}
\end{align}
The lattice sum can be evaluated by using Eq.~(A.8) of~\cite{Nijboer1957}.~
\begin{align}
    \sum_{r=1}^{N_\mathrm{cell}} e^{i \mathbf{k}\cdot \mathbf{R}_{jr}}= \frac{1}{V_\mathrm{cell}} \sum_{\mathbf{G}}(2\pi)^3\delta^{(3)}\left(\mathbf{k}+\mathbf{G}\right)\,e^{i\mathbf{k}\cdot\boldsymbol{\delta}_j} \,, \label{eq: lattice sum identity}
\end{align}
where $\boldsymbol{\delta}_j$ is the  location of the atom hosting the $j$-th orbital in the unit cell that contains the origin of coordinates.~For a given $j$,  $\boldsymbol{\delta}_j=0$ if there exists a $\bar{r}\in\{1,\dots,N_{\textrm{cell}}\}$ such that $\mathbf{R}_{j\bar{r}}=0$, and $\boldsymbol{\delta}_j\neq 0$ otherwise.~In Eq.~(\ref{eq: lattice sum identity}), the sum runs over the reciprocal lattice vectors $\mathbf{G}$.~For each $j$ and $r$, they satisfy $\mathbf{G}\cdot (\mathbf{R}_{jr}-\boldsymbol{\delta}_j) = 2\pi m$, where $m\in\mathds{Z}$. Another useful identity to evaluate the squared modulus of $\widetilde{\psi}_{i\mathbf{k}}$ is
\begin{align}
    (2\pi)^3\Big| \sum_{\mathbf{G}}
    \delta^{(3)}\left(\mathbf{k}+\mathbf{G}\right) \Big|^2 = N_\mathrm{cell} V_\mathrm{cell} \sum_{\mathbf{G}}
    \delta^{(3)}\left(\mathbf{k}+\mathbf{G}\right)\,,
\end{align}
where $N_\mathrm{cell} V_\mathrm{cell}=(2\pi)^3\delta^{(3)}(0)$.~Using Eqs.~(\ref{eq: lattice sum identity}) and (\ref{eq: wave function tb fourier}), we obtain
\begin{align}
    |\widetilde{\psi}_{i\mathbf{k}} (\boldell)|^2 &= {\mathcal{N}_{\mathbf{k}}^2 \over V_\mathrm{cell}} \Big| \sum_{j=1}^n C_{ij}(\mathbf{k}) \widetilde{\varphi}_j(\boldell)  e^{i(\mathbf{k}-\boldell)\cdot \boldsymbol{\delta}_j}\Big|^2\nonumber\\
    &\times\sum_{\mathbf{G}}(2\pi)^3\delta^{(3)}\left(\mathbf{k}-\boldell+\mathbf{G}\right) \,,
    \label{eq:wf2}
\end{align}
and by combining Eq.~(\ref{eq:wf2}) with Eq.~(\ref{eq: response function general}), we obtain the following expression for the response function~$W$:
\begin{widetext}
\begin{align}
    W\left(\boldell, E_e\right) & =   \frac{\mathcal{N}_{\mathbf{k}}^2}{(2\pi)^3}\sum_i \Big| \sum_{j=1}^n C_{ij}(\mathbf{k}) \widetilde{\varphi}_j(\boldell) e^{-i\mathbf{G}^*\cdot \boldsymbol{\delta}_j} \Big|^2\delta\left(E_e  -E_i(\mathbf{k})\right)\Big|_{\mathbf{k} = \boldell-\mathbf{G}^*}  \, . \label{eq: response function tb}
\end{align}
\end{widetext}
Here, we performed the integral over the lattice momentum~$\mathbf{k}$, such that the~$\delta$~function fixes it to~$\mathbf{k} = \boldell-\mathbf{G}^*$, where~$\mathbf{G}^*$ is the unique reciprocal lattice vector that ensures that~$\mathbf{k}$ lies within the first Brillouin zone.~The other terms of the sum over the vectors $\mathbf{G}$ do not contribute.

To evaluate Eq.~(\ref{eq: response function tb}), one needs to specify the wave functions $\varphi_j$, which we will do in Sec.~\ref{sec:tb_calc}.

\subsubsection{Plane wave basis}
\label{sec:DFT}
Let us now express the response function $W$ by using plane waves to write the electron wave function $\psi_{i\mathbf{k}}\left( \mathbf{x}\right)$ as
\begin{align}
 \label{eq: Bloch wave function}
 \psi_{i\mathbf{k}}\left( \mathbf{x} \right)=&\frac{1}{\sqrt{V}}\sum_{\mathbf{G}} u_i\left(\mathbf{k} + \mathbf{G} \right) e^{i \left(\mathbf{k} + \mathbf{G} \right)\cdot \mathbf{x}}\,,
\end{align}
where $\sum_{\mathbf{G}}|u_i\left(\mathbf{k} + \mathbf{G} \right)|^2=1$ for all $i$ and $\mathbf{k}$.~From Eq.~(\ref{eq: Bloch wave function}), we find
\begin{align}
    \widetilde{\psi}_{i\mathbf{k}}(\boldell)=\frac{(2\pi)^3}{\sqrt{V}}\sum_\mathbf{G}u_i(\mathbf{k}+\mathbf{G})\delta^{(3)}(\mathbf{k}+\mathbf{G}-\boldell)\,,
\end{align}
and
\begin{align}
    \left|\widetilde{\psi}_{i\mathbf{k}}(\boldell)\right|^2 =&(2\pi)^3\sum_\mathbf{G}|u_i(\mathbf{k}+\mathbf{G})|^2\delta^{(3)}(\mathbf{k}+\mathbf{G}-\boldell)\,,
\end{align}
where we used $V=(2\pi)^3\delta^{(3)}(0)$.~This result can be directly inserted into Eq.~(\ref{eq: response function general}), which leads to the response function
\begin{align}
    W\left(\boldell, E_e\right) & = V_\mathrm{cell}\sum_i \int_\text{BZ} \frac{\mathrm{d}^3 \mathbf{k}}{(2\pi)^3} \delta\left(E_e - E_i(\mathbf{k})\right) \nonumber\\
    &\times\sum_\mathbf{G}|u_i(\mathbf{k}+\mathbf{G})|^2\delta^{(3)}(\mathbf{k}+\mathbf{G}-\boldell)\, .
\label{eq: response function dft}    
\end{align}
In Sec.~\ref{sec: numerical implementation}, we extract the $u_i(\mathbf{k}+\mathbf{G})$ coefficients and the band structure $E_i(\mathbf{k})$ from state-of-the-art DFT calculations.

\section{Electronic structure calculations for electron ejections in graphene detectors}
\label{sec:electronic}
The equations derived in the previous section refer to  three-dimensional periodic systems.~We are now interested in applying them to the specific case of graphene, which is a single-layer material that is periodic in two dimensions. ~The ``dimensional reduction'' can be performed straightforwardly by means of the replacement specified below,
\begin{align}
V_{\rm cell}\int_{\rm BZ} \frac{{\rm d}^3 \mathbf{k}}{(2\pi)^3} \longrightarrow 
A_{\rm cell}\int_{\rm BZ} \frac{{\rm d}^2 \mathbf{k}}{(2\pi)^2} \,,
\label{eq:3Dto2D}
\end{align}
where $A_{\rm cell}$ is the two-dimensional unit cell of graphene, while $\mathbf{k}$ in the right-(left-)hand-side is a two-dimensional (three-dimensional) lattice vector in the first Brillouin zone.

With a general formalism for electron ejections by DM scattering in graphene detectors in place, we can now focus on the evaluation of the predicted electron ejection rate.~This crucially depends on the response function $W$, which is in turn a function of the initial state electron wave functions.~As a result, numerical evaluation of the predicted electron ejection rate requires detailed electronic structure calculations for graphene. In the following, we perform such electronic structure calculations using two methods: the tight binding approximation, and DFT. From this analysis, DFT will emerge as our recommended framework for electronic structure calculations for DM-electron scattering in graphene-based DM detectors.

\subsection{Tight binding}
\label{sec:tb_calc}

To obtain the graphene response function in the tight binding (TB) approximation, we need to evaluate Eq.~\eqref{eq: response function tb}.
The missing ingredient at this point are the coefficients~$C_{ij}(\mathbf{k})$ that yield the contribution of atomic orbital $j$ in band $i$ to the response function. 
We separate the~$\pi$- and $\sigma$-electrons and write the response function as
\begin{align}
    W(\boldell,E_e) = W_\pi (\boldell,E_e) +\sum_{i=1}^3 W_{\sigma_i}(\boldell,E_e)\, .
\end{align}
In the TB approximation, the coefficients are found as the eigenvectors of the generalized eigenvalue problem in Eq.~\eqref{eq: general eigenvalue problem}.
For a detailed review of the TB approximation in general and for the specific case of graphene, we refer to App.~\ref{app: tight binding review} and~\ref{app: tight binding graphene} respectively.

\subsubsection{$\pi$-electrons}
In case of the $\pi$-electrons in graphene, this eigenvalue problem can be solved analytically, as described in App.~\ref{app: tight binding graphene}.
Therein, the full wavefunction of the~$\pi$-electrons are derived in position and momentum space, which can be found in Eq.~\eqref{eq: wavefunction tb pi position} and~\eqref{eq:psipi2}.
The eigenvalues $E_\pi(\mathbf{k})$ and eigenvectors $\mathbf{C}_\pi$, required for the response function, are given by Eqs.~\eqref{eq: energy eigen values pi} and~\eqref{eq: eigen vectors pi} respectively.
Therefore, the $\pi$-electron contribution to the response function can be written out explicitly.
It can be shown to simplify to
\begin{align}
    W_\pi (\boldell,E_e) &= \frac{\mathcal{N}_\mathbf{k}^2}{(2\pi)^3}\delta(E_e - E_\pi(\mathbf{k})) |\widetilde{\varphi}_{2p_z}(\boldell)|^2\nonumber\\
    &\times \left( 1+\cos(\varphi_\vec{k}-\mathbf{\delta}\cdot \mathbf{G}^*)\right)\bigg|_{\mathbf{k}=\boldell-\vec{G}^*}\, .\label{eq: response function tb pi}
\end{align}
Here, the phase~$\varphi_\mathbf{k}$ and the normalization constant~$\mathcal{N}_\vec{k}$ are given by Eqs.~\eqref{eq: eigen vectors pi} and~\eqref{eq: normalization constant pi} respectively.

\subsubsection{$\sigma$-electrons}
While the procedure for the $\sigma$-electrons is conceptionally identical, the fact that their wavefunctions involve combinations of three atomic orbitals at two atomic sites means that the generalized eigenvalue problem of Eq.~\eqref{eq: general eigenvalue problem} involves $6\times6$ matrices, which can no longer be solved analytically.
Instead, we rely on numerical procedures where we use the \texttt{Eigen} library~\cite{eigen}.

The involved six-dimensional matrices~$\boldsymbol{\mathcal{H}}$ and $\boldsymbol{\mathcal{S}}$ are listed in Eq.\eqref{eq: transfer overlap sigma} of App.~\ref{app: tight binding graphene}.
Using the numerical procedures of the \texttt{Eigen} library, we obtain the eigenvalues or band energies~$E_{\sigma_i}(\mathbf{k})$ as well as the eigenvectors~$\mathbf{C}_{\sigma_i}$.
\footnote{This is a good time to point out the dependence of the normalization constant~$\mathcal{N}_\mathbf{k}$ on the norm of the eigenvectors~$\mathbf{C}$, which is ambiguous.
The numerical eigenvalue routine of \texttt{Eigen} that we use for the $\sigma$-electrons
(namely \texttt{GeneralizedSelfAdjointEigenSolver}) solves the problem $A\mathbf{x} = \lambda B\mathbf{x}$ such that $\mathbf{x}^*B\mathbf{x} = 1$.
In that case,~$\mathcal{N}_\mathbf{k}$ is trivially equal to one, as can be seen from Eq.~\eqref{eq: wave function norm general}.
However, for the analytic solution of the $\pi$-electrons, we had chosen normalized eigenvectors. In that case, $\mathcal{N}_\mathbf{k}\neq 1$, but instead given by Eq.~\eqref{eq: wave function norm general}.}
Finally this allows us to evaluate the $\sigma$-contribution to the response function,
\begin{align}
W_{\sigma_i}(\boldell,E_e) &= \frac{\mathcal{N}_\mathbf{k}^2}{(2\pi)^3}\delta(E_e - E_{\sigma_i}(\mathbf{k})) \nonumber\\
&\times\Bigg|\widetilde{\varphi}_{2s}(\boldell)\left(C_{\sigma_i1} +C_{\sigma_i4} e^{-i \boldsymbol{\delta}\cdot \mathbf{G^*}}\right) \nonumber \\
    &\quad+\widetilde{\varphi}_{2p_x}(\boldell)\left(C_{\sigma_i2} +C_{\sigma_i5} e^{-i \boldsymbol{\delta}\cdot \mathbf{G^*}}\right) \nonumber \\
    &\quad+\widetilde{\varphi}_{2p_y}(\boldell)\left(C_{\sigma_i3} +C_{\sigma_i6} e^{-i \boldsymbol{\delta}\cdot \mathbf{G^*}}\right) \Bigg|^2_{\mathbf{k}=\boldell-\vec{G}^*}\, .\label{eq: response function tb sigma}
\end{align}
By adding up the four distributions given by Eqs.~\eqref{eq: response function tb pi} and~\eqref{eq: response function tb sigma}, we obtain the final TB estimate of the graphene response function.

\begin{figure*}
    \centering
    \includegraphics[width=0.35\textwidth]{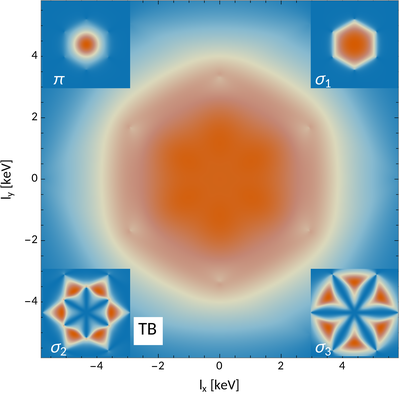}
    \includegraphics[width=0.35\textwidth]{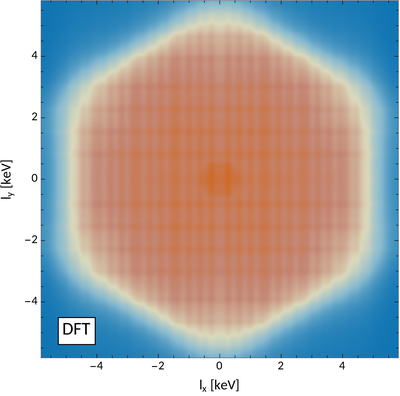}
    \includegraphics[width=0.07\textwidth]{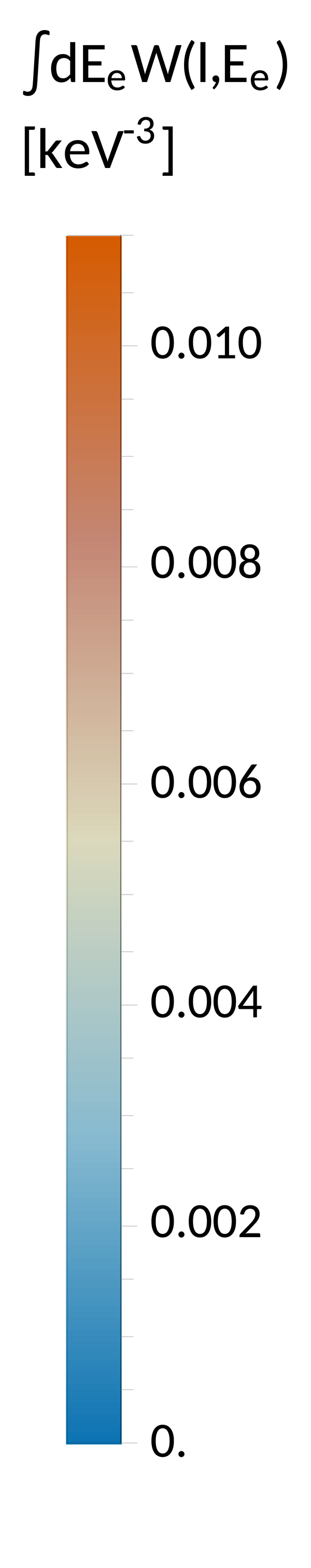}
    \caption{The partially integrated graphene response function evaluated with TB (left) and DFT (right) as a function of $\ell_x$ and $\ell_y$. In the left panel, we also show the contributions of each of the electron bands. We set $\ell_z = 91 \text{ eV}$, such that the vector~$\boldell$ lies almost in the plane of the graphene sheet. The stripe-like structure of the DFT response is an artifact of the grid sampling in the reciprocal space and is integrated out when observables are evaluated.}
    \label{fig: W lx ly plot}
\end{figure*}

Finally, we point out that our TB treatment of the electrons in graphene differs from a previous treatment by Hochberg et al.~\cite{Hochberg:2016ntt}, in particular with regard to the Bloch wavefunctions given by Eq.~\eqref{eq: wave function tb}.
A second crucial difference is our choice for the atomic wavefunctions, which we discuss next.
We present a detailed comparison in App.~\ref{app: comparison hochberg}.

\subsubsection{Atomic wavefunctions}
In order to evaluate the TB estimates of the graphene response function given in Eqs.~\eqref{eq: response function tb pi} and~\eqref{eq: response function tb sigma}, we need to specify the atomic orbitals~$\varphi_i(\mathbf{x})$ (or rather their Fourier transforms~$\widetilde{\varphi}_{i}(\boldell)$) for the electrons in carbon.

We expand the atomic wavefunction into a radial and directional component,~$\varphi_{nlm}(\mathbf{x}) = R_{nl}(r)Y_{l}^m(\hat{\mathbf{x}})$, where~$r=|\mathbf{x}|$ and~$Y_{l}^m(\hat{\mathbf{x}})$ are spherical harmonics.
Following~\cite{Bunge:1993jsz}, we describe the radial part of the atomic orbitals of carbon electrons as linear combinations of Slater-type orbitals (STOs).
\begin{subequations}
 \label{eq: RHF wavefunction}
    \begin{align}
    R_{nl}(r) &= \sum_j C_{nl j} R_\mathrm{STO}(r,Z_{l j},n_{l j})\, .
    \intertext{with}
    R_\mathrm{STO}(r,Z,n) &\equiv a_0^{-3/2}\frac{(2Z)^{n+1/2}}{\sqrt{(2n)!}}\left(\frac{r}{a_0}\right)^{n-1}e^{-\frac{Z r}{a_0}}\, .
\end{align}
\end{subequations}
We include a more detailed description of these wavefunctions, including the values of the different coefficients and the expressions for the Fourier transforms, in App.~\ref{app: RHF wfunctions}.

This choice differs from the previous approach by Hochberg et al.~\cite{Hochberg:2016ntt}, who use re-scaled hydrogenic wavefunctions to approximate the electron wavefunctions in carbon atoms.
We comment on this in App.~\ref{app: comparison hochberg}.

\subsubsection{Limitations}
As described in greater detail in App.~\ref{app: tight binding}, we can reproduce the measured band structure of graphene by adjusting the overlap and transfer parameters of the TB approximation.
However, in particular the overlap parameter, e.g. the parameter~$s$ in Eq.~\eqref{eq: transfer integral matrix}, is given by the overlap integrals of atomic wavefunctions at neighboring atomic sites.
For a given choice of atomic wavefunctions, it is therefore possible to compute~$s$ independently of the band structures.
This gives rise to the issue of self-consistency of this approach.

In~\cite{Hochberg:2016ntt}, the authors use hydrogenic wavefunctions to approximate the atomic orbitals of carbon.
As described in detail in App.~\ref{app: hydrogenic wave functions}, they ensure consistency between the two independent values of the overlap parameters, by re-scaling the effective charge factor~$Z_\mathrm{eff}$.
However, the resulting wavefunctions do not resemble the atomic wavefunctions of carbon, as we describe in App.~\ref{app: atomic wavefunctions}, which is why we chose to use RHF wavefunctions.
While it is possible to perform a similar re-scaling of the RHF wavefunctions to establish consistency with the overlap parameters listed in Tab.~\ref{tab: graphene parameter}, this generally modifies the wavefunctions to the extent that they no longer describe electrons in carbon atoms.
Therefore, it seems to be a limitation of the TB approximation to reconcile the use of realistic atomic wavefunctions with the overlap parameters that reproduce the material's band structure in a fully self-consistent manner.
This issue is not a characteristic of our specific treatment but rather a general feature of the TB approximation itself reflecting the phenomenological nature of this approximation.

\subsection{Density Functional Theory}
\label{sec:DFT_calc}
Having carefully described the features and limitations of the tight-binding approach, we now report on our DFT electronic structure calculations. We start with a brief review of the main assumptions underlying DFT in Sec.~\ref{sec:assumptions}.~In Sec.~\ref{sec:motivation}, we provide a general argument supporting DFT as a framework for electronic structure calculations in the case of DM-induced electron ejections by graphene targets. In Sec.~\ref{sec: numerical implementation}, we describe the details of our specific DFT implementation in a modified version of the {\sffamily QEdark-EFT} code.

\subsubsection{Assumptions}
\label{sec:assumptions}

Density functional theory (DFT)~\cite{Hohenberg/Kohn:1964,PhysRev.140.A1133} (for a review see~\cite{martin_2004}) is a widely used method for calculating the ground-state electronic properties of materials. It allows for explicit treatment of the chemistry and crystal structure without the need for empirically determined input parameters, and provides well-tested and computationally affordable approximations for the many-body electron-electron interactions. In addition, it has been implemented in numerous publically or commercially available computer codes that are convenient to use. 

The theory is based on the Hohenberg-Kohn theorem~\cite{Hohenberg/Kohn:1964} which states that, for electrons in an external potential (in this case provided by the charged atomic nuclei), the total energy is a unique functional of the electron density, with the ground-state density being the one that minimizes the value of this functional. 
The electronic ground state charge density can therefore be obtained variationally. 

In practice the charge density, $n_e(\mathbf{x})$, is written as a sum over the so-called Kohn-Sham wave functions, $\psi_i(\mathbf{x})$~\cite{PhysRev.140.A1133} of a fictitious auxiliary system in which the electrons are non-interacting. This mapping enables convenient computational solution of the many-body Schr\"odinger equation at the expense of an inexact description of the quantum mechanical exchange and correlation terms. These have been obtained numerically using Quantum Monte Carlo for the homogeneous interacting electron gas \cite{Ceperley/Alder:1980}, and a number of well-tested approximations exist that are appropriate for different material systems. An additional widely used and well-established approximation divides the electrons into valence electrons, which are treated explicitly within the DFT calculation, and low-energy core electrons, which are combined with the nuclei in the external potential. This pseudopotential approximation drastically reduces the computational expense and is chemically well founded, since the core electrons are not involved in chemical bonding and are only minimally modified in the solid. 
The choice of pseudopotential, and the numerical and implementational details of the DFT calculation for graphene performed here, are discussed further in Sec.~\ref{sec: numerical implementation}.

\begin{figure*}
    \centering
    \includegraphics[width=0.327\textwidth]{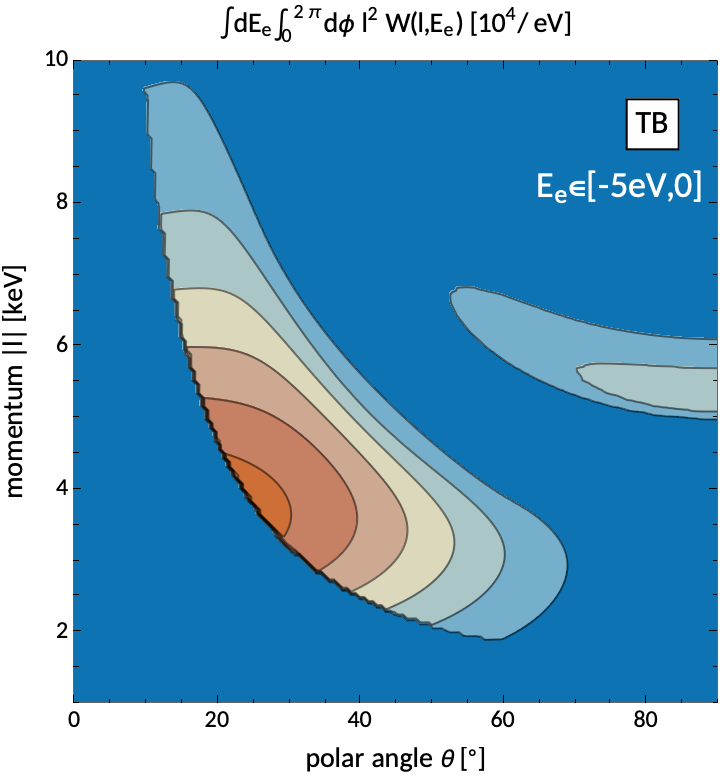}
    \includegraphics[width=0.375\textwidth]{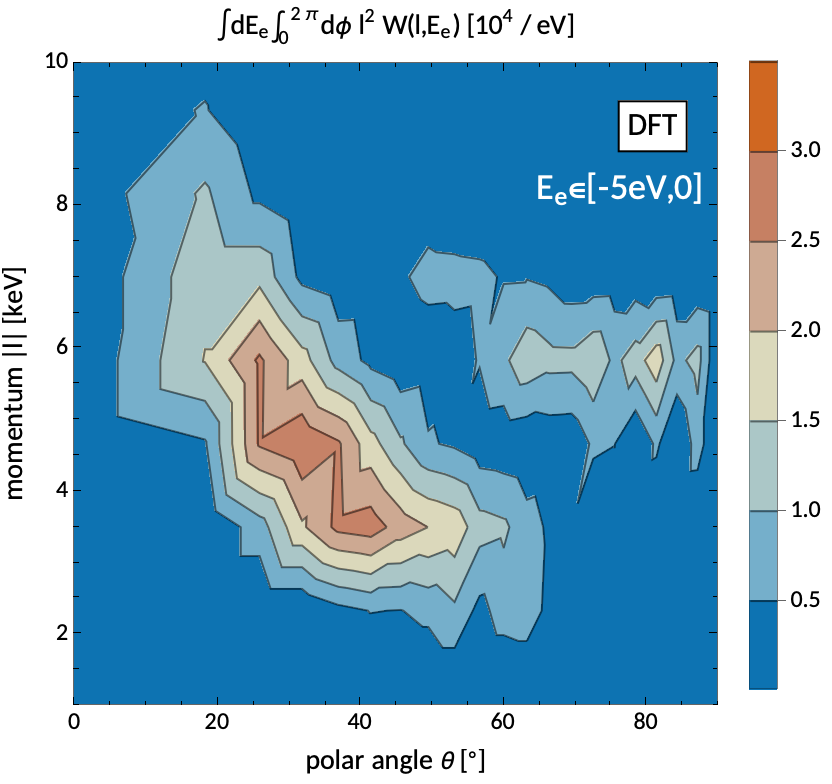}
    \includegraphics[width=0.334\textwidth]{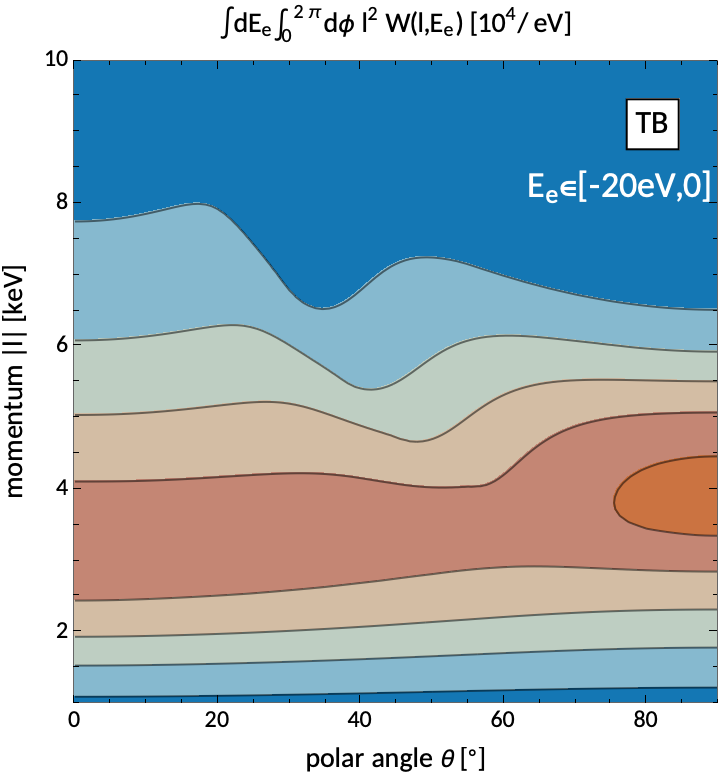}
    \includegraphics[width=0.375\textwidth]{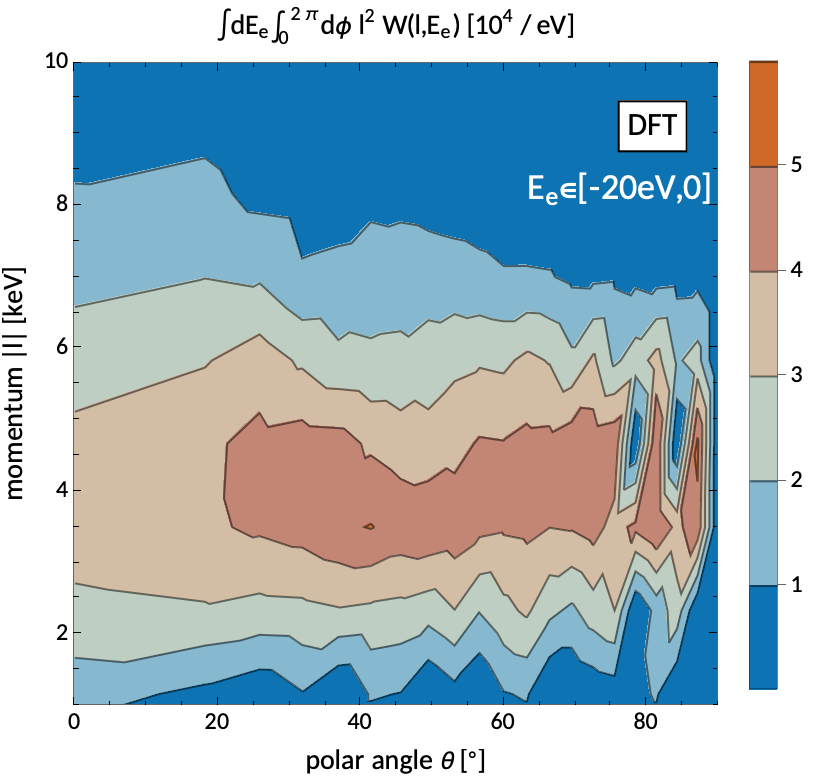}
    \caption{$W(\boldell,E_e)$ integrated between $E_\mathrm{min}<0$ and $0$ and over the azimuthal angle of $\boldell$, $\phi$ for TB (left) and DFT (right). The first row is for $E_\mathrm{min}=-5\,\mathrm{eV}$, including only electrons accessible to low mass DM. The second row has $E_\mathrm{min}=-20\,\mathrm{eV}$ and also includes electrons accessible to heavier DM. As in Fig.~\ref{fig: W lx ly plot}, the distorting effect of the finite sampling of the reciprocal space can be seen for the case of DFT. When evaluating the final state observables, these distortions are washed out.}
    \label{fig:2D plot W}
\end{figure*}

We note that, while the Kohn-Sham wave functions and energies do not formally correspond to true single-electron wave functions and energies (except for the highest occupied level, which provides the ionization energy), in practice their dispersion is usually in remarkable agreement with measured photoelectron spectra. The Kohn-Sham band structure is therefore often treated as a proxy for an effective single-particle band structure in a periodic solid. 
Since the Hohenberg-Kohn theorem describes only the ground-state electron density, however, this is particularly ill-founded for unoccupied conduction-band states. The methodology and results we present here, however, do not rely on any physical interpretation of the Kohn-Sham wavefunctions, since our final electron states are unbound plane-wave states, and, as we show next, our response function is primarily determined by the ground-state charge density rather than the ground-state wavefunctions.

\subsubsection{Motivations}
\label{sec:motivation}

An important observation we can draw from our general electron ejection rate formula is that the graphene response function $W$ is directly related to the ground state electron momentum density, $\rho_e$, defined as the Fourier transform of the electron charge density.~Indeed, an explicit second quantisation calculation allows us to write $\rho_e$ as
\begin{align}
\rho_e(\boldsymbol{\ell}) &= \sum_{i i'} \sum_{\mathbf{G}}\int_{\rm BZ}  \frac{{\rm d}^2 \mathbf{k}}{(2\pi)^2} \, n_{i i'}(\mathbf{k}) \,
u^*_{i}(\mathbf{k}+\mathbf{G})u_{i'}(\mathbf{k}+\mathbf{G}) \nonumber\\
&\times (2\pi)^2\delta^{(2)}(\mathbf{k}+\mathbf{G}-\boldsymbol{\ell}) \,,
\label{eq:EMD}
\end{align}
where
\begin{align}
\, n_{i i'}(\mathbf{k}) = \langle a^{\dagger}_{i\mathbf{k}} a_{i'\mathbf{k}} \rangle 
\end{align}
is the mean ground state occupation number density, while $a^{\dagger}_{i\mathbf{k}}$ and $a_{i'\mathbf{k}}$ are second quantisation creation and annihilation operators associated with the $\psi_{i\mathbf{k}}$ and $\psi_{i'\mathbf{k}}$ Bloch states, respectively~\footnote{Eq.~(\ref{eq:EMD}) for the momentum density $\rho_e$ can be derived from the definition, \begin{align}\rho_e(\boldsymbol{\ell})=\int {\rm d}^3 \mathbf{r}\int {\rm d}^3 \mathbf{r}' e^{-i(\mathbf{r}^{\prime}-\mathbf{r})\cdot \boldsymbol{\ell}}\, \langle \Psi^{\dagger}(\mathbf{r})\Psi(\mathbf{r}')  \rangle\,, \nonumber \end{align}where\begin{align}\Psi(\mathbf{r})=\frac{1}{\sqrt{V}}\sum_i\int _{\rm BZ}\frac{V{\rm d}^2\mathbf{k}}{(2 \pi)^2} \, \psi_{i\mathbf{k}}(\mathbf{r})\,a_{i\mathbf{k}} \,,\nonumber\end{align} angle brackets denote an expectation value on the ground state, while $\psi_{i\mathbf{k}}(\mathbf{r})$ is the Bloch wave function for the initial state electron.}.~In Eq.~(\ref{eq:EMD}), the $i=i'$ diagonal term is directly proportional to the response function $W$. The $i \neq i'$ off-diagonal term describes band mixing effects arising from electron-electron interactions across different bands. While in general $n_{i i'}(\mathbf{k})\neq 0$ for $i\neq i'$, off-diagonal contributions to $\rho_e$ are expected to be sub-leading in the case of graphene, where electron-electron interaction and correlation effects induce variations in the band energies $E_i(\mathbf{k})$ of at most a few \%, see for example Fig.~1 in~\cite{Trevisanutto2008Nov}.

\begin{figure*}
    \centering
    \includegraphics[width=0.48\textwidth]{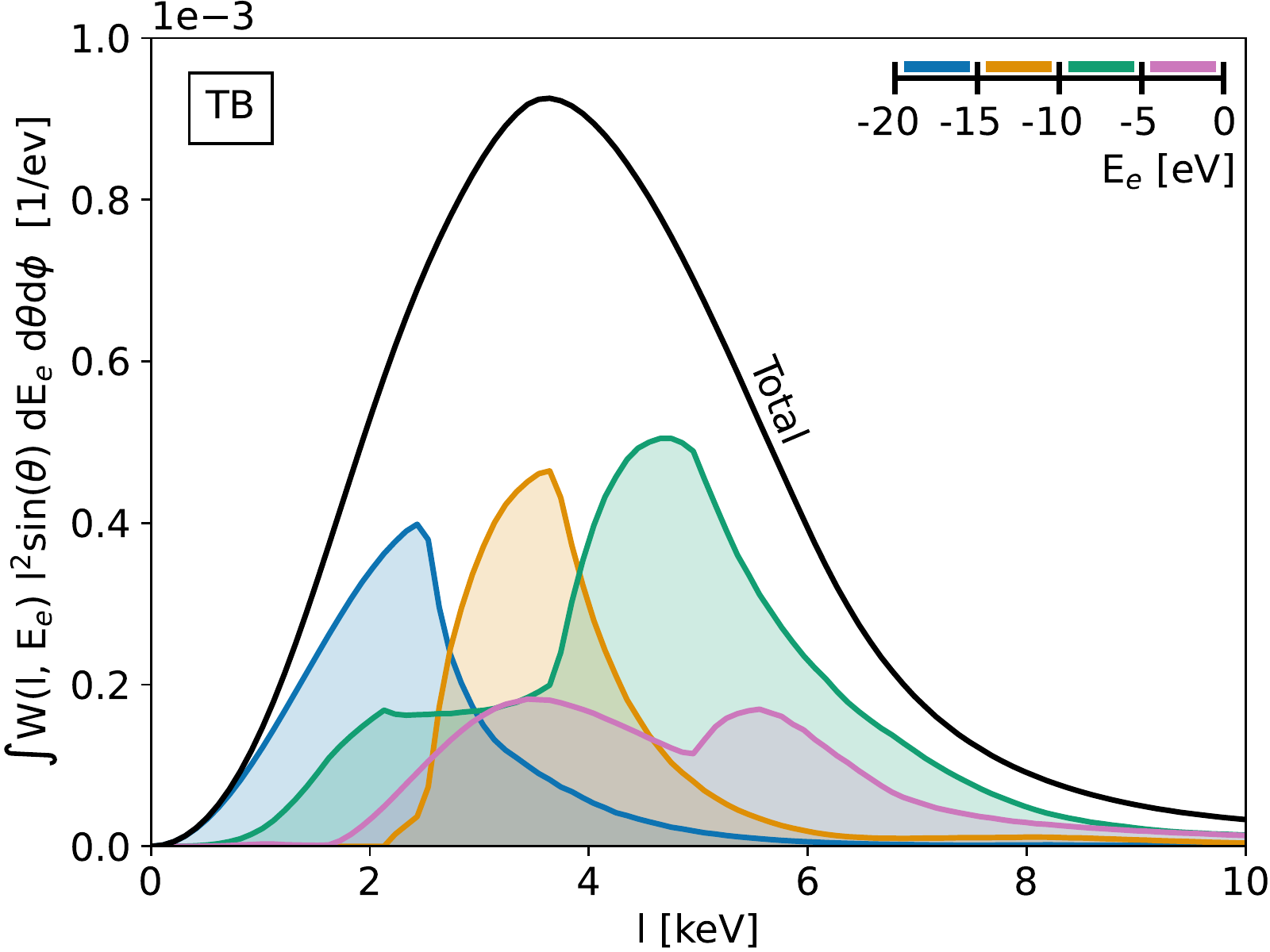}\includegraphics[width=0.48\textwidth]{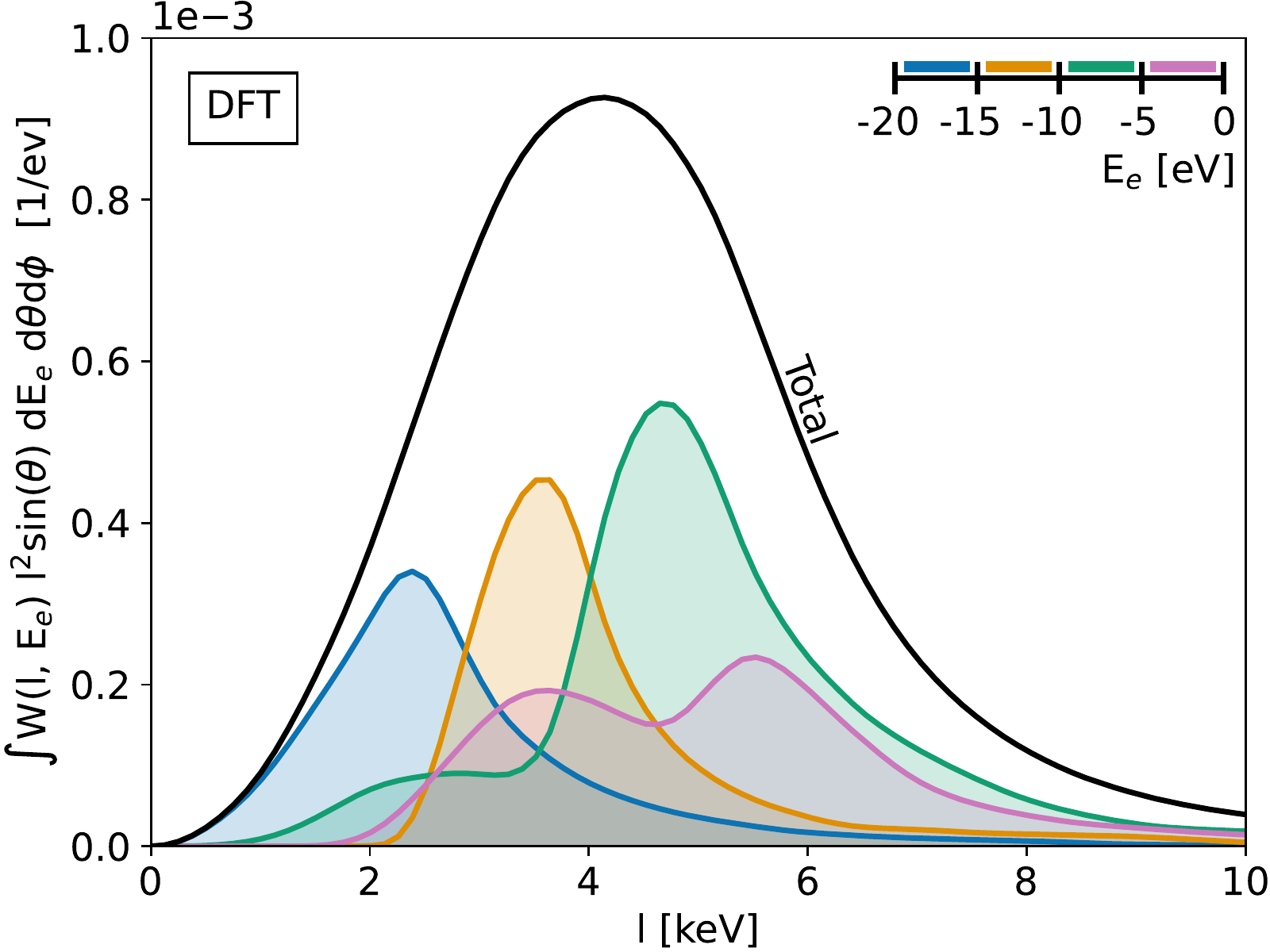}
    \caption{Graphene response function and its dependence on the initial state momentum integrated over various regions of initial state energy and all directions. This plot shows how much different electron energies (that are accessible to different DM candidate masses) contribute to the momentum distribution of the target for TB (left) and DFT (right) simulations.}
    \label{fig: W energy integrated}
\end{figure*}

This is an important observation because it implies that our DFT predictions for $W$ are only marginally affected by the lack of a clear interpretation for the individual Kohn-Sham states, which, in principle, is one of the limitations of a DFT approach.~These states contribute to $W$ mainly through one very specific combination, namely, the Fourier transform of the electron charge density, which, by construction, is self-consistently computed in DFT.

We find this observation a solid argument in favour of DFT as a theoretical framework for computing the graphene response function $W$. This conclusion is also corroborated by the good agreement found between the measured and DFT-calculated graphene Compton profile ~\cite{Feng2019Nov}, which is the longitudinal projection of the electron momentum density, and thus closely related to $W$.

\subsubsection{Numerical implementation}
\label{sec: numerical implementation}

For the numerical evaluation of the graphene responses in the DFT framework, the \texttt{QuantumEspresso v.6.4.1}~\cite{Giannozzi_2009, Giannozzi_2017, doi:10.1063/5.0005082} code was used, and interfaced with \texttt{QEdark-EFT}~\cite{QEdark-EFT}, an extension to the previously established \texttt{QEdark}~\cite{Essig:2015cda} package. Since \texttt{QuantumEspresso} uses a plane-wave basis with periodic boundary conditions, we simulated the graphene sheet as a system containing sheets separated by a large but finite distance, $L_z$. 

For the self-consistent calculations, we used the  \texttt{C.pbe-n-kjpaw\_psl.1.0.0.UPF} pseudopotential provided with the \texttt{QuantumEspresso} package, which includes the $2s^2$ and $2p^2$ electrons in the valence configuration and treats the $1s^2$ electrons as core. The minimal suggested energy cutoff for the plane-wave expansion for this pseudopotential is 40\,Ry for the wave function and 326\,Ry for the charge density. We chose much larger values---2000\,Ry for the wave function cutoff and 16000\,Ry for the charge density cutoff---since for the case of dark-matter induced excitations, we are interested in the high-momentum tails of the electronic wave functions that are usually unimportant for low-energy solid state physics applications.

We used the PBE exchange and correlation functional~\cite{Perdew/Burke/Ernzerhof:1996} with the experimentally measured lattice constant of $a=2.46$ \AA, and sampled reciprocal space with a $16 \times 16 \times 1$ Monkhorst-Pack $k$-point grid, which is sufficient to capture the linearly dispersing Dirac cones at the Fermi level (see Fig.~\ref{fig: band structure}). 
Since carbon is a light atom relativistic effects are minimal and we did not including spin-orbit coupling.

\begin{figure}[t]
\centering
\includegraphics[width=0.45\textwidth]{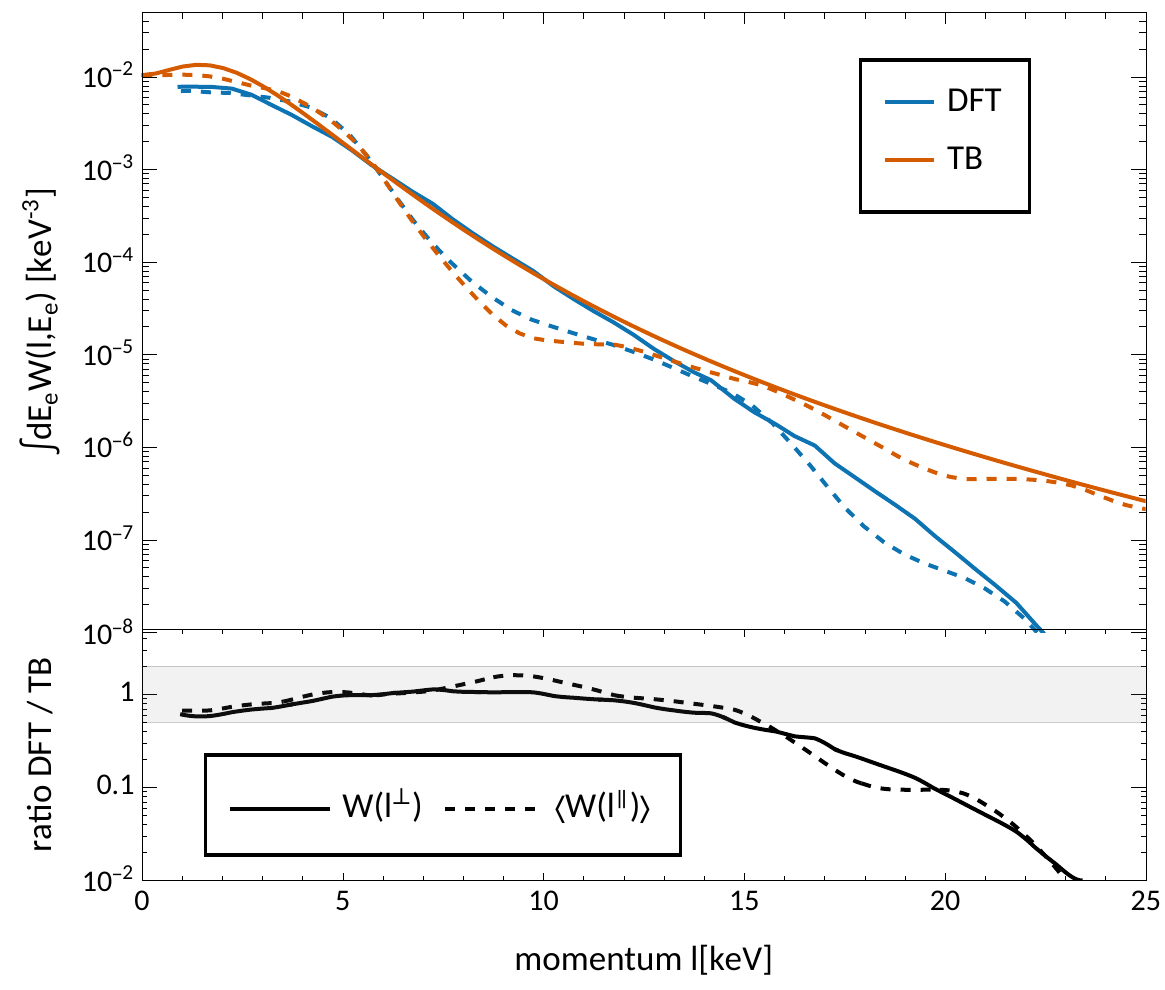}
\caption{Comparison of the partially integrated response function as a function of the initial electron momentum, $\ell$, between DFT (blue) and TB(red). The solid lines illustrate the response function for momenta perpendicular to the graphene sheet. The dashed lines show the dependence of $W$ on momenta parallel to the graphene sheet, where we average over the in-plane directions. The lower panel shows the ratio. As indicated by the gray band, for momenta below $\sim 15$~keV, the two predictions are within a factor of 2. Above, the TB approximation predicts a significantly higher response function.}
\label{fig: W comparison l}
\end{figure}

\begin{figure}[h]
    \centering
      \includegraphics[width=0.45\textwidth]{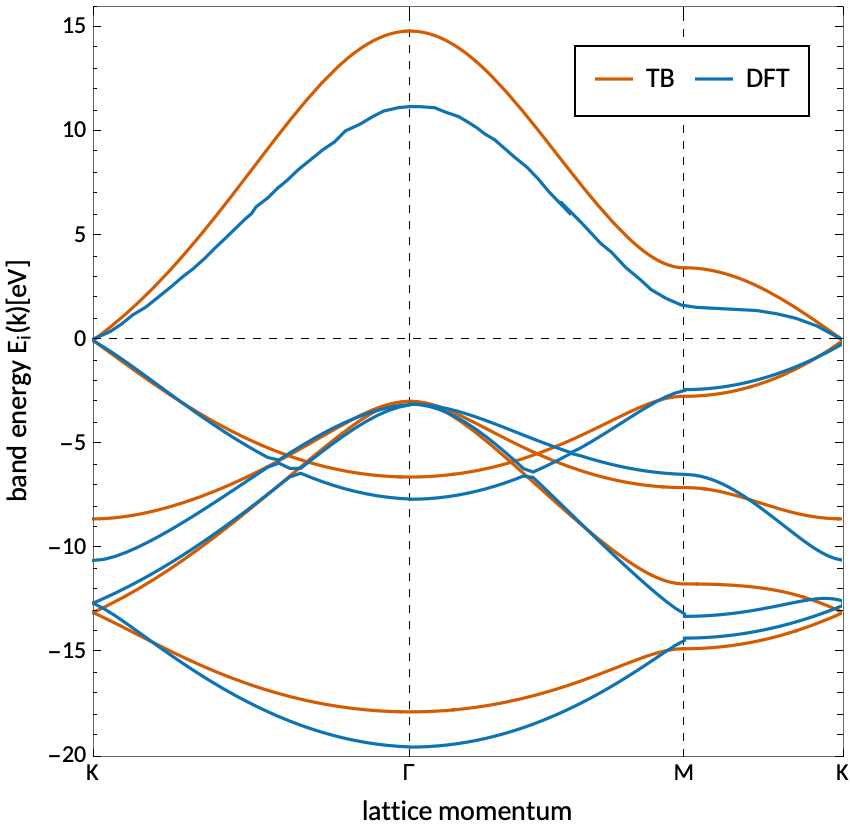}
    \caption{Graphene band structure as calculated from TB and DFT. In the case of DFT, the Dirac cone crossing at K-symmetry points was used to determine a precise value for the Fermi energy of the graphene sheet. We include the valence bands and the conduction band for the~$\pi$-electrons. Further conduction bands are are not shown here, but can be found in Fig.~\ref{fig: energy bands all} for TB.}
    \label{fig: band structure}
\end{figure}

\subsubsection{Discretization}

As a result of the widely spaced periodically repeating graphene sheets required by the periodic boundary conditions of our DFT code, Eq.~(\ref{eq: response function dft}) is discretised and the expression evaluated by \texttt{QEdark-EFT} becomes
\begin{widetext}
\begin{align}
     W\left((\ell_x)_n,(\ell_y)_m,(\ell_z)_o,\left(E_B\right)_l\right) &=\sum_{\mathbf{k},\mathbf{G},i}\frac{\omega_{\mathbf{k}}\left|u_{i}\left(\mathbf{k}+\mathbf{G}\right)\right|^2}{2\delta_{\ell}^3\delta_E}  \Theta\left( 1-\frac{|\left(E_B\right)_l-E_{i\mathbf{k}}|}{\frac{1}{2}\delta_E} \right) \Theta\left( 1-\frac{\left| k_z+G_z-(\ell_z)_o\right|}{\frac{1}{2}\delta_{\ell}} \right)\nonumber\\
     &\times\Theta\left( 1-\frac{\left| k_y+G_y-(\ell_y)_m\right|}{\frac{1}{2}\delta_{\ell}} \right)\Theta\left( 1-\frac{\left| k_x+G_x-(\ell_x)_n\right|}{\frac{1}{2}\delta_{\ell}} \right)\,,
\end{align}
\end{widetext}
where $\delta_E$ and $\delta_{\ell}$ are the bin size in energy and momentum respectively, and $\boldell=\mathbf{k}^\prime-\mathbf{q}$. Subscripts $n$, $m$, $o$ and $l$ denote the index of the corresponding momentum and energy bin, $\omega_\mathbf{k}$ are the weights of the reciprocal lattice k-points and, following the conventions of \texttt{QuantumEspresso}, $\sum_\mathbf{k}\omega_\mathbf{k}=2$. The sum over $i$ runs over $4$ valence bands, and the sum over $\mathbf{G}$ is truncated by 
\begin{align}
    \frac{\left|\mathbf{k}+\mathbf{G}\right|^2}{2m_e}\leq E_\text{cut}
\end{align}
with the value of $E_\text{cut}=27.2\,\mathrm{keV}$. 

\begin{figure*}[t]
    \centering
    \includegraphics[width=0.45\textwidth]{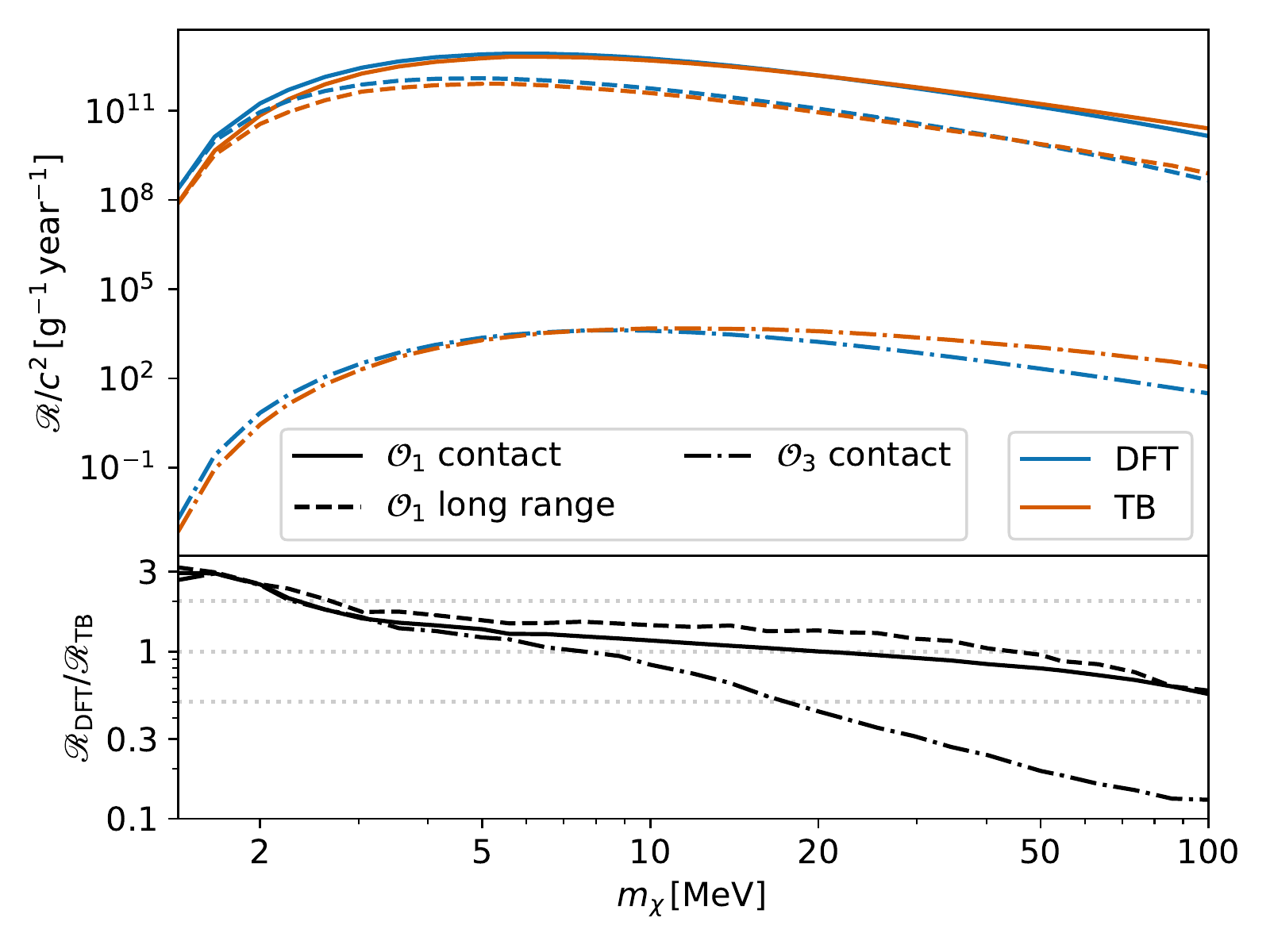}
    \includegraphics[width=0.495\textwidth]{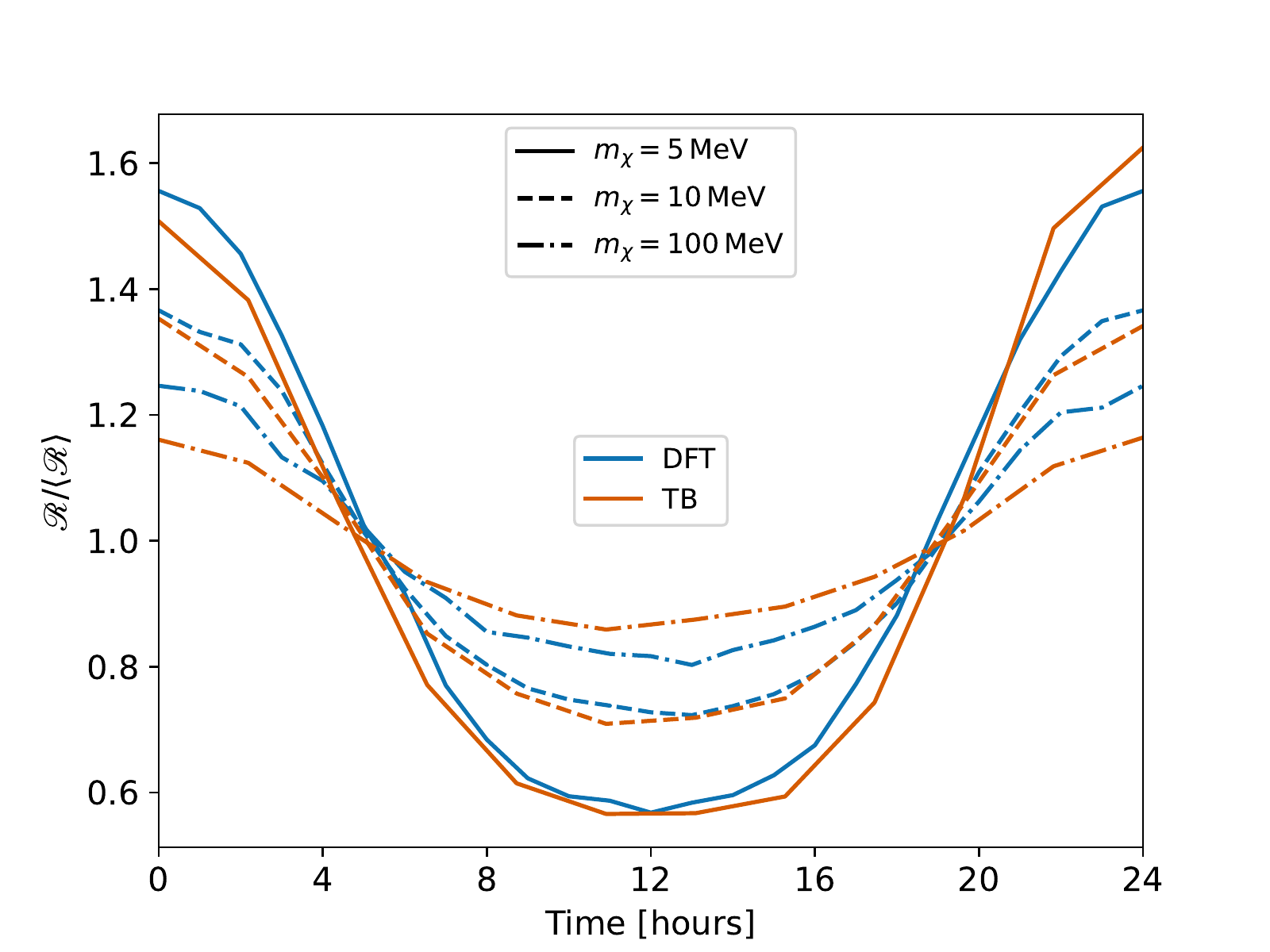}
    \caption{Calculated total time-averaged rate (left) and daily modulation curves (right) of DM-induced electron ejections from graphene obtained with DFT (blue) and TB (red). On the left, we show the rate for $\mathcal{O}_1$ contact type interaction (solid), $\mathcal{O}_1$ long range type interaction (dashed) and $\mathcal{O}_3$ contact type interaction (dash-dotted). The upper panel to the left gives the total rates as a function of DM mass, and the lower panel gives the ratio of the DFT to TB rates. On the right, we show the daily modulation curves for $\mathcal{O}_1$ contact type interaction, where the solid, dashed and dash-dotted curve is for a $5\,\mathrm{MeV}$, $10\,\mathrm{MeV}$ and $100\,\mathrm{MeV}$, respectively.}
    \label{fig: TB vs DFT}
\end{figure*}

\subsection{Comparison of response functions}

In Figs.~\ref{fig: W lx ly plot}-\ref{fig: W comparison l}, we present different ways to visualize the graphene response function obtained with the TB and DFT methods.
In these figures, we integrate the response function defined in Eq.~\eqref{eq: response function general} over all energies~$E_e$ or an interval in~$E_e$ to obtain a function of momentum~$\boldell$ only.

In Fig.~\ref{fig: W lx ly plot}, we depict the response function as a function of $\ell_x$ and $\ell_y$ with $\ell_z\approx 0$.
Both for TB and DFT, we find the characteristic hexagonal shape of the first Brillouin zone of graphene.
In the case of TB, we also depict the individual contributions of the $\pi$ and $\sigma$ electrons.
Note that the distortions of the DFT version originate in the finite grid sampling in reciprocal space and are washed out once we integrate over~$\boldell$ to obtain an observable.

In Fig.~\ref{fig:2D plot W}, we show the response function w.r.t. the angle between the initial state electron momentum and the orientation of the graphene sheet. The response function is integrated over all-electron energies as well as only over electrons within 5\,eV of the Fermi level in order to understand the patterns observed in daily modulation plots for dark matter candidates of various masses. We can see that the DM particles carrying lower kinetic energies will be able to interact preferably with electrons that have their momenta pointing in $\theta \sim$ 40$^\circ$ whereas DM candidates with higher energy will be able to access electrons with momenta pointing in all directions.

In Fig.~\ref{fig: W energy integrated}, we show the contribution of various electron-binding energies to the integrated response function for both TB and DFT. Both approaches show a similar dependence of the momentum contribution integrated over all directions for the selected energy intervals.

In order to facilitate a quantitative comparison of the two approaches, we show the response function as a function of momentum~$\ell^\perp$ (perpendicular to the sheet) and $\ell^\parallel$ (in-sheet momenta) in Fig.~\ref{fig: W comparison l} using a log-scale on the $y$-axis.

Furthermore, the lower panel depicts the ratio of the $W$ functions obtained by DFT and TB.
We can see that for low momenta~$\ell\lesssim 15$ keV, both response functions lie within a factor of 2 of each other.
However, for larger momenta, the TB approximation predicts significantly larger values.

Overall, the Figs.~\ref{fig: W lx ly plot}-\ref{fig: W comparison l} demonstrate that TB and DFT predict the same qualitative features of the graphene response function.
A more quantitative comparison reveals relative deviations between the two approaches that typically do not exceed a factor of 2.
As mentioned above, the exception here is the case of large momenta, where we found larger deviations between the two methods.
The response matrix at large momenta can be sensitive to the electron density close to the atomic nuclei, where the electron orbitals resemble the atomic orbitals the closest. This makes a qualitative comparison or assessment for large momenta difficult as these contributions are smoothened out in the DFT approach.

\section{Case study: daily modulation of the electron ejection rate in a graphene detector}
\label{sec:results}

After our comparative study of the electronic structure calculations of graphene targets in Sec.~\ref{sec:electronic}, we now present and compare the expected electron ejection rates that we obtain using both TB and DFT.
We focus on a hypothetical experimental setting where the electron ejected by an incoming DM particle is recorded independently of the direction of ejection.
We refer to a companion paper (from now onward, Paper II~\cite{PaperII}) for detailed sensitivity studies of different settings for graphene-based DM detectors that are currently in a research and development stage.

Fig.~\ref{fig: TB vs DFT} (left) shows a comparison of the time-averaged TB and DFT rates as a function of the DM particle mass for DM-electron interactions described by the operators $\mathcal{O}_1$ (both contact and long range interactions) and $\mathcal{O}_3$ (contact only) in Tab.~\ref{tab:operators}.

In the case of~$\mathcal{O}_1$, we find that the electron ejection rates predicted by TB and DFT differ by less than a factor of 2 for most DM masses, and up to a factor of 3 at very low masses.
Generally, we find that DFT predicts higher rates at low masses, whereas TB predicts higher rates at high masses.

In the case of $\mathcal{O}_3$ contact type interactions, the quantitive comparison for low masses ($m_\chi \lesssim 20$~MeV) is similar to $\mathcal{O}_1$.
We find larger deviations for heavier masses, however, with the TB approach predicting an $\mathcal{O}(10)$ larger rate at $m_\chi=100$~MeV. 
For this particular operator, large momentum transfers are favoured, and these become kinematically more accessible for larger masses. The difference in rate therefore originates in the graphene response function at large momentum $\boldell$, where TB predicts a higher response than DFT as seen in Fig.~\ref{fig: W comparison l}.
The structure of~$\mathcal{O}_3$ therefore suppresses the contribution of the response function at low momentum, where we have better agreement between the two approaches. Here, the different treatment of the electron density close to the atomic nuclei plays an important role and obstructs a qualitative comparison.
While this is the case of largest deviation between TB and DFT that we present, one should also note, that~$\mathcal{O}_3$ is an extreme but instructive case that does not arise from relativistic DM~models at leading order~\cite{Catena:2022fnk}.

Again comparing the DFT and TB computational frameworks, Fig.~\ref{fig: TB vs DFT} (right) shows the expected daily modulation in the rate of DM-induced electron ejections for three values of the DM particle mass, and for the interaction operator $\mathcal{O}_1$.~In all cases the expected electron ejection rate is divided by the corresponding time-averaged rate, in order to cancel out the $\mathcal{O}(1)$ differences between the DFT and TB rates reported in the left panel of Fig.~\ref{fig: TB vs DFT}.~From Fig.~\ref{fig: TB vs DFT} (right), we conclude that irrespective of the chosen computational framework, the largest relative rate is found when the direction of the DM wind is perpendicular to the graphene sheet. This is due to the fact that the electrons are more spatially constrained in the out-of-plane direction, giving rise to higher momentum contribution to the electron wavefunction and thus increasing the total interaction rate (as discussed further in Paper II). The strong daily modulation we find for the rate of DM-induced electron ejections demonstrates that graphene is well suited for establishing a daily modulation signal characteristic of the directionality of the DM wind.

\begin{figure}[t]
    \centering
    \includegraphics[width=0.485\textwidth]{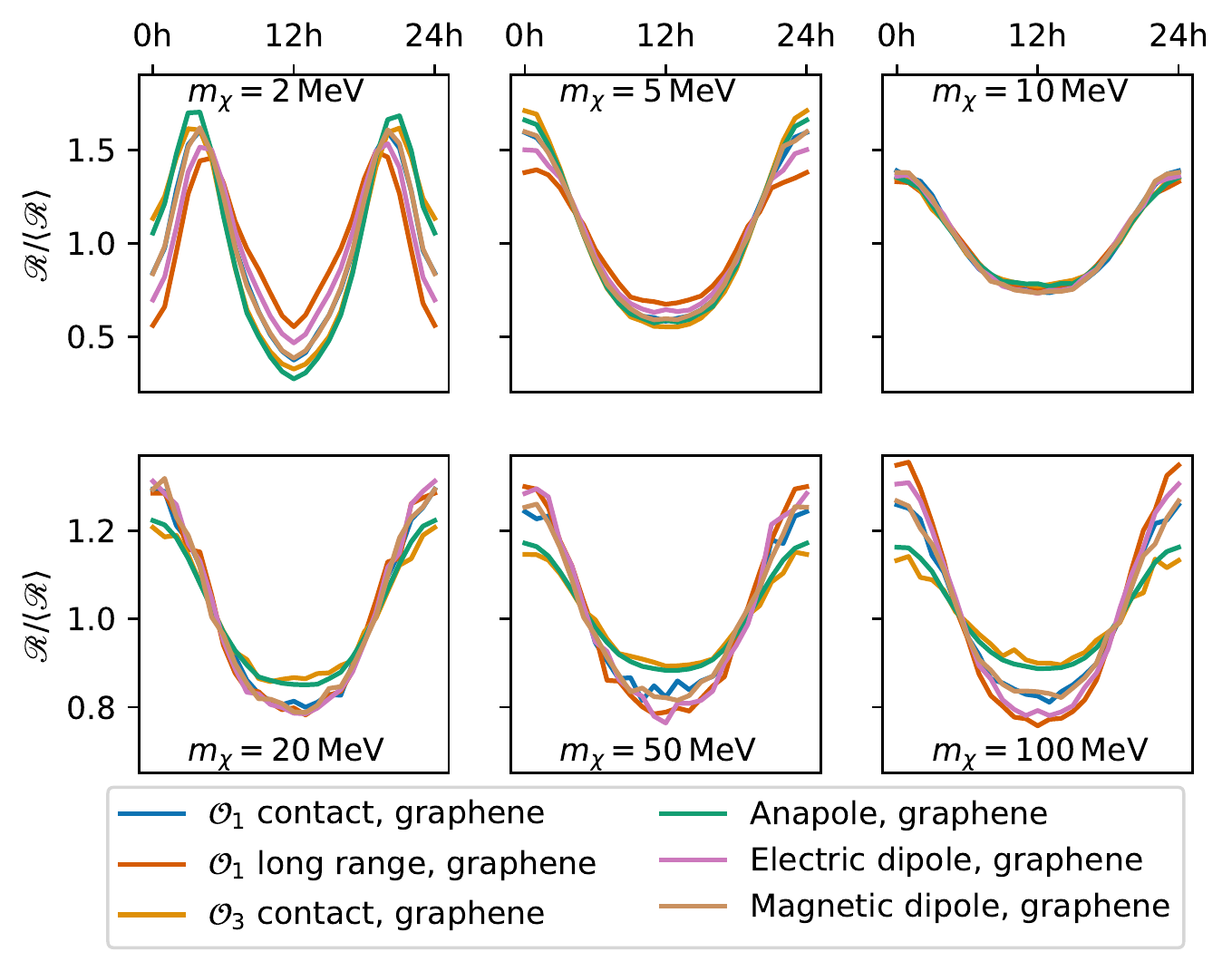}
    \caption{Daily modulation pattern for graphene sheets obtained with DFT. The colors in each panel correspond to the interaction types indicated in the legend, and the top left, top center, top right, bottom left, bottom center and bottom right corresponding to DM masses of $2\,\mathrm{MeV}$, $5\,\mathrm{MeV}$, $10\,\mathrm{MeV}$, $20\,\mathrm{MeV}$, $50\,\mathrm{MeV}$ and $100\,\mathrm{MeV}$, respectively. Note that the $y$-axis differs between the top and bottom plots. Note that for DM masses around $10\,\mathrm{MeV}$, the modulation pattern is similar for all the considered interactions, indicating that the expected result is largely model-independent.}
 \label{fig: daily modulation sheets}
\end{figure}

In Fig.~\ref{fig: daily modulation sheets} we show the daily modulation pattern of electron ejections from graphene for various DM masses and interaction types, now focusing on DFT as a computational framework as the corresponding results from TB are qualitatively similar.~The daily modulation pattern is similar for most of the DM masses and interactions with a maximum at around time=$0$h (when the DM wind is perpendicular to the graphene sheet) and a minimum at around time=$12$h (as for Fig.~\ref{fig: TB vs DFT} (right)). However, for $m_\chi=2\,\mathrm{MeV}$, the maximum is shifted to two peaks around time=$4$h and time=$20$h. This is a consequence of the $2\,\mathrm{MeV}$ DM particle only being able to eject electrons with $E_e$ close to $0$, corresponding to the top two panels in Fig.~\ref{fig:2D plot W}. The location of the peaks around time=$4$h and time=$20$h is due to the DM wind aligning with the peak at $\theta\sim 30^\circ$ and $\ell \sim 4\,\mathrm{keV}$. 

\section{Summary and conclusion}
\label{sec:conclusions}

In this paper, we have investigated two solid-state-physics approaches to modeling DM interactions with graphene-like targets, TB and DFT. 
Below, we summarise the main features of the two methods and the arguments that led us to identify DFT as the preferred framework for modeling the scattering of DM particles by electrons bound in graphene.

Both DFT and TB capture the main characteristics of the band structure of graphene, such as the Dirac cone at the K-symmetry point of the Brillouin zone and valence band energy distributions (Fig.~\ref{fig: band structure}), and predict a response function that reflects the symmetry of the reciprocal space underlying the graphene lattice (Fig.~\ref{fig: W lx ly plot}). The two computational frameworks also predict a qualitatively similar daily modulation pattern in the total rate of electron ejections.
 However, the two methods generally predict electron ejection rates that differ by an~$\mathcal{O}(1)$ factor with TB~(DFT) predicting higher rates for high~(low) DM~masses.

The two approaches employ a different set of approximations that limit their predictivity and region of validity. In order to effectively perform DFT calculations, one usually chooses a radial cutoff to the pseudo-potential and smoothens the core-electron wavefunctions closer to the atomic nucleus. This approximates the electronic structure below that cutoff and suppresses some of the high-momentum contributions to the electronic wavefunction. The total event rate is therefore suppressed when higher energy excitations (of tens of eV) are considered~\cite{Griffin:2021znd}. 

On the other hand, DFT is able to provide a self-consistent calculation of the ground-state electron density, which is the indirect quantity of interest for the DM interactions considered in this work.~Indeed, an important result of this work is our demonstration that in cases where the outgoing electrons can be treated as a plane wave, the DM and electronic contributions to the DM-induced electron ejection rate factorize. The five crystal response functions identified previously in our work~\cite{Catena:2021qsr} then simplify into a single crystal response that is directly proportional to ``diagonal part'' of the Fourier transform of the ground-state electron density.

The TB approach has the advantage of easier implementability and computational affordability.
In its usual low-energy applications, the detailed form of the atomic basis wavefunctions is not explicitly considered. However, since for the case of DM-electron scattering, we are interested in the explicit form of the electronic wavefunctions, one needs to embed the atomic wavefunctions into this framework.

This proves problematic since these atomic wavefunctions are required to satisfy the overlap integrals of TB that reproduce the experimentally observed band structure. The TB approach assumes that a wavefunction satisfying these relations exists, but does not allow us to calculate its form directly. One possible approach for modeling it is to use the hydrogenic wavefunctions and adjust their parameters such that they would satisfy the imposed overlap integrals (as employed in~\cite{Hochberg:2016ntt}). These modified hydrogenic wavefunctions, however, differ significantly from those of real carbon atoms bound within the graphene lattice which is limiting their predictive powers and makes the atomic contribution to the total electronic momenta unreliable. 

Another approach is to use the Roothaan-Hartree-Fock wavefunctions fit to describe unbound carbon atoms instead of hydrogenic wavefunctions as the atomic basis. While this atomic basis does describe the individual carbon atoms better than the hydrogenic wavefunctions, they do not satisfy the overlap integrals given by the structure of the graphene lattice, creating an inconsistency in the implementation of the theory. In order to satisfy these relations, one would have to significantly distort the shape of the atomic orbitals, spoiling the original fit describing the carbon atom.

The problem underlying the form of the atomic orbitals within TB, together with the robust predictive powers of the ground-state electron density of DFT have led us to recommend DFT as the framework of choice 
for graphene-like DM detector modeling. We will further expand this topic and use DFT to obtain predictions for various possible detector setups and DM candidates in the associated Paper~II.

The research software \texttt{Darphene} and an updated version of \texttt{QEdark-EFT} used to obtain the TB and DFT results respectively, will be made publicly available~\cite{Darphene,QEdark-EFT}.

\acknowledgments
The authors thank Yonatan Kahn for valuable discussions and for sharing their code.
R.C. and T.E. acknowledge support from the Knut and Alice Wallenberg project grant Light Dark Matter (Dnr.~KAW 2019.0080). Furthermore, R.C. acknowledges support from individual research grants from the Swedish Research Council, Dnr. 2018-05029 and Dnr. 2022-04299. TE was also supported by the Knut and Alice Wallenberg Foundation (PI, Jan Conrad). N.A.S. and M.M. were supported by the ETH Zürich and by the European Research Council (ERC) under the European Union’s Horizon 2020 research and innovation program Grant Agreement No. 810451.
T.E. thanks the Theoretical Subatomic Physics group at Chalmers University of Technology for its hospitality.

The research presented in this paper made use of the following software packages, libraries, and tools: \texttt{Arb}~\cite{Johansson2017arb}, \texttt{boost}~\cite{boost}, \texttt{Eigen}~\cite{eigen}, \texttt{libphysica}~\cite{libphysica}, \texttt{obscura}~\cite{Emken:2021uzb,obscura}, \texttt{QuantumEspresso}~\cite{Giannozzi_2009, Giannozzi_2017, doi:10.1063/5.0005082}, WebPlotDigitizer~\cite{webplotdigitizer}, and Wolfram Mathematica~\cite{Mathematica}.
Part of the computations were enabled by resources provided by the Swedish National Infrastructure for Computing (SNIC) at the National Supercomputer Centre (NSC).

\appendix

\section{Expanded matrix element}
\label{app: matrix element}
In this appendix, we explicitly give the free particle response function $R_\mathrm{free}$ from Eq.~(\ref{eq: DM material factorisation}). To avoid making the expressions too large, we split $R_\mathrm{free}$ into three separate terms,
\begin{widetext}
\begin{align}
    R_\mathrm{free}=&\overline{\left|\mathcal{M}\right|^2}+ 2 m_e \overline{\Re\left[\mathcal{M}(\nabla_{\boldell}\mathcal{M}^*)_{\boldell=0}\cdot\frac{\mathbf{q}-\mathbf{k}^\prime}{m_e}\right]} + m_e^2 \overline{\left|(\nabla_{\boldell}\mathcal{M})_{\boldell=0}\cdot \frac{\mathbf{q}-\mathbf{k}^\prime}{m_e}\right|^2} \,,
\end{align}
where $\mathbf{q}$ is the momentum transfer, $m_e$ is the electron mass, $\mathbf{k}^\prime$ is the final state electron momentum, and $\boldell$ is the initial state electron momentum. The individual terms can then be expressed as
\begin{align}
    \overline{\left|\mathcal{M}\right|^2} =& c_1^2 + \frac{c_3^2}{4}\left( \frac{\mathbf{q}}{m_e}\times\vPerpEl \right)^2+ \frac{c_7^2}{4}\left(\vPerpEl\right)^2+ \frac{c_{10}^2}{4}\left(\frac{\mathbf{q}}{m_e}\right)^2 +\frac{j_\chi(j_\chi+1)}{12}\Bigg\{ 3 c_4^2+\left(4c_5^2-2c_{12}c_{15}\right)\left( \frac{\mathbf{q}}{m_e}\times\vPerpEl \right)^2 \nonumber\\
    & + c_6^2 \left(\frac{\mathbf{q}}{m_e}\right)^4 + \left(4 c_8^2+2c_{12}^2\right) \left(\vPerpEl\right)^2  + \left(2c_{9}^2+4c_{11}^2+2c_4c_6\right)\left(\frac{\mathbf{q}}{m_e}\right)^2+\left(c_{13}^2+c_{14}^2\right)\left(\frac{\mathbf{q}}{m_e}\right)^2\left(\vPerpEl\right)^2 \nonumber\\
    &+ c_{15}^2\left(\frac{\mathbf{q}}{m_e}\right)^2\left( \frac{\mathbf{q}}{m_e}\times\vPerpEl \right)^2 +2c_{13}c_{14}\left(\frac{\mathbf{q}}{m_e}\cdot \vPerpEl\right)\left(\frac{\mathbf{q}}{m_e}\cdot \vPerpEl\right)\Bigg\}\, ,
\end{align}
where $\vPerpEl=\mathbf{v}-\frac{\mathbf{q}}{2\mu_{\chi e}}-\frac{\boldell}{m_e}$ with $\mathbf{v}$ being the DM initial velocity in the detector rest frame and $\mu_{\chi e}$ the DM-electron reduced mass. $c_i$'s are the effective couplings, and $j_\chi$ is the DM spin which we typically set to 1/2.
\begin{align}
    2 m_e \overline{\Re\left[\mathcal{M}(\nabla_{\mathbf{p}_1}\mathcal{M}^*)_{\mathbf{p}_1=0}\cdot\frac{\mathbf{q}-\mathbf{k}^\prime}{m_e}\right]}=& \left[\frac{c_3^2}{2}\left(\left(\frac{\mathbf{q}}{m_e}\cdot \vPerpEl\right) \frac{\mathbf{q}}{m_e}-\left(\frac{\mathbf{q}}{m_e}\right)^2\vPerpEl\right) - \frac{c_7^2}{2}\vPerpEl\right]\cdot\frac{\mathbf{q}-\mathbf{k}^\prime}{m_e}\nonumber\\
 &+\frac{j_\chi(j_\chi+1)}{6}\Bigg\{ \bigg[\left(4c_5^2+c_{15}^2\left(\frac{\mathbf{q}}{m_e}\right)^2\right)\left(\left(\frac{\mathbf{q}}{m_e}\cdot \vPerpEl\right) \frac{\mathbf{q}}{m_e}-\left(\frac{\mathbf{q}}{m_e}\right)^2\vPerpEl\right)\nonumber\\
 &-\left(4c_8^2+2c_{12}^2+(c_{13}^2+c_{14}^2)\left(\frac{\mathbf{q}}{m_e}\right)^2\right)\vPerpEl\bigg]\cdot\frac{\mathbf{q}-\mathbf{k}^\prime}{m_e}\nonumber\\
 &-2c_{12}c_{15}\left(\left(\frac{\mathbf{q}}{m_e}\cdot \vPerpEl\right) \frac{\mathbf{q}}{m_e}-\left(\frac{\mathbf{q}}{m_e}\right)^2\vPerpEl\right)\cdot\frac{\mathbf{q}-\mathbf{k}^\prime}{m_e}\nonumber\\
 &-2c_{13}c_{14}\left(\frac{\mathbf{q}}{m_e}\cdot \vPerpEl\right)\frac{\mathbf{q}}{m_e}\cdot\frac{\mathbf{q}-\mathbf{k}^\prime}{m_e}\Bigg\}\,,
\end{align}
and 
\begin{align}
    m_e^2 \overline{\left|(\nabla_{\mathbf{p}_1}\mathcal{M})_{\mathbf{p}_1=0}\cdot \frac{\mathbf{q}-\mathbf{k}^\prime}{m_e}\right|^2} =&\left(\frac{c_3^2}{4}\left(\frac{\mathbf{q}}{m_e}\right)^2+\frac{c_7^2}{4}\right)\left(\frac{\mathbf{q}-\mathbf{k}^\prime}{m_e}\right)^2 -\frac{c_3^2}{4}\left(\frac{\mathbf{q}}{m_e}\cdot \frac{\mathbf{q}-\mathbf{k}^\prime}{m_e}\right)^2\nonumber\\
&+\frac{j_\chi(j_\chi+1)}{12}\Bigg\{ \left(\frac{\mathbf{q}-\mathbf{k}^\prime}{m_e}\right)^2 \Bigg[(4c_5^2+c_{13}^2+c_{14}^2-2c_{12}c_{15})\left(\frac{\mathbf{q}}{m_e}\right)^2+4c_8^2+2c_{12}^2 \nonumber\\
&+c_{15}^2\left(\frac{\mathbf{q}}{m_e}\right)^4\Bigg]+ \left(\frac{\mathbf{q}}{m_e}\cdot \frac{\mathbf{q}-\mathbf{k}^\prime}{m_e}\right)^2 \left[ -4c_5^2-c_{15}^2\left(\frac{\mathbf{q}}{m_e}\right)^2+2c_{12}c_{15}+2c_{13}c_{14})\right]  \Bigg\}\, .
\end{align}
\end{widetext}
Furthermore, to rewrite the above equations one can use the following relations
\begin{align}
    \left(\frac{\mathbf{q}}{m_e}\times \vPerpEl\right)^2=&\left(\frac{\mathbf{q}}{m_e}\right)^2\left(\vPerpEl\right)^2-\left(\frac{\mathbf{q}}{m_e}\cdot \vPerpEl\right)^2\,,\\
    \left(\vPerpEl\right)^2|_{\boldell=0}=&\mathbf{v}^2+\frac{\mathbf{q}^2}{4\mu_{\chi e}^2}\frac{m_\chi-m_e}{m_e+m_\chi}-\frac{\Delta E_e}{\mu_{\chi e}}\,,\\
    \left(\vPerpEl\cdot \mathbf{q}\right)|_{\boldell=0}=&\Delta E_e - \frac{\mathbf{q}^2}{2m_e}\,,\\
    \vPerpEl|_{\boldell=0}=&\mathbf{v}-\frac{\mathbf{q}}{2\mu_{\chi e}}\,,
\end{align}
where $\Delta E_e$ is the energy transferred to the target electron.

\section{The lattice structure of graphene}
\label{app: graphene}

\begin{figure*}
    \centering
    \includegraphics[width=0.35\textwidth]{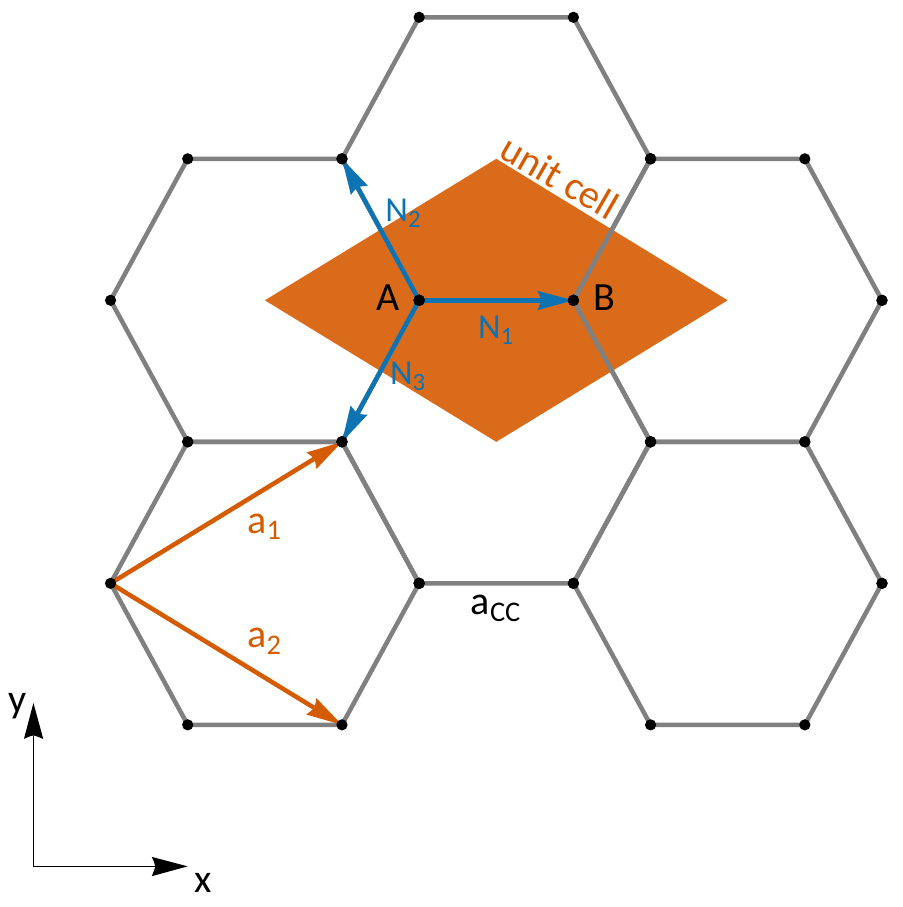}
    \includegraphics[width=0.35\textwidth]{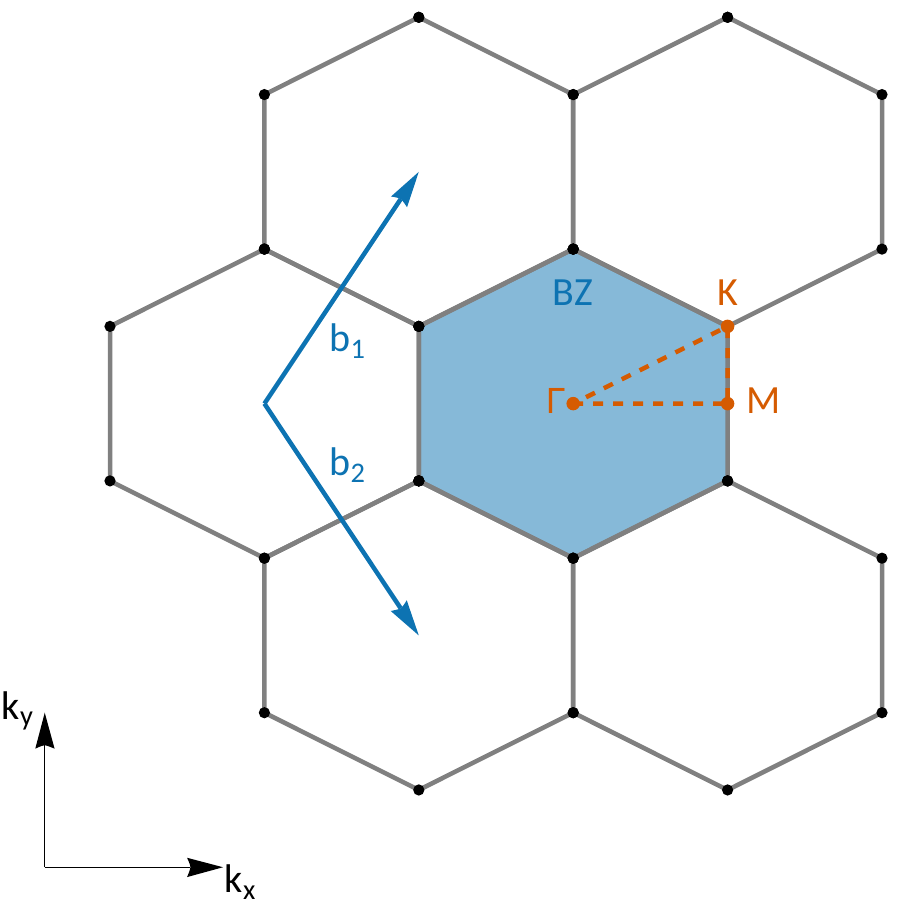}
    \caption{The honeycomb lattice of graphene (left) and the reciprocal lattice (right), together with their respective lattice vectors~$\mathbf{a}_i$ and $\mathbf{b}_i$. The red (blue) shaded area is the unit cell (Brillouin zone) of the (reciprocal) lattice.}
    \label{fig: lattice space}
\end{figure*}

Graphene is a two-dimensional hexagonal lattice of carbon atoms as illustrated in the left panel of Fig.~\ref{fig: lattice space}.
The distance between two neighboring carbon atoms is
\begin{align}
    a_{\rm CC} = 1.42\,\angstrom\, .
\end{align}
The two lattice vectors, which can also be seen in the left panel of fig.~\ref{fig: lattice space} are
\begin{subequations}
\label{eq: lattice vectors}
\begin{align}
    \mathbf{a}_1 &= a \columnvector{3}{\frac{\sqrt{3}}{2}}{\frac{1}{2}}{0}\, ,\quad \mathbf{a}_2 = a\columnvector{3}{\frac{\sqrt{3}}{2}}{-\frac{1}{2}}{0}\, ,
    \intertext{such that}
    a&\equiv | \mathbf{a}_1| = | \mathbf{a}_2| = \sqrt{3} \;a_{\rm CC} \approx 2.46\,\angstrom\, .
\end{align}
\end{subequations}
The same figure also shows the vectors~$\mathbf{N}_i$, pointing to a carbon atom's three closest neighbor atoms,
\begin{subequations}
\label{eq: nearest neighbors}
\begin{align}
    \mathbf{N}_1 &= a_{\rm CC}\columnvector{3}{1}{0}{0} = \columnvector{3}{\frac{a}{\sqrt{3}}}{0}{0}\,,\\
    \mathbf{N}_2 &= a_{\rm CC}\columnvector{3}{-\frac{1}{2}}{\frac{\sqrt{3}}{2}}{0} = \columnvector{3}{-\frac{a}{2\sqrt{3}}}{\frac{a}{2}}{0}\,, \\
    \mathbf{N}_3 &= a_{\rm CC}\columnvector{3}{-\frac{1}{2}}{-\frac{\sqrt{3}}{2}}{0}=  \columnvector{3}{-\frac{a}{2\sqrt{3}}}{-\frac{a}{2}}{0}\, .
\end{align}
\end{subequations}

The lattice vectors~$\mathbf{b}_i$ of the reciprocal lattice, illustrated in the right panel of Fig.~\ref{fig: lattice space}, are
\begin{subequations}
\label{eq: inverse lattice vectors}
\begin{align}
    \mathbf{b}_1 &= b \columnvector{3}{\frac{1}{2}}{\frac{\sqrt{3}}{2}}{0}\, ,\quad \mathbf{b}_2 = b \columnvector{3}{\frac{1}{2}}{-\frac{\sqrt{3}}{2}}{0}\, ,
    \intertext{with}
    b&\equiv | \mathbf{b}_1| = | \mathbf{b}_2| = \frac{4\pi}{\sqrt{3}a}\, .
    \label{eq:b}
\end{align}
\end{subequations}
The figure also shows the three high-symmetry point in the first Brillouin zone~(BZ) (shaded in blue) in~$\mathbf{k}$-space,
\begin{subequations}
\label{eq: high symmetry points}
\begin{align}
    \mathbf{\Gamma} &=\columnvector{3}{0}{0}{0}\, ,\; \mathbf{M} = \columnvector{3}{\frac{2\pi}{3a_{\rm CC}}}{0}{0} =\columnvector{3}{\frac{2\pi}{\sqrt{3}a}}{0}{0}\, ,\\
    \mathbf{K} &=\columnvector{3}{\frac{2\pi}{3a_{\rm CC}}}{\frac{2\pi}{3\sqrt{3}a_{\rm CC}}}{0}=\columnvector{3}{\frac{2\pi}{\sqrt{3}a}}{\frac{2\pi}{3a}}{0}\, .
\end{align}
\end{subequations}

\section{The tight-binding approximation}
\label{app: tight binding}

\subsection{General review}
\label{app: tight binding review}

In this section, we review the tight-binding approximation.
In particular, we emphasize how the energy dispersion~$E_i(\mathbf{k})$, appearing in Eq.~\eqref{eq: energy difference}, and wave function coefficients~$C_{ij}(\mathbf{k})$ of Eq.~\eqref{eq: wave function tb} are evaluated in general using this technique~\cite{saito1998physical}.
The translational symmetry of a lattice should also be reflected in the wave functions. In particular, the electron wave function~$\Psi(\mathbf{x})$ has to satisfy \textit{Bloch's theorem},
\begin{align}
\label{eq: Bloch theorem}
    \mathbf{T}_\mathbf{a} \Psi(\mathbf{x}) \equiv \Psi(\mathbf{x}+\mathbf{a}) = e^{i\mathbf{k}\cdot\mathbf{a}}\Psi(\mathbf{x})\, . 
\end{align}
Here, we introduced the translation operator~$\mathbf{T}_\mathbf{a}$ along one of the lattice vectors~$\mathbf{a}$, and the lattice momentum~$\mathbf{k}$.

One way to write down a generic wave function satisfying Bloch's theorem can be obtained using the \textit{tight binding} approximation. A tight-binding Bloch function~$\Phi_{j\mathbf{k}}(\mathbf{x})$ is an approximation to the system's wave functions which is defined by summing up the wave functions of the $j^{\rm th}$ atomic orbital of isolated atoms at their respective lattice site,
\begin{align}
\label{eq: bloch wave function tb}
    \Phi_{j\mathbf{k}}(\mathbf{x}) &= \frac{1}{\sqrt{N}}\sum_{k=1}^{N}e^{i\mathbf{k}\cdot\mathbf{R}_k}\varphi_j(\mathbf{x}-\mathbf{R}_k)\, ,\quad j = 1,\dots, n\, .
\end{align}
This way, Bloch functions sum up the wave functions~$\varphi_j(\mathbf{x})$ of~$N$ unit cells weighted with a lattice site dependent phase which ensures that Eq.~\eqref{eq: Bloch theorem} is satisfied.
Note that, even if the isolated atomic wave functions~$\varphi_j(\mathbf{x})$ are normalized, the overlaps between neighboring wave functions generally render the Bloch functions as non-normalized.

The actual electron wave functions of the material are linear combinations of the Bloch functions, mixing the different atomic orbitals (but not different lattice momenta),
\begin{align}
\label{eq: wave function general}
    \Psi_{i\mathbf{k}}(\mathbf{x}) = \mathcal{N}_\mathbf{k} \sum_{j = 1}^n C_{ij}(\mathbf{k})\Phi_{j\mathbf{k}}(\mathbf{x})\, ,
\end{align}
where the constant~$\mathcal{N}_\mathbf{k}$ ensures that the wave function~$\Psi_{i\mathbf{k}}(\mathbf{x})$ is normalized.

Using Schr\"odinger's equation, the energy values of these~$n$ states are
\begin{subequations}
    \label{eq: energy eigenvalues}
\begin{align}
    E_i(\mathbf{k}) &= \frac{ \braOket{\Psi_i}{\mathcal{H}}{\Psi_i}}{\braket{\Psi_i}{\Psi_i}}\\
    &= \frac{\int\dd^3\mathbf{x}\;\Psi^*_{i\mathbf{k}}(\mathbf{x})\;\mathcal{H}\;\Psi_{i\mathbf{k}}(\mathbf{x})}{\int\dd^3\mathbf{x}\;\Psi^*_{i\mathbf{k}}(\mathbf{x})\Psi_{i\mathbf{k}}(\mathbf{x})}
\end{align}
\end{subequations}
Substituting Eq.~\eqref{eq: wave function general}, the energy eigenvalues can be expressed as
\begin{subequations}
\label{eq: energy eigenvalues 2}
\begin{align}
    E_i(\mathbf{k}) &= \frac{\sum_{j=1}^n\sum_{j^\prime=1}^n C^*_{ij}(\mathbf{k})C_{ij^\prime}(\mathbf{k})\braOket{\Phi_j}{\mathcal{H}}{\Phi_{j^\prime}}}{\sum_{j=1}^n\sum_{j^\prime=1}^n C^*_{ij}(\mathbf{k})C_{ij^\prime}(\mathbf{k})\braket{\Phi_j}{\Phi_{j^\prime}}}\\
    &\equiv\frac{\mathbf{C}^\dagger_i(\mathbf{k})\cdot \boldsymbol{\mathcal{H}}(\mathbf{k})\cdot\mathbf{C}_i(\mathbf{k})}{\mathbf{C}^\dagger_i(\mathbf{k})\cdot \boldsymbol{\mathcal{S}}(\mathbf{k})\cdot\mathbf{C}_i(\mathbf{k})}\, .
\end{align}
\end{subequations}
Here, we defined the coefficient vectors
\begin{align}
    \mathbf{C}_i(\mathbf{k}) = \columnvector{3}{C_{i1}(\mathbf{k})}{\vdots}{C_{in}(\mathbf{k})}\, ,
\end{align}
as well as the transfer integral matrix~$\boldsymbol{\mathcal{H}}(\mathbf{k})$, and the overlap integral matrix~$\boldsymbol{\mathcal{S}}(\mathbf{k})$.
Their components are defined as
\begin{align}
    \left[\boldsymbol{\mathcal{H}}(\mathbf{k})\right]_{ij} &= \braOket{\Phi_i}{\mathcal{H}}{\Phi_j}\, ,\label{eq: transfer integral matrix}\\
    \left[\boldsymbol{\mathcal{S}}(\mathbf{k})\right]_{ij} &= \braket{\Phi_i}{\Phi_j}\, .\label{eq: overlap integral matrix}
\end{align}

As already seen in Eq.~\eqref{eq: energy eigenvalues 2}, the normalization of~$\Psi_{i\mathbf{k}}(\mathbf{x})$ can be written in terms of the coefficient vector~$\mathbf{C}_i(\mathbf{k})$ and the overlap matrix~$\boldsymbol{\mathcal{S}}(\mathbf{k})$,

\begin{align}
    \braket{\Psi_i}{\Psi_i} &= \mathcal{N}_\mathbf{k}^2\;\mathbf{C}^\dagger_i(\mathbf{k})\cdot \boldsymbol{\mathcal{S}}(\mathbf{k})\cdot\mathbf{C}_i(\mathbf{k})\nonumber\\
    \Rightarrow \mathcal{N}_\mathbf{k} &= \left[ \mathbf{C}^\dagger_i(\mathbf{k})\cdot \boldsymbol{\mathcal{S}}(\mathbf{k})\cdot\mathbf{C}_i(\mathbf{k})\right]^{-1/2}\, .\label{eq: wave function norm general}
\end{align}

The entries of the coefficient vector~$\mathbf{C}_i(\mathbf{k})$ are obtained using the variational principle by minimizing the energy eigenvalues~$E_i(\mathbf{k})$, i.e.
\begin{align}
    \frac{\partial E_i(\mathbf{k})}{\partial C_{ij}^*} &= 0\, .
    \intertext{This equation is equivalent to the following general eigenvalue problem,}
    \left[\boldsymbol{\mathcal{H}}-E_i(\mathbf{k}) \boldsymbol{\mathcal{S}}\right]\cdot\mathbf{C}_{i}(\mathbf{k}) &= 0\, . \label{eq: general eigenvalue problem}
\end{align}
Non-vanishing eigenvectors can only be found, if the \textit{secular equation} applies,
\begin{align}
\label{eq: secular equation}
    \text{det}\left[\boldsymbol{\mathcal{H}}-E_i(\mathbf{k}) \boldsymbol{\mathcal{S}}\right] = 0\, .
\end{align}
For a fixed~$\mathbf{k}$, this polynomial of degree~$n$ can be solved for the~$n$ eigenvalues, i.e. the energy dispersion~$E_i(\mathbf{k})$. 
Furthermore, the eigenvectors determine the electron wave functions~$\Psi_{i\mathbf{k}}(\mathbf{x})$.

Finally, the entries of the transfer integral matrix~$\boldsymbol{\mathcal{H}}(\mathbf{k})$, and the overlap integral matrix~$\boldsymbol{\mathcal{S}}(\mathbf{k})$ are often fixed to specific values, which ensure that the correct band structure of the material of interest is reproduced. These values are obtained either experimentally or from first-principles calculations.

\subsection{Tight-binding approximation for graphene}
\label{app: tight binding graphene}

We will apply the results of the previous section to the case of graphene.
As illustrated in Fig.~\ref{fig: lattice space}, the unit cell of graphene consists of two carbon atoms, denoted~$A$ and~$B$.
The relevant atomic orbitals of the carbon atoms on these locations are~$2s$, $2p_x$,$2p_y$, and $2p_z$ (the~$1s$ orbitals form a low energy core state and do not contribute to the valence band levels).
In graphene, the first three orbitals combine or hybridize and form the so-called~$\sigma$-bands and the~$2p_z$ orbitals hybridize to form the~$\pi$-bands.
As a result, there are $n=2(6)$ $\pi(\sigma)$-bonding atomic orbitals in each unit cell, and Eq.~\eqref{eq: secular equation} constitutes a two-(six-) dimensional eigenvalue problem for the $\pi(\sigma)$~band.
As we will describe in detail, the energy dispersion and wave function for the~$\pi$~electrons can be expressed analytically, whereas the~$\sigma$ electrons require numerical methods to solve the six-dimensional secular equation.

\subsubsection*{\boldmath The $\pi$-electrons}

\begin{figure*}
    \centering
    \includegraphics[width=0.4\textwidth]{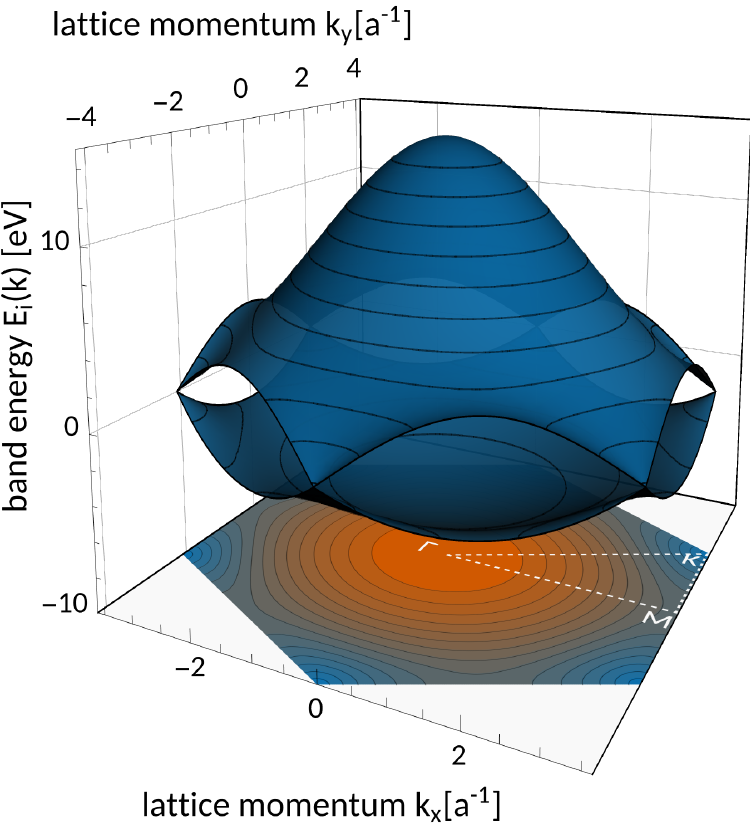}
    \includegraphics[width=0.37\textwidth]{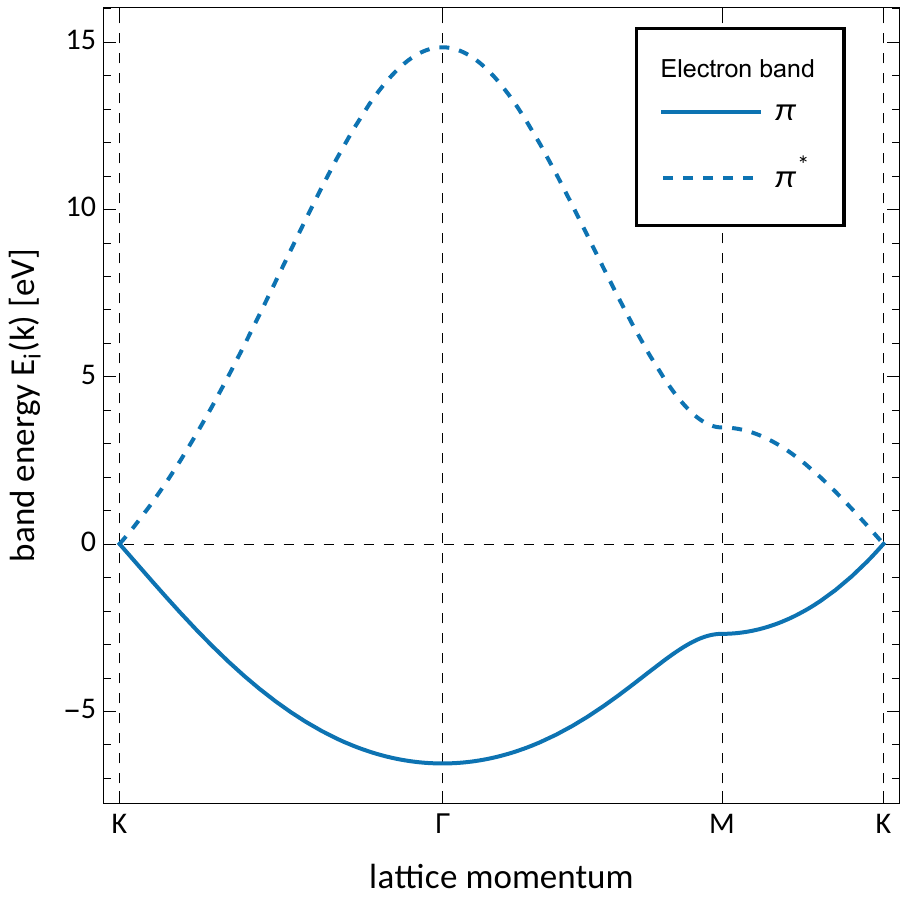}
    \caption{Energy bands of the $\pi$-electrons in graphene evaluated in the tight-binding approximation, see Eq.~\eqref{eq: energy eigen values pi}.}
    \label{fig: energy bands pi}
\end{figure*}
The~$\pi$-electrons are a hybridization of the atomic~$2p_z$ orbitals of carbon.
In order to compute the corresponding energy bands by solving Eq.~\eqref{eq: secular equation}, the main step is to evaluate the transfer integral and overlap integral matrices~$\boldsymbol{\mathcal{H}}(\mathbf{k})$ and~$\boldsymbol{\mathcal{S}}(\mathbf{k})$, which we defined in Eqs.~\eqref{eq: transfer integral matrix} and~\eqref{eq: overlap integral matrix}. 
In this case, they are $2\times 2$ matrices. 
Starting with the former, we substitute Eq.~\eqref{eq: bloch wave function tb} into Eq.~\eqref{eq: transfer integral matrix}.
The diagonal entries read
\begin{align}
\boldsymbol{\mathcal{H}}_{AA} &= \braOket{\Phi_A}{\mathcal{H}}{\Phi_A}\nonumber\\
    &=\frac{1}{N_\mathrm{cell}} \sum_{k,k^\prime=1}^{N_\mathrm{cell}}  e^{i\mathbf{k}\cdot(\mathbf{R}_k-\mathbf{R}_{k^\prime})} \braOket{\varphi_{A}(\mathbf{R}_{k^\prime})}{\mathcal{H}}{\varphi_{A}(\mathbf{R}_k)}\, .
    \intertext{If we only take the terms into account for which~$\mathbf{R}_k=\mathbf{R}_{k^\prime}$ and neglect sub-dominant contributions with $\mathbf{R}_k\neq\mathbf{R}_{k^\prime}$, we find}
\boldsymbol{\mathcal{H}}_{AA}&\approx \epsilon_{2p} \, ,\quad\text{with }\epsilon_{2p}\equiv \braOket{\varphi_{A,k}}{\mathcal{H}}{\varphi_{A,k}}\, .
\end{align}
By analogy, $\mathcal{\boldsymbol{\mathcal{H}}}_{BB}=\epsilon_{2p}$.
For the off-diagonal components, we can do a similar (nearest-neighbor) approximation.
\begin{align}
\boldsymbol{\mathcal{H}}_{AB} &= \braOket{\Phi_A}{\mathcal{H}}{\Phi_B}\nonumber\\
    &= \frac{1}{N_\mathrm{cell}} \sum_{k,k^\prime=1}^{N_\mathrm{cell}}  e^{i\mathbf{k}\cdot(\mathbf{R}_k-\mathbf{R}_{k^\prime})} \braOket{\varphi_{A}(\mathbf{R}_{k^\prime})}{\mathcal{H}}{\varphi_{B}(\mathbf{R}_k)}\, .
    \intertext{Next, we only involve the contributions of the three nearest neighbors of each atom A,}
    &\approx \frac{1}{N_\mathrm{cell}} \sum_{k^\prime=1}^{N_\mathrm{cell}} \sum_{k=1}^3 e^{i\mathbf{k}\cdot((\mathbf{R}_{k^\prime}+\mathbf{N}_{k})-\mathbf{R}_{k^\prime})}\nonumber\\
   &\times \braOket{\varphi_{A}(\mathbf{R}_{k^\prime})}{\mathcal{H}}{\varphi_{B}(\mathbf{R}_{k^\prime}+\mathbf{N}_{k})}
\end{align}
where the three vectors~$\mathbf{N}_k$ are given by Eq.~\eqref{eq: nearest neighbors}. We define the parameter~$t\equiv \braOket{\varphi_{A}(\mathbf{R}_{k^\prime})}{\mathcal{H}}{\varphi_{B}(\mathbf{R}_{k^\prime}+\mathbf{N}_k)}$, which is identical for all~$k$ due to the rotational symmetry of the~$2p_z$ wave function.
This leaves us with
\begin{align}
\boldsymbol{\mathcal{H}}_{AB} &=t \times \sum_{k=1}^3e^{i\mathbf{k}\cdot \mathbf{N}_k} \, .
\end{align}
The other off-diagonal is simply~$\boldsymbol{\mathcal{H}}_{BA}=\boldsymbol{\mathcal{H}}_{AB}^*$.
In summary, the full transfer integral matrix for the $\pi$-electrons in the nearest-neighbor approximation is given by
\begin{align}
    \mathcal{\boldsymbol{\mathcal{H}}} &\approx \begin{pmatrix}
    \epsilon_{2p}   &   t f(\mathbf{k})\\
    t f(\mathbf{k})^*    &\epsilon_{2p} 
    \end{pmatrix}\, .
    \label{eq:hmatrix}
\end{align}
The function~$f(\mathbf{k})$ is defined as
\begin{align}
    f(\mathbf{k}) &\equiv \sum_{k=1}^3e^{i\mathbf{k}\cdot \mathbf{N}_k}\label{eq: f function}\, .
\end{align}
The square of this function can be evaluated as
\begin{align}
    |f(\mathbf{k})|^2 = 3+2\sum_{k=1}^3 \cos(\mathbf{a}_k\cdot \mathbf{k})\, ,\text{ with }\mathbf{a}_3 \equiv \mathbf{a}_2 - \mathbf{a}_1\, .
\end{align}

For the overlap matrix, the previous steps can essentially be repeated to find
\begin{align}
    \mathcal{\boldsymbol{\mathcal{S}}} &\approx \begin{pmatrix}
    1   &   s f(\mathbf{k})\\
    s f(\mathbf{k})^*    &1
    \end{pmatrix}\, ,
\label{eq:smatrix}
\end{align}
where $s\equiv \braket{\varphi_{A}(\mathbf{R}_j)}{\varphi_{B}(\mathbf{R}_j+\mathbf{N}_k)}$.

\paragraph*{Energy bands of $\pi$-electrons}
Given the explicit form of the two matrices, we can show that the corresponding eigenvalues solving the secular equation in Eq.~\eqref{eq: secular equation} are given by
\begin{align}
\label{eq: energy eigen values pi}
    E_{\pi}(\mathbf{k}) = \frac{\epsilon_{2p}\pm t |f(\mathbf{k})|}{1 \pm s |f(\mathbf{k})|}\, .
\end{align}
This energy dispersion is visualized in Fig.~\ref{fig: energy bands pi} for $\mathbf{k}\in\text{BZ}$, as well as along the path between the high-symmetry points given by Eq.~\eqref{eq: high symmetry points}, which is also indicated in the figure.
In order to reproduce the band structure of graphene to a good degree, we used the parameters $\epsilon_{2p_z}=0$ (by convention), $s=0.129$ and $t=-3.033$eV~\cite{saito1998physical}.

Since $t$ is negative, the $`+'$ solution is lower energy and corresponds to the bonding or valence~$\pi$-band, whereas the $`-'$ solution is the anti-bonding or conduction $\pi^*$-band.
The bands are degenerate at the high-symmetry point~$\mathbf{K}$.

Next, we turn our attention towards the~$\mathbf{C}_{i}(\mathbf{k})$ coefficients.
The normalized eigenvectors corresponding to the eigenvalues of Eq.~\eqref{eq: energy eigen values pi} are obtained by solving Eq.~\eqref{eq: general eigenvalue problem}. For the $\pi$-electrons we find
\begin{subequations}
\label{eq: eigen vectors pi}
\begin{align}
    \mathbf{C}_\pi &= \frac{1}{\sqrt{2}}\columnvector{2}{1}{\pm e^{i\varphi_{\mathbf{k}}}}\, ,\\
    \text{with }\varphi_{\mathbf{k}} &= -\arctan{\frac{\text{Im}f(\mathbf{k})}{\text{Re}f(\mathbf{k})}}\, .
\end{align}
\end{subequations}
Hence, the $\pi$-electron wave functions can be written as
\begin{align}
    \Psi_{\pi\mathbf{k}}(\mathbf{x}) = \frac{\mathcal{N}_{\mathbf{k}}}{\sqrt{2N_\mathrm{cell}}}\times &\sum_{k=1}^{N_\mathrm{cell}} \bigg[e^{i\mathbf{k}\cdot \mathbf{R}_k^A}\varphi_{2p_z}(\mathbf{x}-\mathbf{R}_k^A)\nonumber\\
    &+e^{i\varphi_{\mathbf{k}}+i\mathbf{k}\cdot \mathbf{R}_k^B}\varphi_{2p_z}(\mathbf{x}-\mathbf{R}_k^B)\bigg]\, . \label{eq: wavefunction tb pi position}
\end{align}
As opposed to the treatment in~\cite{Hochberg:2016ntt}, we do not perform a nearest-neighbor approximation on this level, as it is not well-defined here.
Instead it is necessary to sum over all~$N$ unit cells.
However, the nearest-neighbor approximation can be applied when computing the norm of~$ \Psi_{\pi\mathbf{k}}(\mathbf{x})$, 
\begin{align}
    &\braket{\Psi_\pi}{\Psi_\pi} = \frac{\mathcal{N}_{\mathbf{k}}^2}{2N_\mathrm{cell}}\sum_{k,{k^\prime}=1}^{N_\mathrm{cell}}\bigg[\nonumber\\
    &e^{i\mathbf{k}\cdot(\mathbf{R}_{k^\prime}^A-\mathbf{R}_k^A)}\braket{\varphi_{2p_z}(\mathbf{x}-\mathbf{R}_k^A)}{\varphi_{2p_z}(\mathbf{x}-\mathbf{R}_{k^\prime}^A)}\nonumber\\
    &+e^{i\mathbf{k}\cdot(\mathbf{R}_{k^\prime}^B-\mathbf{R}_k^A)}\braket{\varphi_{2p_z}(\mathbf{x}-\mathbf{R}_k^A)}{\varphi_{2p_z}(\mathbf{x}-\mathbf{R}_{k^\prime}^B)} e^{i\varphi_{\mathbf{k}}}\nonumber\\
    &+e^{-i\mathbf{k}\cdot(\mathbf{R}_k^B-\mathbf{R}_{k^\prime}^A)}\braket{\varphi_{2p_z}(\mathbf{x}-\mathbf{R}_k^B)}{\varphi_{2p_z}(\mathbf{x}-\mathbf{R}_{k^\prime}^A)}e^{-i\varphi_{\mathbf{k}}}\nonumber\\
    &+e^{-i\mathbf{k}\cdot(\mathbf{R}_k^B-\mathbf{R}_{k^\prime}^B)}\braket{\varphi_{2p_z}(\mathbf{x}-\mathbf{R}_k^B)}{\varphi_{2p_z}(\mathbf{x}-\mathbf{R}_{k^\prime}^B)}\bigg]
\end{align}
Next, we perform the sum over~${k^\prime}$ using the nearest-neighbor approximation (but also taking next-nearest neighbors into account).
In addition to the neighboring contributions, the first and last lines contain terms with~$\mathrm{R}_k=\mathrm{R}_{k^\prime}$, each contributing with~$N_\mathrm{cell}$ to the final sum.
Hence, we find
\begin{align}
    &\braket{\Psi_\pi}{\Psi_\pi} \approx \frac{\mathcal{N}_{\mathbf{k}}^2}{2N_\mathrm{cell}}\Bigg\{ 2N_\mathrm{cell} + \sum_{{k^\prime}=1}^{N_\mathrm{cell}}\sum_{k=1}^3\bigg[\nonumber\\
    &e^{i\mathbf{k}\cdot(\mathbf{R}_{k^\prime}^A+\mathbf{a}_k-\mathbf{R}_{k^\prime}^A)}s^\prime + e^{i\mathbf{k}\cdot(\mathbf{R}_{k^\prime}^A+\mathbf{N}_k-\mathbf{R}_{k^\prime}^A)}s e^{i\varphi_{\mathbf{k}}}\nonumber\\
    &+e^{-i\mathbf{k}\cdot(\mathbf{R}_{k^\prime}^A+\mathbf{N}_k-\mathbf{R}_{k^\prime}^A)}s e^{-i\varphi_{\mathbf{k}}} + e^{-i\mathbf{k}\cdot(\mathbf{R}_{k^\prime}^B+\mathbf{a}_k-\mathbf{R}_{k^\prime}^B)}s^\prime\bigg]\\
    &=\mathcal{N}_{\mathbf{k}}^2\bigg[ 1 + \sum_{k=1}^3 \left( s \cos(\mathbf{k}\cdot\mathbf{N}_k+\varphi_{\mathbf{k}}) + s^\prime \cos(\mathbf{k}\cdot \mathbf{a}_k)\right)\bigg]\, .
\end{align}
Here, $s^\prime$ denotes the overlap integral of the atomic orbitals at next-to-nearest neighboring sites.
We find that the two leading terms are consistent with the general expression of Eq.~\eqref{eq: wave function norm general},
\begin{align}
\braket{\Psi_\pi}{\Psi_\pi} &\approx \mathbf{C}^\dagger_\pi(\mathbf{k})\cdot \boldsymbol{\mathcal{S}}(\mathbf{k})\cdot\mathbf{C}_\pi(\mathbf{k})\nonumber\\
    &=1 + s \sum_{k=1}^3\cos(\mathbf{k}\cdot \mathbf{N}_k+\varphi_{\mathbf{k}})\, ,\label{eq: wave function norm pi 2}
\end{align}
and hence
\begin{align}
   \mathcal{N}_{\mathbf{k}} &= \left[ 1 + s \sum_{k=1}^3\cos(\mathbf{k}\cdot \mathbf{N}_k+\varphi_{\mathbf{k}})\right]^{-1/2} \, . \label{eq: normalization constant pi}
\end{align}

Next, we shift our attention from position space to momentum space.
The Fourier-transformed Bloch wave functions are
\begin{align}
    \widetilde{\Phi}_{\mathbf{k}}(\boldell) &= \frac{1}{\sqrt{N_\mathrm{cell}}}\sum_{k=1}^{N_\mathrm{cell}} e^{i(\boldell+\mathbf{k})\cdot\mathbf{R}_k}\widetilde{\varphi}_j(\boldell)\, ,\\
    \intertext{where~$\boldell$ is the conjugate momentum to~$\mathbf{x}$. Therefore, the $\pi$-electrons' wave functions in momentum space read}
    \widetilde{\Psi}_{\pi\mathbf{k}}(\boldell) &= \frac{\mathcal{N}_{\mathbf{k}}}{\sqrt{2N_\mathrm{cell}}} \widetilde{\varphi}_{2p_z}(\boldell)\sum_{k=1}^{N_\mathrm{cell}}\bigg[e^{i(\boldell+\mathbf{k})\cdot \mathbf{R}_k^A} + e^{i(\boldell+\mathbf{k})\cdot \mathbf{R}_k^B + i \varphi_{\boldell}}\bigg]\, .
    \label{eq:psipi}
\end{align}
Using~$\mathbf{R}_i^B=\mathbf{R}_i^A+\mathbf{N}_1$, we can write this as
\begin{align}
    &= \frac{\mathcal{N}_{\mathbf{k}}}{\sqrt{2N}}\widetilde{\varphi}_{2p_z}(\boldell) \left( 1 + e^{i(\boldell+\mathbf{k})\cdot \mathbf{N}_1 + i \varphi_{\boldell}} \right)\sum_{k=1}^{N_\mathrm{cell}} e^{i(\boldell+\mathbf{k})\cdot \mathbf{R}_k^A}\, .
\end{align}
For the evaluation of the exponential sum, we can follow the steps outlined in Sec.~\ref{ss: general tight binding} and use the identity in Eq.~\eqref{eq: lattice sum identity}.
In the end, we find
\begin{align}
\widetilde{\Psi}_{\pi\mathbf{k}}(\boldell) &= \frac{\mathcal{N}_{\mathbf{k}}}{\sqrt{2N}} \widetilde{\varphi}_{2p_z}(\boldell) \, A^{-1}_{\rm uc} \sum_{\mathbf{G}} \, (2\pi)^2 \delta^{(2)}(\boldell^{\|} + \mathbf{k} - \mathbf{G}) \nonumber \\
&\times \left\{1 + e^{i \left[ \varphi_{\boldell} + (\boldell+\mathbf{k})\cdot \boldsymbol{\delta}\right]}\right\} \,.
\label{eq:psipi2}
\end{align}

\subsubsection*{\boldmath The $\sigma$-electrons}

The~$\sigma$-electrons are in a superposition of the carbon atoms' $2s$, $2\text{p}_x$, and $2\text{p}_y$ orbitals.
Hence, the transfer and overlap matrices are $6\times 6$ matrices, which we can express in terms of four $3\times 3$~sub-matrices,
\begin{subequations}
\label{eq: transfer overlap sigma}
\begin{align}
  \mathcal{\boldsymbol{\mathcal{S}}} &\approx \begin{pmatrix}
  \mathcal{\boldsymbol{\mathcal{S}}}_{AA}   & \mathcal{\boldsymbol{\mathcal{S}}}_{AB}\\
  \mathcal{\boldsymbol{\mathcal{S}}}_{AB} & \mathcal{\boldsymbol{\mathcal{S}}}_{BB}
  \end{pmatrix}\, ,\\
    \mathcal{\boldsymbol{\mathcal{H}}} &\approx \begin{pmatrix}
  \mathcal{\boldsymbol{\mathcal{H}}}_{AA}   & \mathcal{\boldsymbol{\mathcal{H}}}_{AB}\\
  \mathcal{\boldsymbol{\mathcal{H}}}_{AB} & \mathcal{\boldsymbol{\mathcal{H}}}_{BB}
  \end{pmatrix}\, .
\end{align}
The diagonal sub-matrices are diagonal,
\begin{align}
    \mathcal{\boldsymbol{\mathcal{S}}}_{AA}=\mathcal{\boldsymbol{\mathcal{S}}}_{BB} &= \begin{pmatrix}
    1   & 0 & 0\\
    0   & 1 & 0 \\
    0   & 0 & 1
    \end{pmatrix}\\
     \mathcal{\boldsymbol{\mathcal{H}}}_{AA}=\mathcal{\boldsymbol{\mathcal{H}}}_{BB} &= \begin{pmatrix}
    \epsilon_{2s}   & 0 & 0\\
    0   & \epsilon_{2p} & 0 \\
    0   & 0 & \epsilon_{2p}
    \end{pmatrix}\, ,
\end{align}
with~$\epsilon_{2s}=-8.87 \text{eV}$ and~$\epsilon_{2p} = 0$.
The off-diagonal matrices are given by
\begin{align}
    \mathcal{\boldsymbol{\mathcal{S}}}_{AB} &=  \begin{pmatrix}
    \mathcal{S}_{ss}  &   \mathcal{S}_{s p_x}  &   \mathcal{S}_{s p_y}  \\
    -\mathcal{S}_{s p_x}  &   \mathcal{S}_{p_x p_x}  &   \mathcal{S}_{p_x p_y}  \\
    -\mathcal{S}_{s p_y}  &   \mathcal{S}_{p_x p_y}  &   \mathcal{S}_{p_y p_y}  
    \end{pmatrix}\, ,
\end{align}
with entries
\begin{widetext}
\begin{align}
\mathcal{S}_{ss} &=S_{ss}\left(e^{i k_{1} a_\mathrm{CC}}+2 e^{-i k_{1} a_\mathrm{CC} / 2} \cos \left(\frac{\sqrt{3} k_{2} a_\mathrm{CC}}{2}\right)\right)\, , \\
\mathcal{S}_{sp_x} &=S_{sp}\left(-e^{i k_{1} a_\mathrm{CC}}+e^{-i k_{1} a_\mathrm{CC} / 2} \cos \left(\frac{\sqrt{3} k_{2} a_\mathrm{CC}}{2}\right)\right)\, , \\
\mathcal{S}_{sp_{y}} &=-i \sqrt{3} S_{sp} e^{-i k_{1} a_\mathrm{CC} / 2} \sin \left(\frac{\sqrt{3} k_{2} a_\mathrm{CC}}{2}\right) \, ,\\
\mathcal{S}_{p_{x} p_{x}} &=-S_{\sigma} e^{i k_{1} a_\mathrm{CC}}+\frac{\left(3 S_{\pi}-S_{\sigma}\right)}{2} e^{-i k_{1} a_\mathrm{CC} / 2} \cos \left(\frac{\sqrt{3} k_{2} a_\mathrm{CC}}{2}\right) \, ,\\
\mathcal{S}_{p_{x} p_{y}} &=\frac{i \sqrt{3}}{2}\left(S_{\sigma}+S_{\pi}\right) e^{-i k_{1} a_\mathrm{CC} / 2} \sin \left(\frac{\sqrt{3} k_{2} a_\mathrm{CC}}{2}\right) \, ,\\
\mathcal{S}_{p_{y} p_{y}} &=S_{\pi} e^{i k_{1} a_\mathrm{CC}}+\frac{\left(S_{\pi}-3 S_{\sigma}\right)}{2} e^{-i k_{1} a_\mathrm{CC} / 2} \cos \left(\frac{\sqrt{3} k_{2} a_\mathrm{CC}}{2}\right)\, .
\end{align}
\end{widetext}
\end{subequations}
The other off-diagonal sub-matrix is given by the conjugate matrix, i.e.~$\mathcal{\boldsymbol{\mathcal{S}}}_{BA} = \mathcal{\boldsymbol{\mathcal{S}}}_{AB}^\dagger$.

The off-diagonal sub-matrices~$\mathcal{\boldsymbol{\mathcal{H}}}_{AB}$ and $\mathcal{\boldsymbol{\mathcal{H}}}_{BA}$ are obtained by replacing $\mathcal{S}\rightarrow \mathcal{H}$ and $S\rightarrow H$.
The numerical parameters are given in Table~\ref{tab: graphene parameter}.

\begin{table}[h!]
    \centering
    \begin{tabular}{|ll|ll|}
    \hline
    $\mathcal{S}$ &   value   &   $\mathcal{H}$ & value [eV]\\
    \hline
    \hline
    $s$   &   0.129 & $t$ & -3.033\\
    $s^\prime$    &  0.0087 & $\epsilon_{2s}$ &-8.868 \\
     &&$\epsilon_{2p}$ &  0.0\\
    $S_{ss}$  &   0.212& $H_{ss}$    &-6.769  \\
    $S_{sp}$  &   0.16& $H_{sp}$    &-5.580  \\
    $S_{\sigma}$  &   0.146& $H_{\sigma}$    &-5.037 \\
    $S_{\pi}$  &   0.129& $H_{\pi}$    & -3.033 \\
    \hline
    \end{tabular}
    \caption{Parameters of the overlap and transfer matrices for the $\sigma$-electrons}
    \label{tab: graphene parameter}
\end{table}

\begin{figure}[h!]
    \centering
    \includegraphics[width=0.45\textwidth]{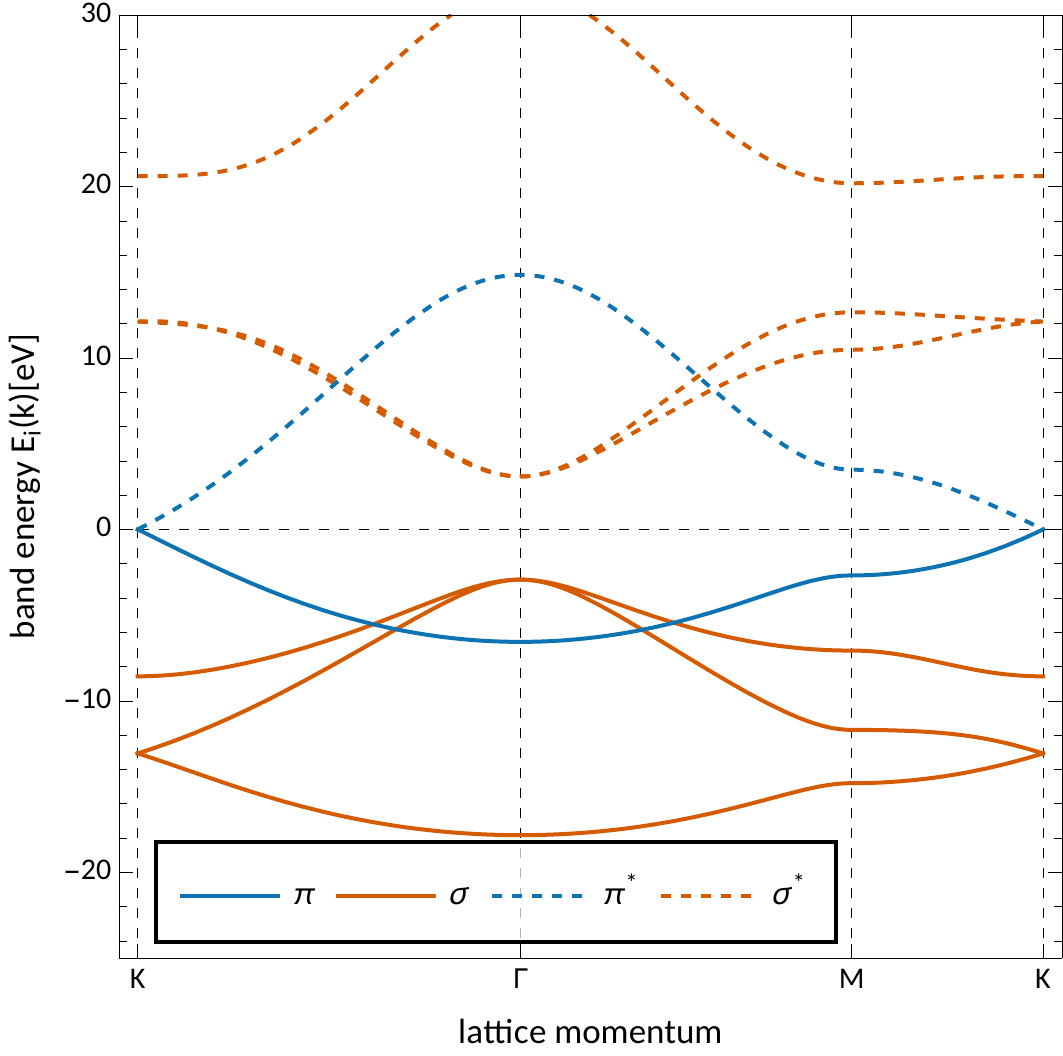}
    \caption{Energy bands of the $\pi$ and $\sigma$-electrons obtained using the tight-binding approximation. Solid lines show the valence bands, dashed lines are the conduction bands.}
    \label{fig: energy bands all}
\end{figure}

The energy bands of the $\sigma$-electrons are obtained by solving the secular equation i.e. Eq.~\eqref{eq: secular equation}.
Since we are dealing with~$6\times6$ matrices, we use the numerical functionality of the \texttt{Eigen} library~\cite{eigen}.
The resulting energy bands for the~$\sigma$-electrons are depicted in Fig.~\ref{fig: energy bands all}.
The same \texttt{Eigen} function that computes the eigenvalues of the matrices in Eq.~\eqref{eq: transfer overlap sigma}, i.e. the energy bands of the~$\sigma$-electrons, by solving the secular equation also result in the six-dimensional eigenvectors~$\mathbf{C}_{\sigma_i}(\mathbf{k})$ and therefore the wave functions according to Eq.~\eqref{eq: wave function general}.
\begin{subequations}
\label{eq: bloch wave functions sigma}
\begin{align}
    \Psi_{\sigma_i\mathbf{k}}(,\mathbf{x}) =\mathcal{N}_{\mathbf{k}}\sum_{j=1}^6 C_{\sigma_i j}(\mathbf{k})\Phi_{j\mathbf{k}}(\mathbf{x})\, ,\;  (i=1,2,3)\, .
\end{align}
The six Bloch wave functions are given by
\begin{align}
    \Phi_{1\mathbf{k}}(\mathbf{x}) &= \frac{1}{\sqrt{N_\mathrm{cell}}}\sum_{k=1}^{N_\mathrm{cell}} e^{i\mathbf{k}\cdot \mathbf{R}_k^A}\varphi_{2s}(\mathbf{x}-\mathbf{R}_k^A)\, ,\\ 
    \Phi_{2\mathbf{k}}(\mathbf{x}) &= \frac{1}{\sqrt{N_\mathrm{cell}}}\sum_{k=1}^{N_\mathrm{cell}} e^{i\mathbf{k}\cdot \mathbf{R}_k^A}\varphi_{2p_x}(\mathbf{x}-\mathbf{R}_k^A)\, ,\\ 
    \Phi_{3\mathbf{k}}(\mathbf{x}) &= \frac{1}{\sqrt{N_\mathrm{cell}}}\sum_{k=1}^{N_\mathrm{cell}} e^{i\mathbf{k}\cdot \mathbf{R}_k^A}\varphi_{2p_y}(\mathbf{x}-\mathbf{R}_k^A)\, ,\\ 
     \Phi_{4\mathbf{k}}(\mathbf{x}) &= \frac{1}{\sqrt{N_\mathrm{cell}}}\sum_{k=1}^{N_\mathrm{cell}} e^{i\mathbf{k}\cdot \mathbf{R}_k^B}\varphi_{2s}(\mathbf{x}-\mathbf{R}_k^B)\, ,\\ 
    \Phi_{5\mathbf{k}}(\mathbf{x}) &= \frac{1}{\sqrt{N_\mathrm{cell}}}\sum_{k=1}^{N_\mathrm{cell}} e^{i\mathbf{k}\cdot \mathbf{R}_k^B}\varphi_{2p_x}(\mathbf{x}-\mathbf{R}_k^B)\, ,\\ 
    \Phi_{6\mathbf{k}}(\mathbf{x}) &= \frac{1}{\sqrt{N_\mathrm{cell}}}\sum_{k=1}^{N_\mathrm{cell}} e^{i\mathbf{k}\cdot \mathbf{R}_k^B}\varphi_{2p_y}(\mathbf{x}-\mathbf{R}_k^B)\, .
\end{align}
\end{subequations}
Finally, the Fourier transform of the wave function is given by
\begin{align}
    \widetilde{\Psi}_{\sigma_i\mathbf{k}}(\boldell) &= \frac{\mathcal{N}_{\mathbf{k}}}{\sqrt{N_\mathrm{cell}}} \sum_{k=1}^{N_\mathrm{cell}}\nonumber\\
    &\Bigg\{\widetilde{\varphi}_{2s}(\boldell)\bigg[C_{\sigma_i 1}e^{i(\boldell+\mathbf{k})\cdot \mathbf{R}_k^A} +C_{\sigma_i 4} e^{i(\boldell+\mathbf{k})\cdot \mathbf{R}_k^B}\bigg] \nonumber \\
    &\quad+\widetilde{\varphi}_{2p_x}(\boldell)\bigg[C_{\sigma_i 2}e^{i(\boldell+\mathbf{k})\cdot \mathbf{R}_k^A} +C_{\sigma_i 5} e^{i(\boldell+\mathbf{k})\cdot \mathbf{R}_k^B}\bigg] \nonumber \\
    &\quad+\widetilde{\varphi}_{2p_y}(\boldell)\bigg[C_{\sigma_i 3}e^{i(\boldell+\mathbf{k})\cdot \mathbf{R}_k^A} +C_{\sigma_i 6} e^{i(\boldell+\mathbf{k})\cdot \mathbf{R}_k^B}\bigg] \Bigg\} \\
    &= \frac{\mathcal{N}_{\mathbf{k}}}{\sqrt{N_\mathrm{cell}}} \sum_{k=1}^{N_\mathrm{cell}} e^{i(\boldell+\mathbf{k})\cdot \mathbf{R}_k^A} \nonumber\\
    &\Bigg\{\widetilde{\varphi}_{2s}(\boldell)\bigg[C_{\sigma_i 1} +C_{\sigma_i 4} e^{i(\boldell+\mathbf{k})\cdot \boldsymbol{\delta}}\bigg] \nonumber \\
    &\quad+\widetilde{\varphi}_{2p_x}(\boldell)\bigg[C_{\sigma_i 2} +C_{\sigma_i 5} e^{i(\boldell+\mathbf{k})\cdot \boldsymbol{\delta}}\bigg] \nonumber \\
    &\quad+\widetilde{\varphi}_{2p_y}(\boldell)\bigg[C_{\sigma_i 3} +C_{\sigma_i 6} e^{i(\boldell+\mathbf{k})\cdot \boldsymbol{\delta}}\bigg] \Bigg\}\, .
\end{align}
Here, we again used $\mathbf{R}_j^B = \mathbf{R}_j^A + \boldsymbol{\delta}$.
Just like in the case of the $\pi$-electrons, we use Eq.~\eqref{eq: lattice sum identity} to express the exponential sum in terms of a sum over the reciprocal lattice vectors~$\mathbf{G}$,
\begin{align}
    \widetilde{\Psi}_{\sigma_i\mathbf{k}}(\boldell) &=\frac{\mathcal{N}_{\mathbf{k}}}{\sqrt{N}} A_{\rm uc}^{-1} \, \sum_{\mathbf{G}} \, (2\pi)^2 \delta^{(2)}(\boldell^{\|} + \mathbf{k} - \mathbf{G})\nonumber\\
    &\quad\Bigg\{\widetilde{\varphi}_{2s}(\boldell)\bigg[C_{\sigma_i 1} +C_{\sigma_i 4} e^{i(\boldell+\mathbf{k})\cdot \boldsymbol{\delta}}\bigg] \nonumber \\
    &\quad+\widetilde{\varphi}_{2p_x}(\boldell)\bigg[C_{\sigma_i 2} +C_{\sigma_i 5} e^{i(\boldell+\mathbf{k})\cdot \boldsymbol{\delta}}\bigg] \nonumber \\
    &\quad+\widetilde{\varphi}_{2p_y}(\boldell)\bigg[C_{\sigma_i 3} +C_{\sigma_i 6} e^{i(\boldell+\mathbf{k})\cdot \boldsymbol{\delta}}\bigg] \Bigg\} \, . \label{eq:psisigma2}
\end{align}

\section{Atomic wavefunctions}
\label{app: atomic wavefunctions}

The evaluation of the graphene response function requires a specific form of the atomic wavefunctions of carbon~$\varphi_i(\mathbf{x})$ (and its Fourier transform~$\widetilde{\varphi}_i(\boldell)$).
We present results for two particular choices.
As proposed by Hochberg et al.~\cite{Hochberg:2016ntt}, we start by approximating the wavefunctions of carbon with hydrogenic wave functions with a re-scaled ~$Z_\mathrm{eff}$ factor.
We improve upon this choice by using Roothaan-Hartree-Fock (RHF) wavefunctions for the ground states of carbon~\cite{Bunge:1993jsz}.
In this appendix, we summarize the explicit wave functions both in position and momentum space and present a comparison.

The wave function of the atomic state $(n,\ell,m)$ in position space is given by
\begin{align}
   \varphi_{n l m}(\mathbf{x}) &= R_{nl}(r) Y_l^m(\hat{\mathbf{x}})\, ,\label{eq: atomic wavefunction fourier}
\end{align}
where~$Y_l^m(\hat{\mathbf{x}})$ are spherical harmonics, and $R_{nl}(r)$ is the radial component of the wave function.

Regardless of the choice of form for the radial component, the corresponding Fourier-transformed wave function in momentum space $\widetilde{\varphi}_i(\boldell)$ for a given position space wave function $\varphi_i(\mathbf{x})$ is defined as
\begin{align}
    \widetilde{\varphi}_i(\boldell) = \int \dd^3\mathbf{x}\,\varphi_i(\mathbf{x}) e^{-i\boldell\cdot\mathbf{x}}\, , \label{eq: atomic wavefunction general momentum}
\end{align}
which fixes the normalization of the wave function to 
\begin{align}
    \int \dd^3 \mathbf{x} \;|\varphi_i(\mathbf{x})|^2 &=1\, , \quad    \int \frac{\dd^3 \boldell}{(2\pi)^3}\; |\widetilde{\varphi}_i(\boldell)|^2 = 1\, . \label{eq: wavefunction normalization}
\end{align}

For the evaluation of the graphene response function using the TB approximation, the required wavefunctions are those of the atomic orbitals~$2s$ and~$2p$ in the environment of a carbon atom.
Additionally for the ~$2p$ orbitals, the crystal structure of graphene gives rise to the $2p_x$, $2p_y$, and $2p_z$ orbitals.
The corresponding wavefunctions are given by
\begin{subequations}
    \label{eq: 2pi orbitals}
\begin{align}
    \varphi_{2p_i}(\vec{x}) & = R_{2p}(r) Y_i(\hat{\vec{x}})\, , \;\text{with}\\
    Y_i(\hat{\vec{x}}) &\equiv \sqrt{\frac{3}{4\pi}}\frac{x_i}{r}\, .
\end{align}
\end{subequations}

Similarly, the relevant momentum wave functions can be written as
   
\begin{align}
    \widetilde{\varphi}_{n l m}(\boldell) &= \chi_{nl}(\ell) Y_l^m(\hat{\boldell})\, , \label{eq: atomic wavefunction general}
\intertext{and hence for the $2p_i$ states we find}
\widetilde{\varphi}_{2p_i}(\boldell) &= \chi_{2p}(\ell) Y_i(\hat{\boldell})\, . \label{eq: 2pi orbitals momentum}
\end{align}

\begin{figure*}
    \includegraphics[width=0.45\textwidth]{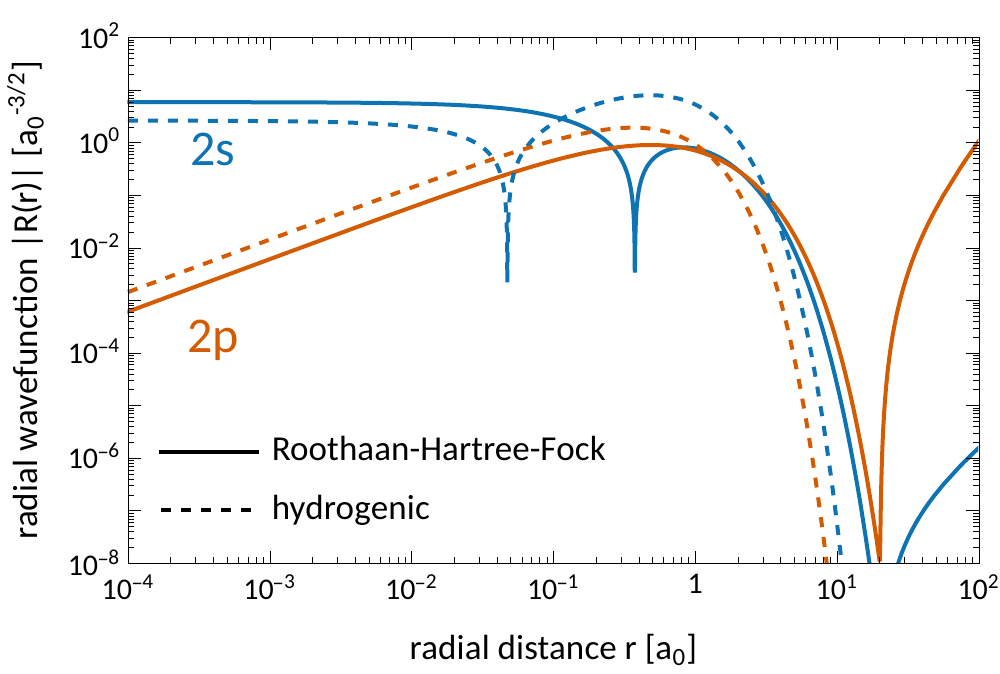}
    \includegraphics[width=0.45\textwidth]{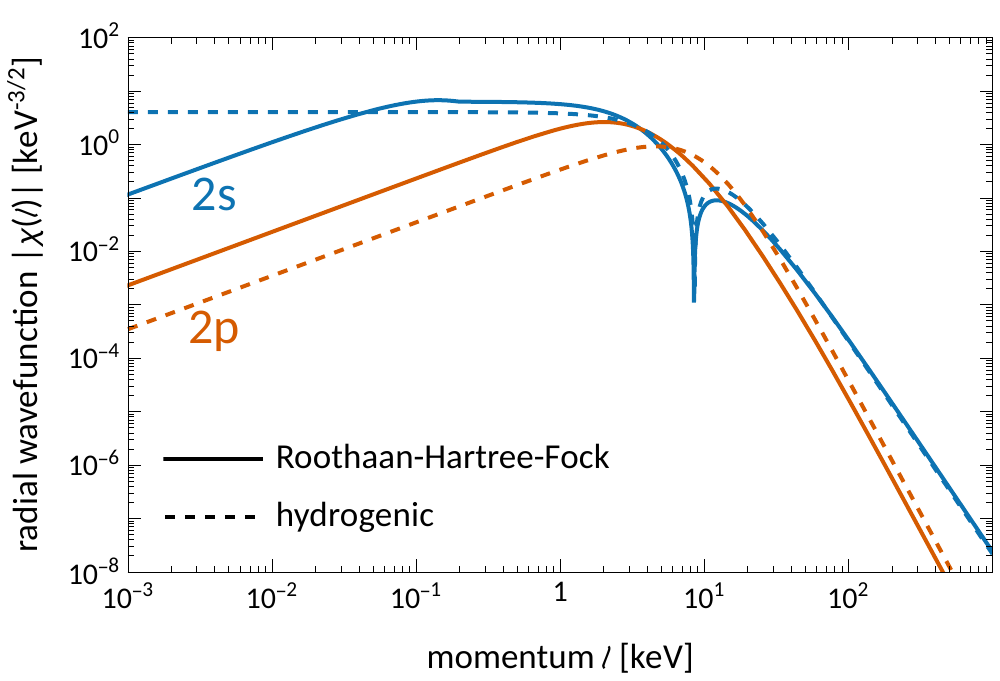}
    \caption{Comparison of the hydrogenic and the RHF wavefunctions for the $2s$ and $2p$ orbitals of carbon. The left (right) panel shows the radial wavefunction~$R_{nl}(r)$ ($\chi_{nl}(\ell)$) in position (momentum) space. Note that the values of $Z_\mathrm{eff}$ for the hydrogenic wavefunctions were tuned to reproduce the overlap integrals of graphene, see Eq.~\eqref{eq: zeff values}.}
    \label{fig: carbon wavefunctions}
\end{figure*}

\subsection{Hydrogenic wave functions}
\label{app: hydrogenic wave functions}

We list the hydrogenic wavefunctions proposed in~\cite{Hochberg:2016ntt} to approximate the groundstate wave functions of carbon atoms.
In position space, the wave functions are given by
\begin{subequations}
\label{eq: hydrogenic wave functions position}
\begin{align}
\varphi_{2s}(\mathbf{x}) &= \sqrt{\frac{(Z_\mathrm{eff}^{2s})^3}{56\pi a_0^3}} \left(1-\frac{Z_\mathrm{eff}^{2s}r}{a_0}\right) e^{-Z_\mathrm{eff}^{2s}r/(2a_0)}\, ,\\
\varphi_{2p_x}(\mathbf{x}) &= \sqrt{\frac{(Z_\mathrm{eff}^{2p_{x/y}})^5}{32\pi a_0^3}} \frac{r}{a_0} e^{-Z_\mathrm{eff}^{2p_{x/y}}r/(2a_0)}\sin\theta\cos\varphi\, ,\\
\varphi_{2p_y}(\mathbf{x}) &= \sqrt{\frac{(Z_\mathrm{eff}^{2p_{x/y}})^5}{32\pi a_0^3}} \frac{r}{a_0} e^{-Z_\mathrm{eff}^{2p_{x/y}}r/(2a_0)}\sin\theta\sin\varphi\, ,\\
\varphi_{2p_z}(\mathbf{x}) &= \sqrt{\frac{(Z_\mathrm{eff}^{2p_z})^5}{32\pi a_0^3}} \frac{r}{a_0} e^{-Z_\mathrm{eff}^{2p_z}r/(2a_0)}\cos\theta\, .
\end{align}
\end{subequations}
 Following~\cite{Hochberg:2016ntt}, the effective charge~$Z_\mathrm{eff}$ parameters are determined to reproduce the overlap integrals for graphene listed in Tab.~\ref{tab: graphene parameter},
  \begin{align}
     Z_\mathrm{eff}^{2s}&= 4.59\, , \quad    Z_\mathrm{eff}^{2p_{x/y}} = 5.49\, , \quad   Z_\mathrm{eff}^{2p_z} = 4.02\, .  \label{eq: zeff values}
 \end{align}
 While this improves the self-consistency of the TB formalism, the resulting wave functions are not close to those of the more accurate Roothaan-Hartree-Fock wavefunctions for carbon atoms, as seen in Fig.~\ref{fig: carbon wavefunctions}.
 
The momentum space wave functions required to describe the electrons in graphene can be approximated as
\begin{subequations}
\label{eq: hydrogenic wave functions momentum}
\begin{align}
    \widetilde{\varphi}_{2s}(\boldell) &= \sqrt{8\pi}\left(Z_\mathrm{eff}^{2s}\right)^{5/2} a_0^{3/2}\frac{a_0^2|\boldell|^2-\left(Z^{2s}_{\rm eff}/2\right)^2}{\left(a_0^2|\boldell|^2+(Z^{2s}_{\rm eff}/2)^2\right)^3}\, ,\\
    \widetilde{\varphi}_{2p_x}(\boldell) &\approx \sqrt{8\pi}\left(Z_\mathrm{eff}^{2p_{x/y}}\right)^{7/2}a_0^{3/2}\frac{a_0\ell_x}{\left(a_0^2|\boldell|^2+(Z^{2p_{x/y}}_{\rm eff}/2)^2\right)^3}\, ,\\
    \widetilde{\varphi}_{2p_y}(\boldell) &\approx \sqrt{8\pi}\left(Z_\mathrm{eff}^{2p_{x/y}}\right)^{7/2}a_0^{3/2}\frac{a_0\ell_y}{\left(a_0^2|\boldell|^2+(Z^{2p_{x/y}}_{\rm eff}/2)^2\right)^3}\, ,\\
     \widetilde{\varphi}_{2p_z}(\boldell) &\approx \sqrt{8\pi}\left(Z_\mathrm{eff}^{2p_z}\right)^{7/2}a_0^{3/2}\frac{a_0\ell_z}{\left(a_0^2|\boldell|^2+(Z^{2p_{z}}_{\rm eff}/2)^2\right)^3}\, ,
\end{align}
\end{subequations}
where only the expression for~$\widetilde{\varphi}_{2s}(\boldell)$ is exact.

\subsection{Roothaan-Hartree-Fock wavefunctions}
\label{app: RHF wfunctions}

\begin{table}[h!]
\begin{tabular}{|l|l|l|}
\multicolumn{3}{l}{\textbf{2s}}\\
\hline
$n_{lj}$    & $Z_{lj}$& $C_{nlj}$ \\
\hline
\hline
1   &   8.4936  &   -0.071727   \\
1   &   4.8788  &   0.438307    \\
3   &   15.466  &   -0.000383   \\
2   &   7.05   &    -0.091194   \\
2   &   2.264  &   -0.393105    \\
2   &   1.4747  &   -0.579121   \\
2   &   1.1639 &     -0.126067  \\
\hline
\end{tabular}\hspace{1cm}
\begin{tabular}{|l|l|l|}
\multicolumn{3}{l}{\textbf{2p}}\\
\hline
$n_{lj}$    & $Z_{lj}$& $C_{nlj}$ \\
\hline
\hline
2   &   7.05    &   0.006977    \\
2   &   3.2275  &   0.070877 \\
2   &   2.1908  &   0.230802 \\
2   &   1.4413  &   0.411931 \\
2   &   1.0242  &   0.350701 \\
\hline
\end{tabular}
\caption{Coefficients of RHF wavefunctions as defined in Eq.~\eqref{eq: RHF wavefunction} for the $2s$ (left) and $2p$ (right) orbital of carbon. Values taken from~\cite{Bunge:1993jsz}.}
\label{tab:RHF coefficients}
\end{table}

Instead of re-scaled hydrogenic wavefunctions, we recommend using Roothaan-Hartree-Fock (RHF) wave functions that can be found in~\cite{Bunge:1993jsz}.
The radial part of the RHF wavefunction is given in Eq.~\eqref{eq: RHF wavefunction} as a linear combination of Slater-type orbitals (STOs).
We repeat the expression here for convenience.
\begin{subequations}
    \begin{align}
    R_{nl}(r) &= \sum_j C_{nl j} R_\mathrm{STO}(r,Z_{l j},n_{l j})\, .
    \intertext{An STO is defined as}
    R_\mathrm{STO}(r,Z,n) &\equiv a_0^{-3/2}\frac{(2Z)^{n+1/2}}{\sqrt{(2n)!}}\left(\frac{r}{a_0}\right)^{n-1}e^{-\frac{Z r}{a_0}}\, .
\end{align}
\end{subequations}
Finally, the RHF~coefficients $C_{nl j}$ as well as the parameters $n_{nl}$ and $Z_{nl}$ for carbon are tabulated in~\cite{Bunge:1993jsz}, and summarized in Tab.~\ref{tab:RHF coefficients} for convenience.

Moving on to momentum space, the wave function is obtained via Eq.~\eqref{eq: atomic wavefunction fourier}.
Using the plane-wave expansion of the exponential, we can write the radial part of the wave function in Eq.~\eqref{eq: atomic wavefunction general momentum} as the spherical Bessel transform of $R_{nl}(r)$,
\begin{align}
\chi_{nl}(\ell) &= 4\pi i^l \int \dd r\; r^2 R_{nl}(r) j_l(\ell r)\, ,
\end{align}
where $j_l(x)$ is the spherical Bessel function.

\begin{widetext}
    Evaluating this expression for the RHF wavefunction given in Eq.~\eqref{eq: RHF wavefunction} yields
\begin{align}
    \chi_{nl}(\ell) &= \sum_j C_{nl j} \left[\frac{2\pi a_0}{Z_{l j}} \right]^{3/2} 2^{n_{l j}-l}\left[\frac{i a_0 \ell}{Z_{l j}}\right]^l \frac{(n_{l j}+l+1)!}{\sqrt{(2n_{l j})!}} \frac{{}_2F_1\left(\frac{1}{2}(2+l+n_{l j}),\frac{1}{2}(3+l+n_{l j}),\frac{3}{2}+l,-\left(\frac{a_0 \ell}{Z_{l j}}\right)^2\right)}{\Gamma\left(\frac{3}{2}+l\right)}\, ,
\end{align}
where  ${}_2F_1\left(a,b,c,z\right)$ is the hypergeometric function.
\end{widetext}
Using these expressions, we can evaluate all relevant atomic orbitals for the graphene response function by applying Eq.~\eqref{eq: 2pi orbitals momentum}.

When we compare the hydrogenic to the RHF wavefunctions in Fig.~\ref{fig: carbon wavefunctions}, we find the hydrogenic wavefunctions to be a poor match to the RHF wavefunctions, which are accepted as a good description for the atomic carbon groundstate wave functions.
Instead, we conclude the necessity to use actual carbon atomic wavefunctions.
This conclusion becomes even more robust when comparing the graphene response functions computed with TB and DFT.

\section{Comparison of our TB treatment to Hochberg et al. (2017)}
\label{app: comparison hochberg}

The first study of graphene targets for sub-GeV DM~searches was published by Hochberg et al.~\cite{Hochberg:2016ntt}.
Therein, the authors chose a semi-analytic approach to describe the electron wavefunctions in graphene based on the tight-binding (TB) approximation.
In reproducing their results, we noticed a number of deviations to our TB treatment of the electron wavefunctions in graphene.
\begin{itemize}
    \item The modeling of the Bloch wavefunctions in the above-mentioned work does not satisfy the Bloch's theorem given in Eq.~\eqref{eq: Bloch theorem}. This also gives rise to a different normalization factor~$\mathcal{N}_\mathbf{k}$.
    \item While the hydrogenic wavefunctions proposed in~\cite{Hochberg:2016ntt} give overlap integrals in agreement with the TB theory, they differ significantly from more accurate RHF atomic wavefunctions of carbon atoms, as shown in App.~\ref{app: atomic wavefunctions}.
    \item After a careful evaluation of the formula for the DM~induced electron ejection rate, we find an extra factor of 1/2 coming from normalizing to the number of unit cells instead of the number of carbon atoms in the system.
\end{itemize}

In this appendix, we review how we improved the TB treatment of graphene wavefunctions and how the updated electron ejection rate compares to the one presented by Hochberg et al.
\footnote{We note that we were able to reproduce the signal energy spectra reported in~\cite{Hochberg:2016ntt} by accounting for all differences.}

\subsection{Bloch states and their normalization}
The ``Bloch states'' proposed in~\cite{Hochberg:2016ntt} are given by
\begin{subequations}
\label{eq: Bloch wave function Hochberg}
\begin{align}
    \Phi_{A\vec{k}}(\mathbf{x}) &= \varphi_A(\mathbf{x})\, ,\\ 
    \Phi_{B\vec{k}}(\mathbf{x}) &=\sum_{k=1}^3 e^{i \vec{k}\cdot \mathbf{N}_k} \varphi_B(\mathbf{x}-\mathbf{N}_k)\, .
\end{align}
\end{subequations}
Compared to the expression of Eq.~\eqref{eq: bloch wave function tb}, these states, while capturing the nearest-neighbour approximation, do not formally correspond to Bloch wavefunctions.
Consequently, they do not satisfy Bloch's theorem or rigorously describe a periodic system, see Eq.~\eqref{eq: Bloch theorem}.

One consequence of this choice of Bloch states is a deviating normalization factor~$\mathcal{N}_\mathbf{k}$.
As an example, using the Bloch states by Hochberg et al., it is possible to compute the \emph{exact} normalization factor for the~$\pi$-electrons,
\begin{align}
    \mathcal{N}_{\boldell} &=\left[2+s\sum_{k=1}^3\cos(\boldell\cdot \mathbf{N}_k+\varphi_{\boldell})+s^\prime \sum_{k=1}^3\cos(\boldell\cdot \mathbf{a}_k)\right]^{-1/2}\, .
\end{align}
While this factor correctly normalizes the electron wavefunctions involving the Bloch states of~\eqref{eq: Bloch wave function Hochberg}, it deviates from our respective expression for the $\pi$-electrons in Eq.~\eqref{eq: normalization constant pi}. 
Furthermore, our normalization constant is consistent with Eq.~\eqref{eq: wave function norm general}, i.e. the general expression for the normalization constant for Bloch wavefunctions as given by Eq.~\eqref{eq: bloch wave function tb}.
Similar arguments hold for the $\sigma$-electrons.

\subsection{Atomic wavefunctions}
In~\cite{Hochberg:2016ntt}, Hochberg et al. propose to describe the atomic wavefunctions of carbon by using hydrogenic wavefunctions with a re-scaled~$Z_\mathrm{eff}$ factor.
The re-scaling ensures that the overlap integrals of the wavefunctions are consistent with the TB parameters that reproduce the band structure of graphene listed in Tab.~\ref{tab: graphene parameter}.
For completeness, we list these wavefunctions in App.~\ref{app: hydrogenic wave functions}.

In contrast, we model the carbon wavefunctions using Roothaan-Hartree-Fock (RHF) wavefunctions~\cite{Bunge:1993jsz}, which we summarize in App.~\ref{app: RHF wfunctions}.
By comparison, we find that the re-scaled hydrogenic wavefunctions are a poor approximation for the required ground state wavefunctions of atomic electrons in carbon, as can be seen in Fig.~\ref{fig: carbon wavefunctions}.

While the RHF wavefunctions are a better description of carbon electrons, it should be noted that their overlap integrals are not consistent with the overlap parameters for graphene in Tab.~\ref{tab: graphene parameter}.
A re-scaling of the RHF wavefunctions similar to the approach by Hochberg et al. is possible, but spoils the accurate description of electrons in atomic carbon.
However, this seems to be a general feature of evaluating electron wavefunctions in the TB approximation.

\subsection{Electron ejection rate}

\begin{figure}[t!]
    \centering
    \includegraphics[width=0.45\textwidth]{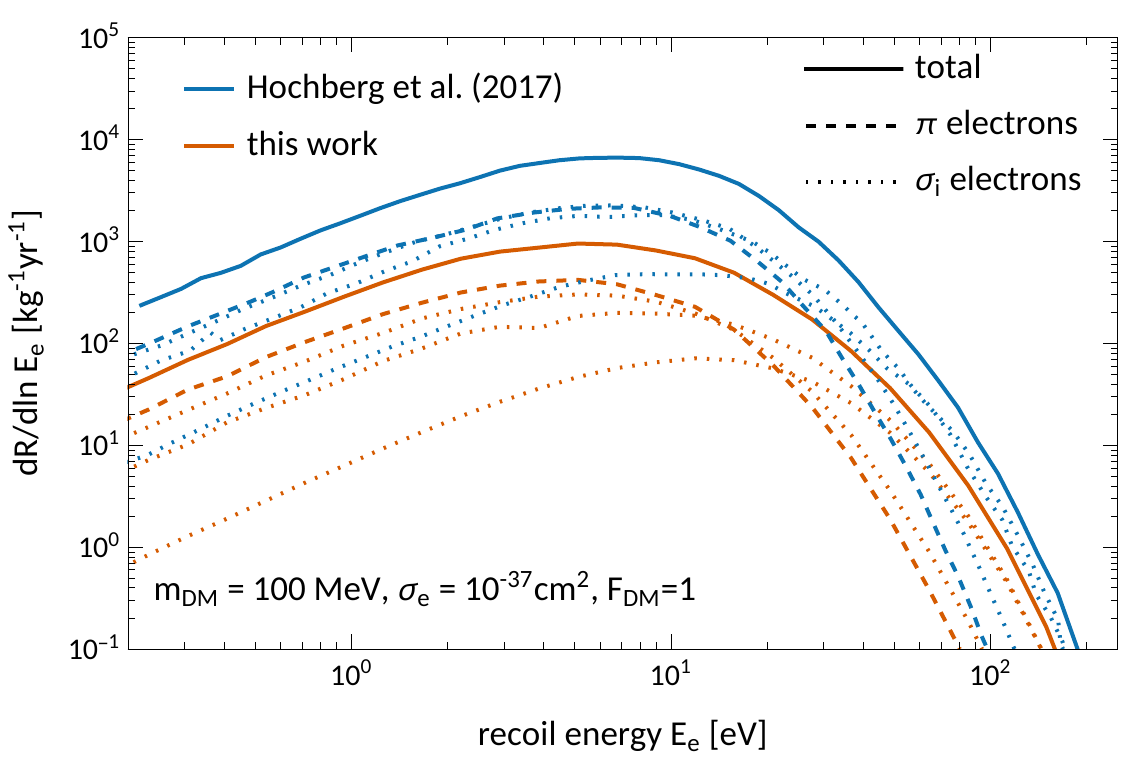}
    \caption{Comparison of the energy spectrum between our TB results and the results by Hochberg et al. (spectrum digitized from Fig. 1 of~\cite{Hochberg:2016ntt}). The solid lines show the total spectrum, whereas the dashed (dotted) lines show the contributions of the $\pi$- ($\sigma$-) electrons. For this figure, we used the SHM parameters from~\cite{Hochberg:2016ntt} in order to facilitate the comparison (Hochberg et al. use $v_0=220\text{ km/s}$ as opposed to our choice of $v_0=238\text{ km/s}$.)}
    \label{fig: energy spectrum comparison}
\end{figure}

In~\cite{Hochberg:2016ntt}, the total rate of DM-induced electron ejections in graphene is given as
\begin{align}
    R&=2 \frac{\rho_\chi}{m_\chi}N_C A_\mathrm{uc}\sum_i\int\frac{\dd^2\boldell}{(2\pi)^2}\int\dd^3\mathbf{v} g(\mathbf{v}) v\sigma_i(\boldell)\, ,
    \label{eq:rate Hochberg}
\intertext{where}
    v\sigma_i(\boldell) &=\frac{\bar{\sigma}_e}{\mu^2}\int \frac{\dd^3 \mathbf{k}_f}{(2 \pi)^{3}} \frac{\dd^3 \mathbf{q}}{4 \pi}\left|F_{\mathrm{DM}}(q)\right|^{2}\left|\widetilde{\Psi}_{i}\left(\boldell, \mathbf{q}-\mathbf{k}_{f}\right)\right|^{2} \nonumber\\
& \times \delta\left(\frac{k_{f}^{2}}{2 m_{e}}-E_{i}(\boldell)+\Phi+\frac{q^{2}}{2 m_{\chi}}-\mathbf{q} \cdot \mathbf{v}\right)\, .
\label{eq:sigmav Hochberg}
\end{align}
Here, we use the notation of Hochberg et al.
This needs to be compared to our Eqs.~\eqref{eq:rate_general} and~\eqref{eq: response function general}.
Here, we use the replacement Eq.~\eqref{eq:3Dto2D} for two-dimensional targets, and further identify
\begin{align}
R_\mathrm{free}\rightarrow \frac{16\pi m_e^2m_\chi^2}{\mu^2_{e\chi }} \bar{\sigma}_e |F_\mathrm{DM}(q)|^2
\end{align}
to facilitate the comparison. 
We find agreement between our expressions with one exception. Instead of the number of carbon atoms~$N_C$, we find that the electron ejection rate is proportional to the number of unit cells~$N_\mathrm{cell}$. Hence our expressions for the electron ejection rates differ by a factor of 2.

In Fig.~\ref{fig: energy spectrum comparison}, we compare the energy spectrum for a DM~particle of 100~MeV mass using the TB approach presented by Hochberg et al. and compare it to the improved version presented in this work. Note that the interaction model used by Hochberg et al. corresponds to~$\mathcal{O}_1$ interactions in our general framework.
We find that our predicted spectrum is about one order of magnitude lower than the one presented in~\cite{Hochberg:2016ntt}.

\bibliography{ref,ref2,bibliography,Nicola,references}
\end{document}